\documentclass{article}

\usepackage{amsmath,amsfonts,amssymb}
\usepackage{amsthm}
\usepackage{dsfont}

\usepackage{graphicx} 
\usepackage{float}
\usepackage{booktabs}

\usepackage{subcaption}

\newtheorem{theorem}{Theorem}

\usepackage{xcolor}
\usepackage[hidelinks]{hyperref}
\usepackage{bm}

\def\IP{\mathbb{P}}
\def\IR{\mathbb{R}}

\def\bF{\mathbf{F}}
\def\bV{\mathbf{V}}

\def\bb{\mathbf{b}}

\def\bx{\mathbf{x}}

\def\bz{\bm{z}}
\def\b0{\bm{0}}

\def\R{\mathcal{R}}

\hypersetup{
    colorlinks=true,
    linkcolor=blue, 
    citecolor=blue, 
    urlcolor=blue, 
}

\usepackage{tcolorbox}

\usepackage{tikz}
\usetikzlibrary{shapes,arrows}
\usetikzlibrary{positioning}
\tikzstyle{cloud} = [draw,
    ellipse,
    fill=red!20, 
    node distance=0.87cm,
    minimum height=2em]
\tikzstyle{line} = [draw, -latex']
\usetikzlibrary{shapes.symbols,shapes.callouts,patterns}
\usetikzlibrary{calc}
\usetikzlibrary{arrows.meta}
\definecolor{MyDarkRed}{RGB}{139, 0, 0}
\definecolor{MyDarkGreen}{RGB}{0, 100, 0}
\usepackage{pgfplots}
\pgfplotsset{compat=1.18}

\def\R{\mathcal{R}}

\usepackage[resetlabels]{multibib}
\newcites{app}{Supplementary material references}

\usepackage{geometry}
\usepackage{authblk}
\usepackage{orcidlink}

\title{Introduced cases, spread of infection in a community and travel control measures}
\author[1,*]{Julien Arino\,\orcidlink{0000-0001-6409-5027}}
\author[1]{Clotilde Djuikem\,\orcidlink{0000-0003-4815-743X}}
\author[2]{Amy Hurford\,\orcidlink{0000-0001-6461-5686}}
\affil[1]{Department of Mathematics, University of Manitoba, Winnipeg, Manitoba, Canada}
\affil[2]{Departments of Biology and Mathematics \& Statistics, Memorial University of Newfoundland and Labrador, St John's, Newfoundland and Labrador, Canada}
\affil[*]{Corresponding author: \href{mailto:julien.arino@umanitoba.ca}{julien.arino@umanitoba.ca}}
\date{\today}

\begin{document}


\newpage
\setcounter{page}{1}
\maketitle

\begin{abstract}
Travel restrictions, border screening and quarantine were widely used during the COVID-19 pandemic to prevent or delay the spatial spread of SARS-CoV-2, yet their effectiveness varied greatly across jurisdictions.
To investigate how the timing and combination of border control measures interact with pathogen introductions and local transmission, we formulate a coupled Exporter--Importer epidemic model in both deterministic (ordinary differential equations) and stochastic (continuous-time Markov chain and multitype branching process approximation) frameworks.
The model distinguishes latently infected, detected infectious and undetected infectious individuals, and explicitly separates recently arrived travellers in an introduction layer from the destination community.
We derive analytical expressions for the control and basic reproduction numbers, final epidemic sizes and probabilities of outbreak extinction, and complement them with stochastic simulations.
We show that pre-departure screening in the Exporter alone cannot prevent outbreaks in an Importer community with reproduction number larger than unity because latently infected travellers cross borders undetected.
Post-arrival quarantine and screening in the introduction layer, as well as travel interruptions, can substantially reduce the attack rate in the Importer, but only when implemented with high stringency and triggered early.
Triggering border interventions based on local detections in the Importer almost always acts too late because silent introductions precede local case detection, whereas triggers informed by surveillance in the Exporter provide a wider window of action.
Effective protection of the Importer ultimately requires combining proactive, Exporter-informed border measures with local public health interventions that keep the community reproduction number near or below one.

\end{abstract}

\section{Introduction}
During the recent COVID-19 pandemic, governments the world over took public health measures to curtail the spread of pathogens.
These measure ranged in consequences from relatively benign to society-altering, for instance with the imposition of stay-at-home orders for entire countries.
Among the myriad of measures used, several aimed to curtail the spatio-temporal spread of the pathogen, namely, travel interruptions, enhanced border control and quarantine.
However, the efficacy of these measures is debatable and was indeed debated \cite{GrepinHoLiEtAl2020}.

In a few instances, travel control measures were used successfully. 
In Canada, Newfoundland and Labrador required that non-residents submit travel declaration forms and have a plan to self-isolate for 14 days immediately after arrival \cite{MohammadiCojocaruArinoHurford2023}; these public health measures resulted in some of the lowest incidence rates in Canada \cite{HurfordMartignoniLoredoOstiAnokyeEtAl2022}.
Likewise, Nunavut imposed a two-weeks pre-travel quarantine, which had to be spent in airport hotels in three Canadian cities (Edmonton, Ottawa and Winnipeg) prior to travel to the territory.
As a consequence, incidences were also low in Nunavut, with no COVID-19 cases reported in Nunavut until November 2020, nine months after the COVID-19 pandemic began \cite{VilchesAbdollahiCiprianoEtAl2022}.
Manitoba banned non-essential travel to areas of the province north of 53 degrees north except for residents \cite{CarinoQuillEtAl2022,Manitoba2020NorthernOrder}, while Yukon restricted non-essential travel into the territory, strictly controlled its few road and air entry points and required a mandatory 14-day self-isolation upon arrival \cite{McPheeKnowlesHoffmanKanary}. 
Prince Edward Island did not experience any substantial community outbreaks prior to June 2021 \cite{HurfordMartignoniLoredoOstiAnokyeEtAl2022}, but for travellers arriving by road, the island can only be accessed by one bridge.
While not physically barricaded to essential supply chains, this bridge functioned as a heavily controlled checkpoint where non-essential travel from outside the province was strictly forbidden or subjected to a mandatory 14-day quarantine \cite{DaSilva2023,NAO2020PEI}.

For all these Canadian provinces and territories, travel restriction measures worked relatively well. 
However, all these locations are characterised by a large degree of isolation of the communities, with limited and easily controllable road access.
Some nations, many of which are islands, also had similar results, for example: China, Hong Kong, Taiwan, Vietnam, Iceland, Australia, New Zealand, Singapore, Bhutan, South Korea and 11 of 25 Pacific Island countries and territories that had not reported a single case of COVID-19 by 15 February 2021 \cite{gay2021pacific,sun2023covid,WeberKopfVaha,wikipediaCOVDzero}. 
Tuvalu did not report a COVID-19 case in border quarantine until 20 May 2022, well after vaccination of the population had occurred \cite{who2022tuvalu}.
So there is clearly a case to be made for travel control measures.

However, these measures are disruptive and should be used wisely and with discrimination.
Prior to the COVID-19 pandemic, the International Health Regulations (2005) stated in Article 43 \cite{WHO-IHR-2005} that ``Regulations shall not preclude States Parties from implementing health measures, in accordance with their relevant national law and obligations under international law, in response to specific public health risks or public health emergencies of international concern'' with the proviso that ``such measures shall not be more restrictive of international traffic and not more invasive or intrusive to persons than reasonably available alternatives that would achieve the appropriate level of health protection''.
Amendments to the IHR were adopted in 2024, but they do not affect this article \cite{LeePiper2025}.
In practice, most nations or subnational jurisdictions that implemented travel control measures saw very little success, as these piecemeal bans were often implemented too late and failed to stop the introduction of new variants \cite{grepin2023effectiveness, habibi2020do}.

Yet travel control measures were taken time and time again, for instance when new variants of concern emerged.
In the latter case, countries would take country specific measures; for instance, Canada banned direct flights from India and Pakistan for 30 days in April 2021 \cite{GovernmentofCanadaCBSA}.
In view of this, when travel control measures were useful, they fall under Article 43, but in most instances, they were contrary to this recommendation since they hindered traffic without achieving much protection.

It is however evident that the ``spatialisation'' of the spread of a pathogen relies on its introduction into locations from which it is heretofore absent.
These introductions rely on the movement of individuals harbouring the pathogen.
So, naively, slowing or forbidding such movement slows or interdicts introductions.
Fittingly, many of the measures used during the COVID-19 pandemic date back to the plagues of the Middle Ages: quarantines, travel certificates and plague walls.
However, preventing \emph{all} introductions is, for all intents and purposes, \emph{impossible}.
Even in a world that was much less connected than it is today, movement suppression measures have failed repeatedly; a well-known example is the 27 kilometres long ``Mur de la peste'' (plague wall) built in 1721 to contain the spread of the 1720 plague outbreak that was ravaging Marseille \cite{ermus2023great}.
Other control measures dating back to the plagues of the Middle Ages, such as quarantine or health certificates, seem to have been more efficacious \cite{kosciejew2023documentary}.

So border control measures sometime work, sometime do not.
They are, in any event, highly disruptive.
Our aim here is to gain a better understanding of the pathogen introduction process, in particular in relation to travel control measures.
In previous work \cite{MohammadiCojocaruArinoHurford2023}, some of the authors used statistical models to consider the risk of importation into Newfoundland and Labrador.
There is a need to connect these importation rates with the risk and expected characteristics of an outbreak occurring in destination population to understand how travel measures combine with other public health measures  \cite{HurfordMartignoniLoredoOstiAnokyeEtAl2022} and inform best combinations of measures to manage a pandemic \cite{MartignoniArinoHurford2024}. 
One of the coauthors also investigated issues related to case introductions in several papers \cite{ArinoBoelleMillikenPortet2021,ArinoBajeuxPortetWatmough2020}, while the issue of the duration of the stochastic phase following an introduction was considered in \cite{ArinoMilliken2022b}.
Other work on the subject includes \cite{MilwidGabrieleRivetOgdenEtAl2024}, who modelled the number of importations arriving at the four Canadian airports that accepted international arrivals in 2021: Toronto Pearson, Montreal Trudeau, Vancouver International and Calgary International.

Here, we focus in on the efficacy of travel control measures in relation to timing of these measures compared to that of an outbreak stemming from introductions.
This paper is organised as follows.
In Section~\ref{sec:model}, we present the introduction model, in both deterministic (ODE) and stochastic (CTMC) forms; we also introduce the measures used to study outbreaks.
Properties of the components of the model are then investigated both mathematically and computationally; in Section~\ref{sec:results-exporter}, we consider the Exporter; Section~\ref{sec:results-importer} is devoted to the Importer when considered in isolation; the full model is studied in Section~\ref{sec:results-exporter-importer}.
The effect of travel control measures, i.e., the main focus of the paper, is considered in Section~\ref{sec:travel-control}.
A Discussion concludes the work. 
Computations and proofs are grouped in appendices.

\section{The mathematical model(s)}
\label{sec:model}

The model framework is a variation on a model in \cite{ArinoBoelleMillikenPortet2021}.
It is a metapopulation-type model considering two locations, the \emph{Exporter} and the \emph{Importer}, with movement only occurring from the Exporter to the Importer.
In Section~\ref{sec:model-base}, we describe the model in each location.
We then specify how introductions happen and formulate the ordinary differential equations version of the model in Section~\ref{sec:model-introductions}.
The continuous time Markov chain equivalent to the model and its multitype branching process approximation are presented in Sections~\ref{sec:model-CTMC} and \ref{sec:model-MBPA}, respectively.

\subsection{The base model in locations}
\label{sec:model-base}

In both locations, the base models used are a variation on the SLIAR model of \cite{ArinoBrauerPvdDWatmoughWu2006}.
Instead of focusing on the nature of infectious cases (symptomatic or asymptomatic) as in \cite{ArinoBrauerPvdDWatmoughWu2006}, the model considers \emph{detected} and \emph{undetected} infectious individuals.

The numbers of individuals susceptible, latently infected, infectious and detected, infectious and not detected and recovered/removed are denoted $S(t)$, $L(t)$, $D(t)$, $U(t)$ and $R(t)$, respectively.
(Time dependence is omitted unless required.)
Upon contact with $D$ and $U$ individuals, susceptible individuals become infected at the rate $\beta S(U+\xi D)$, first going through a period of latency.
The parameter $\xi$ is typically assumed to be less than 1, since public health recommendations were for detected individuals to stay home, wear masks, etc., lowering their infectiousness to others.

The latent period has an exponentially distributed duration with mean $1/\varepsilon$ (progression rate $\varepsilon$).
At the end of their latency period, a proportion $p$ of individuals progress to the detected compartment, with the remaining $1-p$ moving to the undetected compartment.
The detection parameter $p$ is critical to the system.
Detection can occur at any stage during the infection and recovery/removal process, but the model assumes that, except in the case of pre-departure screening (see Section~\ref{sec:model-introductions}), the ``decision'' happens when individuals finish their latent period.
So, in essence, $p$ aggregates several ``sub-$p$'' that encode for flows from, for instance, from $U$ to $D$.
The absence of such flows means that the dynamics of $U$ is a little different from what it would be were such flows present.
Namely, we would expect that the detection rate on exit from the latent stage would be smaller than $p$, leading to a higher flow into $U$, followed by a higher outflow rate out of $U$ than what is used here because of additional flows.

After spending an average $1/\gamma$ time units infectious, individuals move to the recovered compartments $R$, provided they survive death due to the disease, which occurs for a proportion $\delta$ of individuals leaving the $D$ and $U$ compartments.

\subsection{Modelling mobility-induced introductions}
\label{sec:model-introductions}
\begin{figure}[htbp]
    \centering
    \def\hstep{*3}
    \begin{tikzpicture}[scale=0.8, 
    every node/.style={transform shape},
    auto,
    box/.style={minimum width={width("N-1")+2pt},
    draw, rectangle}]
    \node [box] at (0\hstep,3) (SE) {$S_E$};
    \node [box] at (1\hstep,3) (LE) {$L_E$};
    \node [box] at (2\hstep,4.5) (DE) {$D_E$};
    \node [box] at (2\hstep,1.5) (UE) {$U_E$};
    \node [box] at (3\hstep,3) (RE) {$R_E$};
    \node [box, fill=blue!50] at (1\hstep,9) (LI) {$L_{I}$};
    \node [box, fill=red!50] at (2\hstep,7.5) (DI) {$D_{I}$};
    \node [box, fill=blue!50] at (2\hstep,10.5) (UI) {$U_{I}$};
    \node [box] at (0\hstep,13.5) (SC) {$S_C$};
    \node [box] at (1\hstep,13.5) (LC) {$L_{C}$};
    \node [box] at (2\hstep,15) (DC) {$D_{C}$};
    \node [box] at (2\hstep,12) (UXC) {$U_{C}$};
    \node [box] at (3\hstep,13.5) (RC) {$R_{C}$};
    \node[rotate=90] at (-1,2.5) {\Large\textbf{Exporter}};
    \node[rotate=90] at (-1,11.25) {\Large\textbf{Importer}};
    \draw [black,rounded corners] (-0.65,1) -- (-0.65,5) -- (3\hstep+0.7,5) -- (3\hstep+0.7,1) -- cycle;
    \draw [black,rounded corners] (-0.65,6.9) -- (-0.65,16) -- (3\hstep+0.7,16) -- (3\hstep+0.7,6.9) -- cycle;
    \draw [black,rounded corners, dotted] (-0.55,11.25) -- (3\hstep+0.6,11.25) -- (3\hstep+0.6,7.05) -- (-0.55,7.05)  -- cycle node at (2.75, 11) {\textbf{Introduction layer}};
    \draw [black,rounded corners, dotted] (-0.55,11.25) -- (3\hstep+0.6,11.25) -- (3\hstep+0.6,15.9) -- (-0.55,15.9)  -- cycle node at (2.2, 11.5) {\textbf{Community}};
    \path [line,thick] (SE) to node [above, sloped] (TextNode) {$\Phi_E S_E$} (LE);
    \path [line,thick] (LE) to node [midway,above,sloped] (TextNode) {$p_E\varepsilon_E L_E$} (DE);
    \path [line,thick] (LE) to node [midway,above,sloped] (TextNode) {$(1-p_E)\varepsilon_E L_E$} (UE);
    \path [line,thin,dotted] (UE) to node [midway,above,sloped] (TextNode) {$\sigma_Em_UU_E$} (DE);
    \path [line,thick] (UE) to node [midway,above,sloped] (TextNode) {$(1-\delta_E)\gamma_E U_E$} (RE);
    \path [line,thick] (DE) to node [midway,above,sloped] (TextNode) {$(1-\delta_E)\gamma_E D_E$}  (RE);
    \path [line,dashed] (LE) to node [midway, above, sloped] (TextNode) {$m_L L_E$} (LI);
    \path [line,dashed] (DE) to node [midway, above, sloped] (TextNode) {$m_D D_E$} (DI);
    \path [line,dashed, bend right=20] (UE) to node [midway, below, sloped] (TextNode) {$(1-\sigma_E)m_U U_E$} (UI);
    \path [line,thick] (LI) to node [midway,below,sloped] (TextNode) {$p_{I}\varepsilon_I L_{I}$} (DI);
    \path [line,thick] (LI) to node [midway,above,sloped] (TextNode) {$(1-p_{I})\varepsilon_I L_{I}$} (UI);
    \path [line,thick] (DI) -- (3\hstep,7.5) -- node [midway, above,sloped] (TextNode) {$(1-\delta_I)\gamma_I D_{I}$} (RC);
    \path [line,thick] (UI) to node [midway,below,sloped] (TextNode) {$(1-\delta_I)\gamma_I U_{I}$} (RC);
    \path [line,thick] (SC) to node [midway, above, sloped] (TextNode) {$\Phi_C S_C$} (LC);
    \path [line,thick] (LC) to node [midway, above, sloped] (TextNode) {$p_C\varepsilon_C L_C$} (DC);
    \path [line,thick] (LC) to node [midway,above,sloped] (TextNode) {$(1-p_C)\varepsilon_C L_{C}$} (UXC);
    \path [line,thick] (DC) to node [midway, above,sloped] (TextNode) {$(1-\delta_C)\gamma_C D_{C}$} (RC);
    \path [line,thick] (UXC) to node [midway, above,sloped] (TextNode) {$(1-\delta_C)\gamma_C U_{C}$} (RC);
    \end{tikzpicture}
    \caption{The Exporter-Importer model.
    Blue compartments are those on which quarantine can have an effect, while red ones would be absent in case of pre-travel screening.
    Dashed arrows are flows that are important to introductions.}
    \label{fig:base_model}
\end{figure}
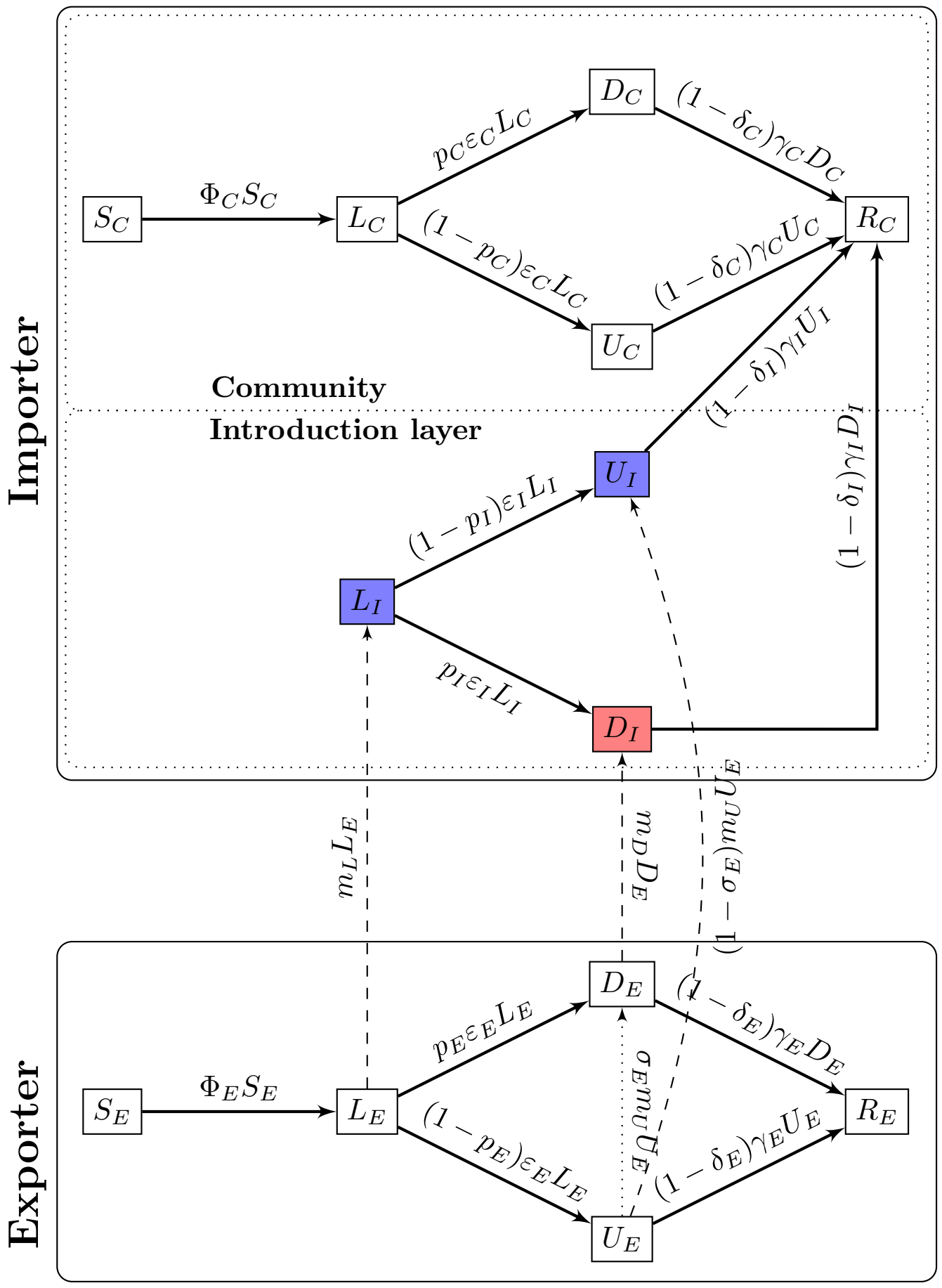

Spatial structure is incorporated as in \cite{ArinoBoelleMillikenPortet2021} by using the simplified metapopulation structure shown in Figure~\ref{fig:base_model}.
There are two locations, the Exporter and the Importer; movement is one directional from the Exporter to the Importer because we focus on the initial transition of the infectious pathogen from one location to another without any ``interference'' from other processes, as would be the case in a similar model with bi-directional movement.
The Importer is further subdivided between the Community and a so-called Introduction layer; see details below.

In the Exporter, the index $E$ is used for state variables. 
Individuals in compartments $X=\{L_E,D_E,U_E\}$ travel from the Exporter to the Importer at the \emph{per capita} rates $m_{X}$, which could be time- or state-dependent but are typically constant here.
One exception is $m_D$: during the early stages of the COVID-19 pandemic, $m_D$ was typically zero, as detected individuals were supposed to stay home and were not allowed to board flights.
However, when considering smaller jurisdictions such as towns, such movement was possible.

In practice, the only flows between the Exporter and the Importer that we consider are those related to infected individuals, i.e., $L_E$, $D_E$ and $U_E$.
In principle, we should also consider flows of $S_E$ and $R_E$ individuals, but these flows 
\begin{itemize}
    \item do not play a role in the mathematical computations in the paper (whether they are present or not, they play no role in the reproduction numbers, for instance); 
    \item disrupt the computational analysis, since they greatly decrease the susceptible population in the exporter (reducing the outbreaks there) and increase the number of infected individuals in the importer (increasing the severity of outbreaks there).
\end{itemize}
By not considering movement of the susceptible and immune populations, we are in effect assuming that when such movements happen, they are compensated by reciproqual movement from the Importer, i.e., that the system is at a demographic equilibrium.

The force of infection $\Phi_E$ in the Exporter takes the form
\begin{equation}\label{eq:force-infection-SE}
    \Phi_E=\beta_E(\xi_ED_E+U_E).
\end{equation}

The model incorporates pre-departure screening. 
Because the detection parameter $p_E$ amalgamates characteristics of the test and screening effort in the Exporter, we use a separate parameter $\sigma_E$.
Similarly to $p_E$, $\sigma_E\in[0,1]$.
When $\sigma_E$ is positive, i.e., when pre-departure screening is active, then it is assumed that $\sigma_E\geq p_E$: contrary to $p_E$, here it is only the characteristics of the test that apply.
Undetected $U_E$ individuals ``caught'' by pre-departure screening are not allowed to travel and thus enter the $D_E$ compartment.

In the Importer, we consider two subpopulations: the \emph{introduction} layer (called \emph{importation} layer in \cite{ArinoBoelleMillikenPortet2021}) and the \emph{community}.
In a mechanism reminiscent of resident-travellers (or movers-stayers) models \cite{ArinoVdD2003a}, although individuals in the introduction layer and the community mix and interact together, their membership to either group is preserved, albeit temporarily for those in the introduction layer.

When infected individuals arrive in the Importer from the Exporter, they are ``singled out'' by being placed in the introduction layer, with numbers there denoted $L_I$, $D_I$ and $U_I$. 
In these compartments, they interact with members of the community, potentially leading to new infections when those members are susceptible $S_C$.
They follow the same infected disease life history as individuals in other compartments, except that recovery is into the community recovered compartment $R_C$, since at that point, their disease status is of no consequence to the community.
This ``accounting trick'' allows to consider the precise break up between introductions and community generated cases. 

The force of infection acting on susceptible individuals $S_C$ in the community takes the form
\begin{equation}
    \Phi_C = 
    \beta_{I}(\xi_{I}D_I+U_I)
    +\beta_{C}(\xi_{C}D_C+U_C).
\end{equation}
Here, $\beta_{X}$, $X\in\{I,C\}$, is the transmission coefficient when susceptible individuals $S_C$ meet infectious individuals from group $X$ and $\xi_X$ is the reduction of infectiousness due to awareness of one's disease status.

\subsection{Ordinary differential equations model}
\label{sec:model-ODE}
The model takes the following form in the exporter,
\begin{subequations}
\label{sys:general-model}
\begin{align}
    S_E' &= -\Phi_ES_E 
    \label{sys:general-model-dSE} \\
    L_E' &= \Phi_ES_E-(\varepsilon_E+m_L)L_E
    \label{sys:general-model-dLE} \\
    D_E' &= p_E\varepsilon_EL_E+\sigma_Em_UU_E-(\gamma_E+m_D)D_E
    \label{sys:general-model-dDE} \\
    U_E' &= (1-p_E)\varepsilon_EL_E-(\gamma_E+m_U)U_E
    \label{sys:general-model-dUE} \\
    R_E' &= (1-\delta_E)\gamma_E(D_E+U_E),
    \label{sys:general-model-dRE}
    \intertext{while the dynamics in the introduction layer is governed by}
    L_I' &= m_LL_E-\varepsilon_IL_I 
    \label{sys:general-model-dLI} \\
    D_I' &= m_DD_E+p_I\varepsilon_IL_I
    -\gamma_ID_I
    \label{sys:general-model-dDI} \\
    U_I' &= (1-\sigma_E)m_UU_E+(1-p_I)\varepsilon_IL_I
    -\gamma_IU_I
    \label{sys:general-model-dUI} 
    \intertext{and that of the community is driven by}
    S_C' &= -\Phi_CS_C 
    \label{sys:general-model-dSC} \\
    L_C' &= \Phi_CS_C-\varepsilon_CL_C
    \label{sys:general-model-dLC} \\
    D_C' &= p_C\varepsilon_CL_C-\gamma_CD_C
    \label{sys:general-model-dDC} \\
    U_C' &= (1-p_C)\varepsilon_CL_C
    -\gamma_CU_C
    \label{sys:general-model-dUC} \\
    R_{C}' &= (1-\delta_C)\gamma_{C}(D_C+U_C)
    +(1-\delta_I)\gamma_I(D_I+U_I).
    \label{sys:general-model-dRC}
\end{align}
\end{subequations}
System~\eqref{sys:general-model} is considered with nonnegative initial conditions. 
To avoid a trivial case, we further assume that $L_E(0)+D_E(0)+U_E(0)>0$.

Properties of \eqref{sys:general-model} are detailed later, but before proceeding further, let us note here that \eqref{sys:general-model} is an \emph{epidemic} model.
As such, all infected variables ($L_E$, $D_E$, $U_E$, $L_I$, $D_I$, $U_I$, $L_C$, $D_C$ and $U_C$) invariably go to zero as $t\to\infty$ and the alternatives distinguished by the basic reproduction number(s) are whether an epidemic takes place, which is characterised by a temporary increase in infections, or not.

\subsection{Continuous time Markov chain model}
\label{sec:model-CTMC}

\begin{table}[htbp]
\centering
\begin{tabular}{lll}
\toprule
Event & Transition   & Transition rates \\
\midrule
Infection of $S_E$ by $D_E$ &  $\to(S_E-1,L_E+1)$ & $\beta_E \xi_E S_E D_E$\\
Infection of $S_E$ by $U_E$ &  $\to(S_E-1,L_E+1)$ & $\beta_E S_E U_E$\\
End of incubation in $L_E$ (detected) & $\to(L_E-1,D_E+1)$ & $p_E\varepsilon_E L_E  $ \\
End of incubation in $L_E$ & $\to(L_E-1,U_E+1)$ & $(1-p_E)\varepsilon_E L_E  $ \\
Detection outbound $U_E$ & $\to(U_E-1,D_E+1)$ & $\sigma_Em_U U_E  $ \\
Recovery in  $D_E$ & $\to(D_E-1,R_E+1)$ & $(1-\delta_E)\gamma_E D_E  $ \\
Recovery in  $U_E$ & $\to(U_E-1,R_E+1)$ & $(1-\delta_E)\gamma_E U_E  $ \\
Death due to disease of $D_E$ & $\to(D_E-1)$ & $\delta_E\gamma_E D_E  $ \\
Death due to disease of $U_E$ & $\to(U_E-1)$ & $\delta_E\gamma_E U_E  $ \\
\midrule
Movement of $L_E$ & $\to(L_E-1,L_I+1)$ & $m_L L_E  $\\
Movement of $D_E$ & $\to(D_E-1,D_I+1)$ & $m_D D_E  $\\
Movement of $U_E$ & $\to(U_E-1,U_I+1)$ & $(1-\sigma_E)m_U U_E  $\\
\midrule
End of incubation in  $L_I$ (detected) & $\to(L_I-1,D_I+1)$ & $p_I\varepsilon_I L_I  $ \\
End of incubation in  $L_I$ & $\to(L_I-1,U_I+1)$ & $(1-p_I)\varepsilon_I L_I  $ \\
Recovery in  $D_I$ & $\to(D_I-1,R_C+1)$ & $(1-\delta_I)\gamma_I D_I  $ \\
Recovery in  $U_I$ & $\to(U_I-1,R_C+1)$ & $(1-\delta_I)\gamma_I U_I  $ \\
Death due to disease of $D_I$ & $\to(D_I-1)$ & $\delta_I\gamma_I D_I  $ \\
Death due to disease of $U_I$ & $\to(U_I-1)$ & $\delta_I\gamma_I U_I  $ \\
\midrule
Infection of $S_C$ by $D_C$ &  $\to(S_C-1,L_C+1)$ & $\beta_{C} \xi_{C} S_C D_C$\\
Infection of $S_C$ by $U_C$ &  $\to(S_C-1,L_C+1)$ & $\beta_{C} S_C U_C$\\

Infection of $S_C$ by $D_I$ &  $\to(S_C-1,L_C+1)$ & $\beta_{I} \xi_{I} S_C D_I$\\
Infection of $S_C$ by $U_I$ &  $\to(S_C-1,L_C+1)$ & $\beta_{I} S_C U_I$\\
End of incubation in  $L_C$ (detected) & $\to(L_C-1,D_C+1)$ & $p_C\varepsilon_C L_C  $ \\
End of incubation in  $L_C$ & $\to(L_C-1,U_C+1)$ & $(1-p_C)\varepsilon_C L_C  $ \\
Recovery in  $D_C$ & $\to(D_C-1,R_C+1)$ & $(1-\delta_C)\gamma_C D_C  $ \\
Recovery in  $U_C$ & $\to(U_C-1,R_C+1)$ & $(1-\delta_C)\gamma_C U_C  $ \\
Death due to disease of $D_C$ & $\to(D_C-1)$ & $\delta_C\gamma_C D_C  $ \\
Death due to disease of $U_C$ & $\to(U_C-1)$ & $\delta_C\gamma_C U_C  $ \\
\bottomrule
\end{tabular}
\caption{Reaction rates for the Exporter-Importer CTMC model with introduction layer.}
\label{tab:transition-prop}
\end{table}

Let $T$ denote the set of all finite-dimensional vectors whose components are nonnegative integers. 
The CTMC model related to the deterministic model~\eqref{sys:general-model} of Exporter-Importer is given by
\begin{equation}\label{eq:CTMC}
    \mathbf{X}_t=(S_E,S_C,L_E, L_I, L_C, D_E, D_I, D_C, U_E, U_I, U_C,R_E,R_C), \; t\in \mathbb{R}_+,
\end{equation}
and is characterised by transition probabilities from state \(\mathbf{k}\) to \(\mathbf{j}\neq\mathbf{k}\),

\begin{equation}\label{eq:transition-pro}
    \mathbb{P}(\mathbf{X}(t+\Delta t)=\mathbf{j}|\mathbf{X}(t)=\mathbf{k})=\sigma(\mathbf{k},\mathbf{j})\Delta t + o(\Delta t)
\end{equation}
with transition rates \(\sigma(\mathbf{k},\mathbf{j})\) given in Table~\ref{tab:transition-prop}, in which the resulting state $\mathbf{j}$ is shown.

\subsection{Multitype branching process approximation of the CTMC}
\label{sec:model-MBPA}

To analytically approximate the CTMC $\mathbf{X}_t$ near the disease-free equilibrium of the deterministic system \eqref{sys:general-model}, we use a multitype branching process approximation (MBPA). 
The full formulation of the probability generating functions (p.g.f.) and the backward Kolmogorov system governing this theoretical threshold behaviour is detailed in Appendix~\ref{app:CTMC-MBPA}.

\subsection{Characterisation of epidemic events and outbreaks}
\label{sec:model-types-epidemic}
In stochastic realisations of the CTMC, we distinguish between the \emph{underlying} (or \emph{unobservable}) epidemic across all infected compartments and the \emph{observable} epidemic seen by public health authorities through the detected compartments $D_E$, $D_I$ and $D_C$.
To delineate individual outbreaks in time, we set a threshold $\tau$ for the maximum time allowed between two successive monitored events for them to belong to the same outbreak; whenever two successive events are separated by more than $\tau$ time units, a new outbreak begins.
Depending on which CTMC transitions are monitored, this defines three types of outbreaks: \emph{observable outbreaks} (monitoring only detection transitions $L\to D$, plus $D_E\to D_I$ in the Importer), \emph{unobservable outbreaks} (monitoring all case activations $L\to\{D,U\}$, plus $D_E\to D_I$ and $U_E\to U_I$ in the Importer) and \emph{infection outbreaks} (monitoring new infection transitions $S\to L$).
When detection is imperfect ($p<1$), gaps greater than $\tau$ can occur between successive detected cases even while underlying transmission continues uninterrupted, so a single unobservable outbreak can fragment into multiple observable outbreaks.
Full definitions and an illustration (Figure~\ref{fig:inter-outbreak-times}) are provided in Appendix~\ref{app:characterisation-outbreaks}.
In practice, we set $\tau = 14$~days in our numerical simulations, as was used for instance by the Public Health Agency of Canada \cite{PHAC2020CaseContact}, based on evidence from \cite{LauerGrantzBiEtAl2020} estimating that 99\% of symptomatic SARS-CoV-2 infections develop symptoms within 14~days of exposure.

\section{Results in the Exporter}
\label{sec:results-exporter}
Although this is not the main focus of the work, understanding the behaviour of the disease in the Exporter is useful, in particular because it allows us to demonstrate the use of the different tools we use to characterise outbreaks in an otherwise well understood context.

\subsection{Deterministic considerations (Exporter)}
\label{sec:results-exporter-deterministic}
The deterministic model for the Exporter consists in \eqref{sys:general-model-dSE}--\eqref{sys:general-model-dRE} considered without any change.
First, we find the control reproduction number for the Exporter component to be
\begin{equation}\label{eq:RcE}
\R_c^E = 
\frac{\varepsilon_E}{\varepsilon_E+m_L}
\left(
\frac{\xi_E p_E}{\gamma_E+m_D}
+
\frac{1-p_E}{\gamma_E+m_U}
\left[1+\frac{\xi_E\sigma_E m_U}{\gamma_E+m_D}\right]
\right)
\beta_ES_E(0),
\end{equation}
where the term \emph{control} refers to the presence of exit screening.
Details of the computation are given in Appendix~\ref{app:R0-and-final-sizes-exporter}.
The factor $\varepsilon_E/(\varepsilon_E+m_L)$ represents the probability that a latent individual in $L_E$ progresses to an infectious stage before leaving the Exporter.
With probability $p_E$, they enter the detected compartment $D_E$, where they have relative infectiousness $\xi_E$ and remain in the Exporter for a mean time $1/(\gamma_E+m_D)$.
With probability $1-p_E$, they enter the undetected compartment $U_E$, where they spend a mean time $1/(\gamma_E+m_U)$; a fraction $\sigma_E m_U/(\gamma_E+m_U)$ of these individuals are detected by pre-departure screening when attempting to travel and transferred into $D_E$, where they remain in the Exporter for an additional mean time $1/(\gamma_E+m_D)$ with relative infectiousness $\xi_E$.
In the absence of pre-departure screening ($\sigma_E=0$), we obtain as a special case the basic reproduction number for the Exporter,
\begin{equation}\label{eq:R0E}
\R_0^E = \left.\R_c^E\right|_{\sigma_E=0} =
\frac{\varepsilon_E}{\varepsilon_E+m_L}
\left(
\frac{\xi_E p_E}{\gamma_E+m_D}
+
\frac{1-p_E}{\gamma_E+m_U}
\right)
\beta_ES_E(0).
\end{equation}

Because there is no flow back from the Importer to the Exporter, the values \eqref{eq:RcE} and \eqref{eq:R0E} derived here are independent of the rest of the system and can be taken ``as is''.
That is, one does not have to assume that the Exporter be isolated for these values to be valid, as is often the case in general metapopulations, where there is a location-specific and a general reproduction number, with the former computed by setting movement rates to zero.
It is useful nonetheless to consider $\left.\R_0^E\right|_{m_X=0}$, i.e., the reproduction number in isolation.

\begin{figure}[htbp]
    \centering
    \includegraphics[width=0.6\linewidth]{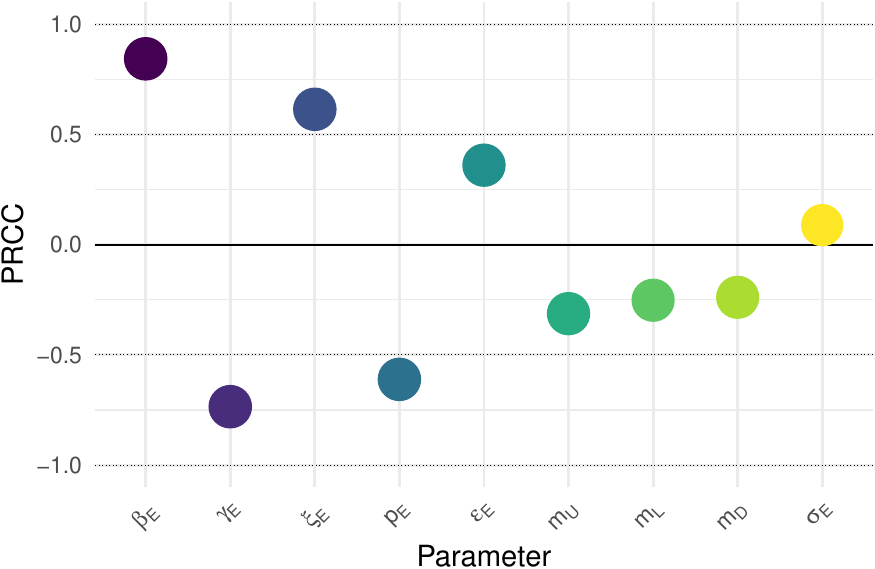}
    \caption{Partial rank correlation coefficients (PRCC) sensitivity analysis of the control reproduction number $\R_c^E$ in the Exporter.}
    \label{fig:sensitivity-R0E}
\end{figure}

In terms of effect of parameters on $\R_c^E$, a global sensitivity analysis is shown in Figure~\ref{fig:sensitivity-R0E}.
Note that this figure is conditioned on obtaining values of $\R_c^E\in[0.8,5]$, i.e., we require parameters used in the sensitivity analysis to yield reasonable $\R_c^E$ values in order to be considered.
Here and for later similar analyses, parameter space is sampled using a Sobol method \cite{iooss2026sensitivity,sobol2001global}. 
This leads to requiring a much finer sampling than is actually used. 
For instance, in Figure~\ref{fig:sensitivity-R0E}, partial rank correlation coefficients are computed using 131,384 samples, but this required generating a Sobol sample with 200,000 points, with the remaining points discarded because they led to unreasonable $\R_c^E$ values.

Upon inspection of Figure~\ref{fig:sensitivity-R0E}, we observe that, as could be expected because of the linear relation linking them, $\beta_E$ influences $\R_c^E$ the most.
Also, most dependencies are as usual in this type of model.
We also note that the rates $m_L$, $m_U$ and $m_D$ at which individuals leave the Exporter play a negative role, albeit not as important as disease characteristics.
This is because movement out of the Exporter is unidirectional and removes infected individuals ($L_E$, $D_E$ and $U_E$) from the local population, acting as an additional removal process that shortens the mean residence times $1/(\varepsilon_E+m_L)$, $1/(\gamma_E+m_D)$ and $1/(\gamma_E+m_U)$ during which infected individuals can progress or transmit the infection in the Exporter.
On the other hand, screening increases the reproduction number, as it increases the $D_E$ population.

We also derive a final size relation, which takes the form
\begin{equation}\label{eq:final-size-Exporter}
\ln\left(\frac{S_E(0)}{S_E(\infty)}\right) =
\frac{\R_c^E}{S_E(0)}(S_E(0)-S_E(\infty))+\beta_E\bb\bV_E^{-1}\bx(0),
\end{equation}
where $\R_c^E$ (which reduces to $\R_0^E$ when $\sigma_E=0$) and $\bb\bV_E^{-1}\bx(0)$ are defined by \eqref{eq:RcE} and \eqref{eq:final-size-exporter-bVx} in Appendix~\ref{app:R0-and-final-sizes-exporter}.
Solving numerically for $S_E(\infty)$ allows to compute for instance the so-called \emph{attack rate} $(S_E(0)-S_E(\infty))/S_E(0)$, i.e., the fraction (often expressed as a percentage) of the susceptible population infected during the course of the epidemic. 
The final size and consequently the attack rate, are driven almost entirely by the reproduction number $\R_c^E$.
For instance, a PRCC sensitivity analysis of the attack rate in the Exporter was conducted using the same sample points as those utilised in Figure~\ref{fig:sensitivity-R0E}. 
Results (not shown) were virtually indistinguishable from those in Figure~\ref{fig:sensitivity-R0E}, with the maximum difference between computed partial rank correlation coefficients being 0.0002.
Interestingly, $\sigma_E$ has a very small (slightly positive) PRCC in both Figure~\ref{fig:sensitivity-R0E} and the final size sensitivity: although pre-departure screening retains detected $U_E$ travellers in the Exporter as $D_E$ individuals, its contribution in \eqref{eq:RcE} and \eqref{eq:final-size-exporter-bVx} is multiplied by both the small movement rate $m_U$ and the reduced infectiousness $\xi_E$.

\begin{figure}[htbp]
    \centering
    \includegraphics[width=0.6\textwidth]{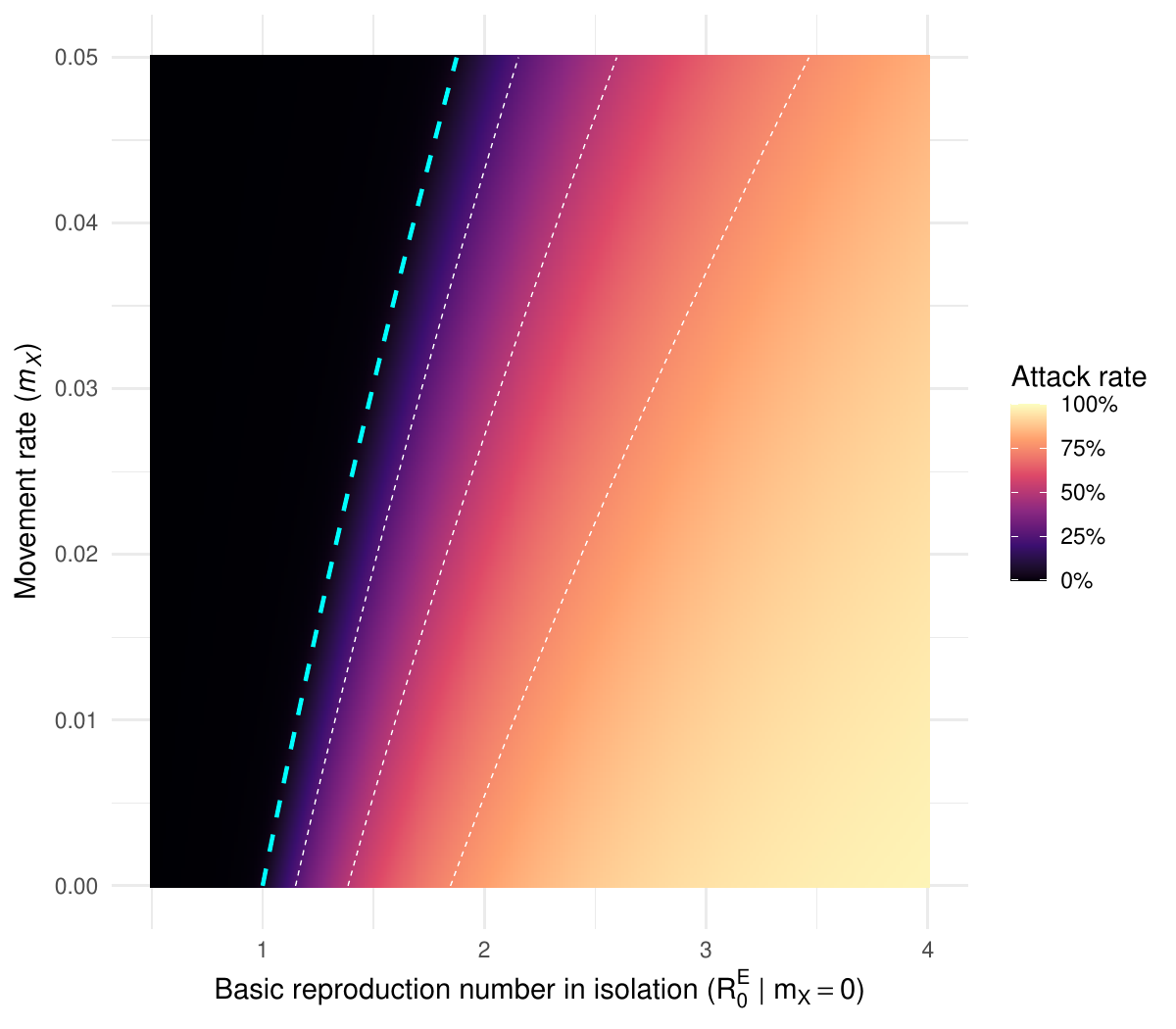}
    \caption{Exporter attack rate as a function of the reproduction number in isolation  and the rate of movement of individuals from the Exporter to the Importer. 
    	The blue curve shows where $\R_0^E=1$.}
    \label{fig:heatmap-finalsize-exporter-vary-movement}
\end{figure}

Figure~\ref{fig:heatmap-finalsize-exporter-vary-movement} shows how the attack rate in the Exporter varies as a function of the basic reproduction number in isolation $\R_0^E|m_X=0$ and the movement rate. 
The $\R_0^E=1$ curve meets $\R_0^E|m_X=0$ when the latter is unity and there is no movement but then shifts to the right as $m_X$ increases.
Figure~\ref{fig:heatmap-finalsize-exporter-vary-movement} confirms an intuitively understood fact: if there is no reintroduction of cases, then allowing those present to leave lessens the severity of the infection locally, since this acts as a ``pressure release'' of the infection in the Exporter.
The behaviour here is also classic: since \eqref{sys:general-model} is an epidemic model, when $\R_0^E<1$, solutions decrease to the disease-free continuum of equilibria.

\subsection{Continuous time Markov chain (Exporter)}
\label{sec:results-exporter-CTMC}
In the case of the Exporter model, the CTMC corresponding to \eqref{sys:general-model-dSE}--\eqref{sys:general-model-dRE} uses only states $S_E$, $L_E$, $D_E$, $U_E$ and $R_E$ and transitions in the first two blocks of rows in Table~\ref{tab:transition-prop}, except that the transitions corresponding to movement out of the Exporter (the second block of rows) take the form $\to(X_E-1)$, $X\in\{L,D,U\}$, instead of $\to(X_E-1,X_I+1)$ as listed in Table~\ref{tab:transition-prop}.

\begin{figure}[htbp]
    \centering
    \includegraphics[width=0.5\textwidth]{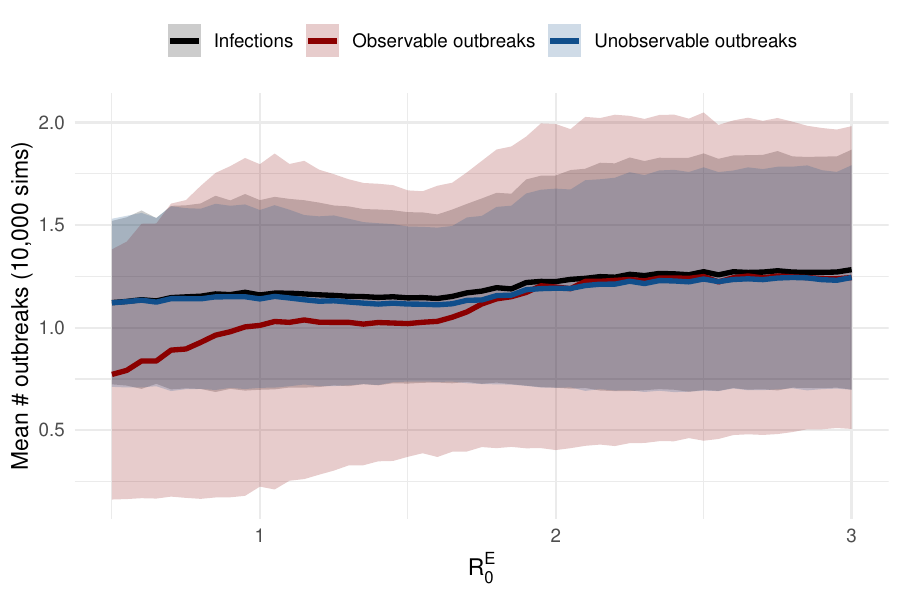}
    \caption{Mean number of outbreaks of different types in the Exporter with an initial condition $S_E(0)=9,999$, $L_E(0)=1$ and all other compartments empty.}
    \label{fig:mean-number-outbreaks-exporter}        
\end{figure}
In Figure~\ref{fig:mean-number-outbreaks-exporter}, we show the mean number of outbreaks (as defined in Section~\ref{sec:model-types-epidemic} and Appendix~\ref{app:characterisation-outbreaks}) as a function of the basic reproduction number $\R_0^E$ in the Exporter, when the initial condition is $S_E(0)=9,999$, $L_E(0)=1$ and all other variables zero.
Remark that the average number of observable outbreaks is increasing, while unobservable outbreaks (based on all infectious activations, detected and undetected) and outbreaks based on new infections remain roughly constant and slightly larger than 1, with the former slightly smaller than the latter.

\subsection{Multitype branching process approximation (Exporter)}
\label{sec:results-exporter-MBPA}
To calculate the analytical probability of an outbreak resulting from a small introduction size, we construct the multitype branching process specifically associated with the Exporter.
The formal definitions of the location-specific generating functions and the derivation of the theoretical outbreak bounds (Theorem~\ref{th:exists-fix-pt-exporter}) are detailed in Appendix~\ref{app:CTMC-MBPA-exporter}.
\begin{figure}[htbp]
    \centering
    \begin{subfigure}{0.49\textwidth}
        \includegraphics[width=\linewidth]{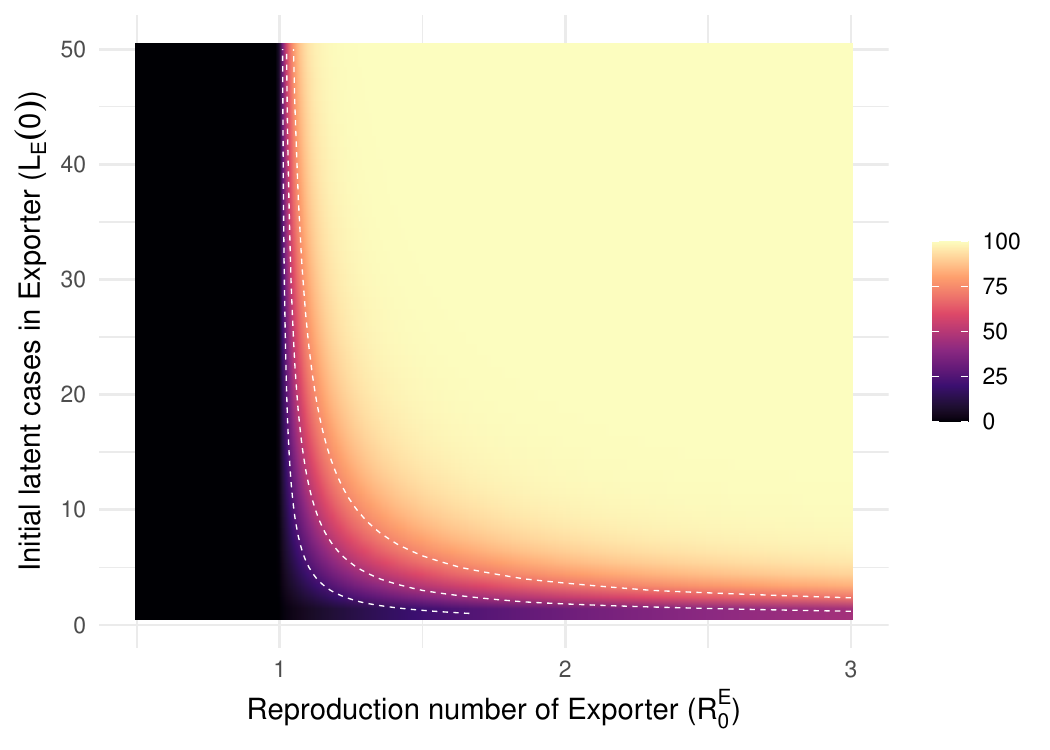}
        \caption{MBPA}
        \label{fig:proba-outbreak-exporter-MBPA}    
    \end{subfigure}
    \begin{subfigure}{0.49\textwidth}
        \includegraphics[width=\linewidth]{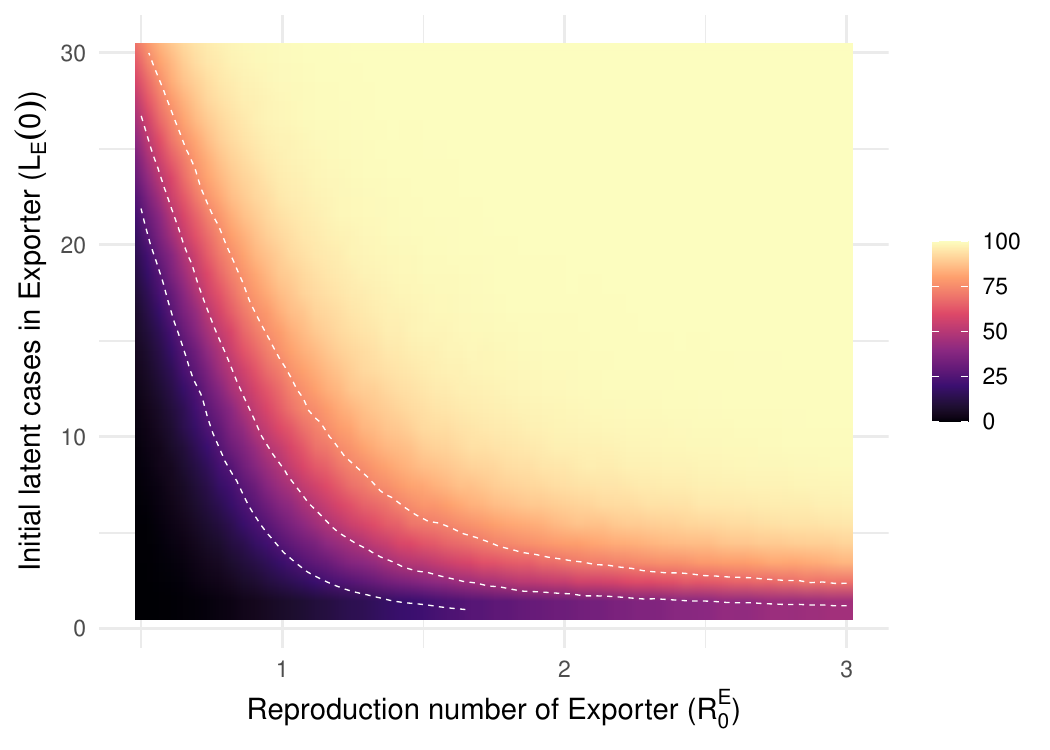}
        \caption{CTMC}
        \label{fig:proba-outbreak-exporter-CTMC}    
    \end{subfigure}
    \caption{Probability of an outbreak in the Exporter computed (a) theoretically using the multitype branching process approximation (MBPA) and (b) empirically via stochastic CTMC simulations.}
    \label{fig:proba-outbreak-exporter}
\end{figure}

Figure~\ref{fig:proba-outbreak-exporter} shows the probability of an outbreak in the Exporter as a function of the reproduction number $\R_0^E$ and the initial number of latently infected individuals $L_E(0)$.
This also serves to illustrate a fundamental difference between the results obtained using the MBPA (Figure~\ref{fig:proba-outbreak-exporter-MBPA}) and the CTMC (Figure~\ref{fig:proba-outbreak-exporter-CTMC}).
The MBPA gives an exact result that can be computed deterministically from Theorem~\ref{th:exists-fix-pt-exporter}. 
On the other hand, the result does not cover the case $\R_0^E<1$ and is valid only close to the disease-free equilibrium, as can be inferred from the increasing discrepancy as one considers larger and larger initial number of latent individuals.
This is where the CTMC comes in handy, since it allows to computationally evaluate the probability of an outbreak.
(Note that some analytical results can also be obtained from the CTMC but this is beyond the scope of the present paper.)
However, in order to obtain good information from the CTMC, one has to run a large number of realisations at each point in parameter space being considered.

As can be expected, the initial number of infections plays an extremely important role in determining whether an outbreak takes place.
While the MBPA does not allow to conclude when the initial infections are large of $\R_0^E$ is small than one, both methods show that the probability of an outbreak is moderate if the initial number of cases is low, but increases rapidly as soon as that number increases.
This is all the more striking in Figure~\ref{fig:proba-outbreak-exporter-CTMC}, where we see that when the initial introduction size in the Exporter is large, even very small $\R_0^E$ see outbreaks.

\section{Results in the Importer in isolation}
\label{sec:results-importer}

We now focus briefly on the Importer when there is no movement between locations, i.e., when $m_L=m_D=m_U=0$.
This particular case is useful because it corresponds to the situation that prevails if a total interruption of travel is enforced.


We start with properties of the deterministic version of the Importer subsystem.
We find the basic reproduction number in the Importer in isolation,
\begin{equation}\label{eq:R0C}
    \R_0^C = \frac{\beta_{C}}{\gamma_C}\left(p_C\xi_{C}+1-p_C\right)S_C(0)
\end{equation}
and final size relation
\begin{equation}\label{eq:final-size-Importer}
    \ln\left(\frac{S_C(0)}{S_C(\infty)}\right) =
    \frac{\R_0^C}{S_C(0)}\left(S_C(0)-S_C(\infty)\right)
    +\beta_{C}\bb\bV_I^{-1}\bx(0),
\end{equation}
where $\beta_C\bb\bV_I^{-1}\bx(0)$ takes the form
\begin{equation}
\label{eq:final-size-importer-bVx}
\beta_C\bb\bV_I^{-1}\bx(0) = 
\frac{\beta_I}{\gamma_I}
\left[
(p_IL_I(0)+D_I(0))\xi_I+
(1-p_I)L_I(0)+U_I(0)
\right].    
\end{equation}
See Appendix~\ref{app:R0-and-final-sizes-importer} for details of the computation of $\R_0^C$ and the final size relation.
Note that to obtain \eqref{eq:final-size-importer-bVx}, we assumed that $\bx(0)=(L_I(0),D_I(0),U_I(0),0,0,0)$, i.e., there are initially no infected individuals in the community.
It is easy to obtain a final size relation involving the positive initial numbers of infected individuals in the community, but this is outside the scope of this work, where we focus on the case where the community starts without cases.

It is interesting to note that $\beta_I$ and $\xi_I$ do not appear in \eqref{eq:R0C}.
While this may at first seem counter-intuitive, it is easy to explain.
Indeed, while $L_I$, $D_I$ and $U_I$ contribute to infection, the absence of movement means that they do not replenish.
Therefore, their capacity to trigger an epidemic is limited to their initiating the infection process in the community.
This is seen in the final size relation \eqref{eq:final-size-Importer}.

\begin{figure}[htbp]
    \centering
    \begin{subfigure}{0.49\textwidth}
        \includegraphics[width=\linewidth]{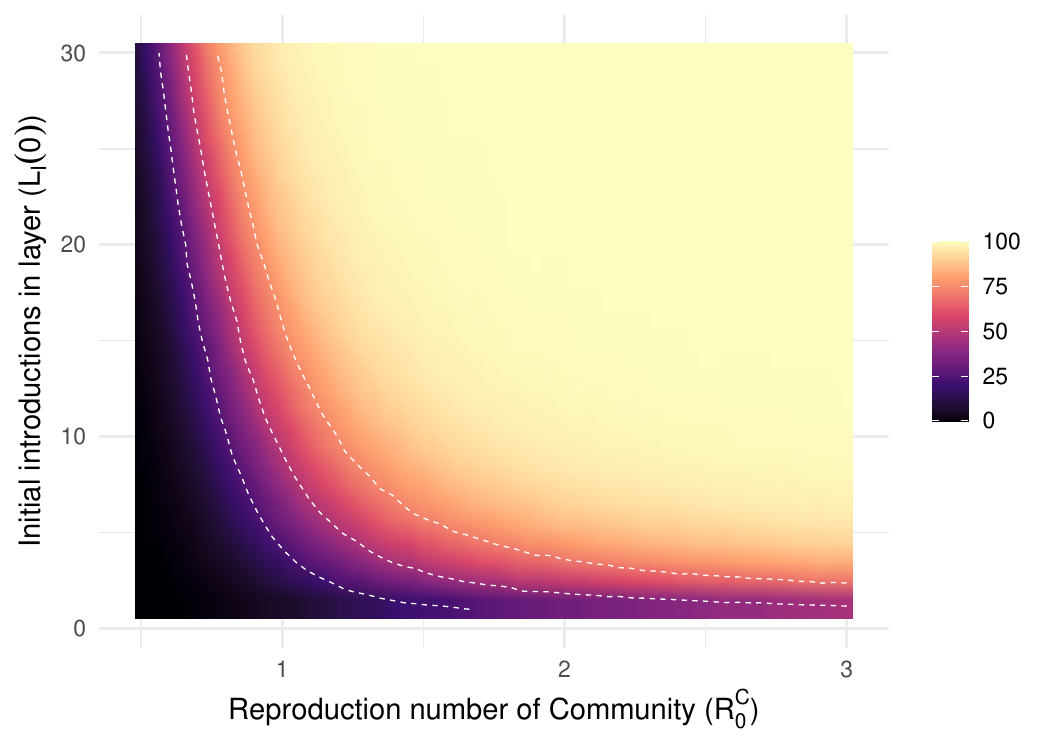}
        \caption{Function of $L_I(0)$}
        \label{fig:proba-outbreak-importer-ctmc}
    \end{subfigure}
    \begin{subfigure}{0.49\textwidth}
        \includegraphics[width=\linewidth]{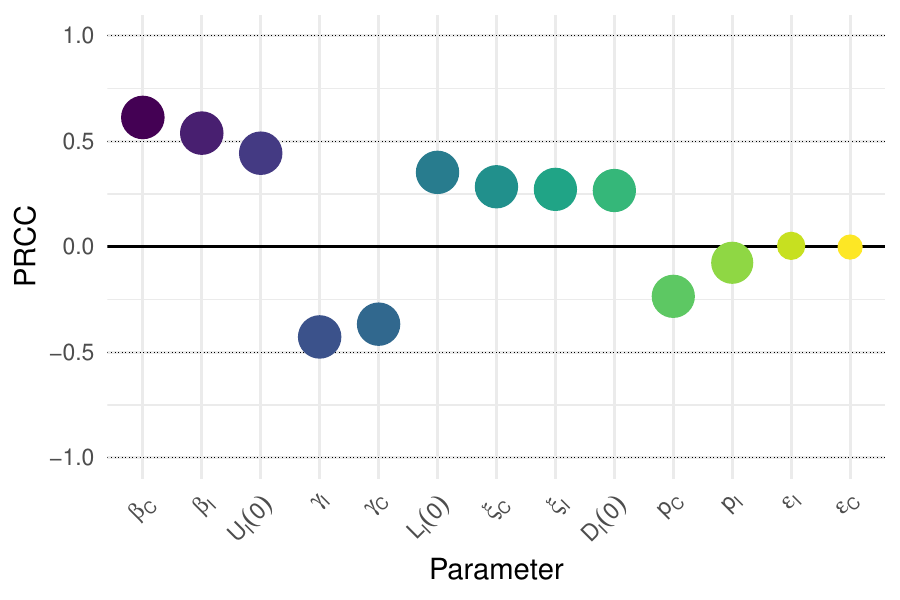}
        \caption{PRCC}
        \label{fig:proba-outbreak-importer-mbpa}
    \end{subfigure}
    \caption{Percentage of simulations leading to an outbreak in the Importer as a function of (a) the intensity $\R_0^C$ of local control and the initial number $L_I(0)$ of latently infected individuals and (b) various parameters through a sensitivity analysis.}
    \label{fig:proba-outbreak-importer}
\end{figure}
We then use the continuous-time Markov chain version of the Importer subsystem in Figure~\ref{fig:proba-outbreak-importer-ctmc}, where we observe that the picture looks very similar to that seen in Figure~\ref{fig:proba-outbreak-exporter}.
For very low introduction numbers, the estimated probability of an outbreak is relatively low, even when $\R_0^C$ is high.
However, for values of $\R_0^C$ larger than 1, this probability increases rapidly.
For instance, when $\R_0^C=2$, an initial $L_I(0)=1$ leads to an outbreak in about 33\% of cases, while $L_I(0)=6$ leads to outbreaks in about 90\% of cases.

The formal multitype branching process approximation mathematically formulated for the Importer to estimate exact theoretical outbreak bounds (Theorem~\ref{th:exists-fix-pt-importer}) is fully established in Appendix~\ref{app:CTMC-MBPA-importer}.
In Figure~\ref{fig:proba-outbreak-importer-mbpa}, we use Theorem~\ref{th:exists-fix-pt-importer} to compute the probability of a disease outbreak depending on the values of parameters, which we in turn use to compute the partial rank correlation coefficients of sensitivity of outbreak probability.
The parameters $\beta$, $\gamma$ and $\xi$ corresponding to those that played the most role in variations of $\R_0^E$ (Figure~\ref{fig:sensitivity-R0E}), are influential in the probability of an outbreak, together with, mixed in order of importance, $U_I(0)$, $L_I(0)$ and $D_I(0)$.

\section{Results in the Exporter-Importer model}
\label{sec:results-exporter-importer}
We now return to the full model \eqref{sys:general-model}, its CTMC equivalent and the latter's branching process approximation.
\subsection{Deterministic considerations (full model)}
\label{sec:results-exporter-importer-deterministic}
For the general Exporter-Importer model, we find the basic reproduction number
\begin{equation}\label{eq:R0}
    \mathcal{R}_0 = \max\left\{\mathcal{R}_0^E, \mathcal{R}_0^C\right\}.
\end{equation}
This is established in Appendix~\ref{app:R0-and-final-sizes-exporter-importer}.
In that appendix, the final size relation \eqref{eq:final-size-Importer-coupled} is also derived for the general Exporter-Importer model \eqref{sys:general-model}.
As shown in Appendix~\ref{app:supercritical-vs-subcritical} (equations \eqref{eq:supercritical-DeltaSC}--\eqref{eq:subcritical-DeltaSC}), this relation exhibits two distinct regimes depending on local control in the Importer: when $\R_0^C > 1$, the deterministic final size $S_C(0)-S_C(\infty)$ stays close to the unforced community epidemic size for any positive cumulative importation load $\Lambda_{E\to I} > 0$, whereas when $\R_0^C < 1$, the cumulative number of community infections scales linearly with $\Lambda_{E\to I}$ via $(1-\R_0^C)^{-1}$.

\begin{figure}[htbp]
  \begin{subfigure}[b]{0.48\textwidth}
    \centering
    \includegraphics[width=\linewidth]{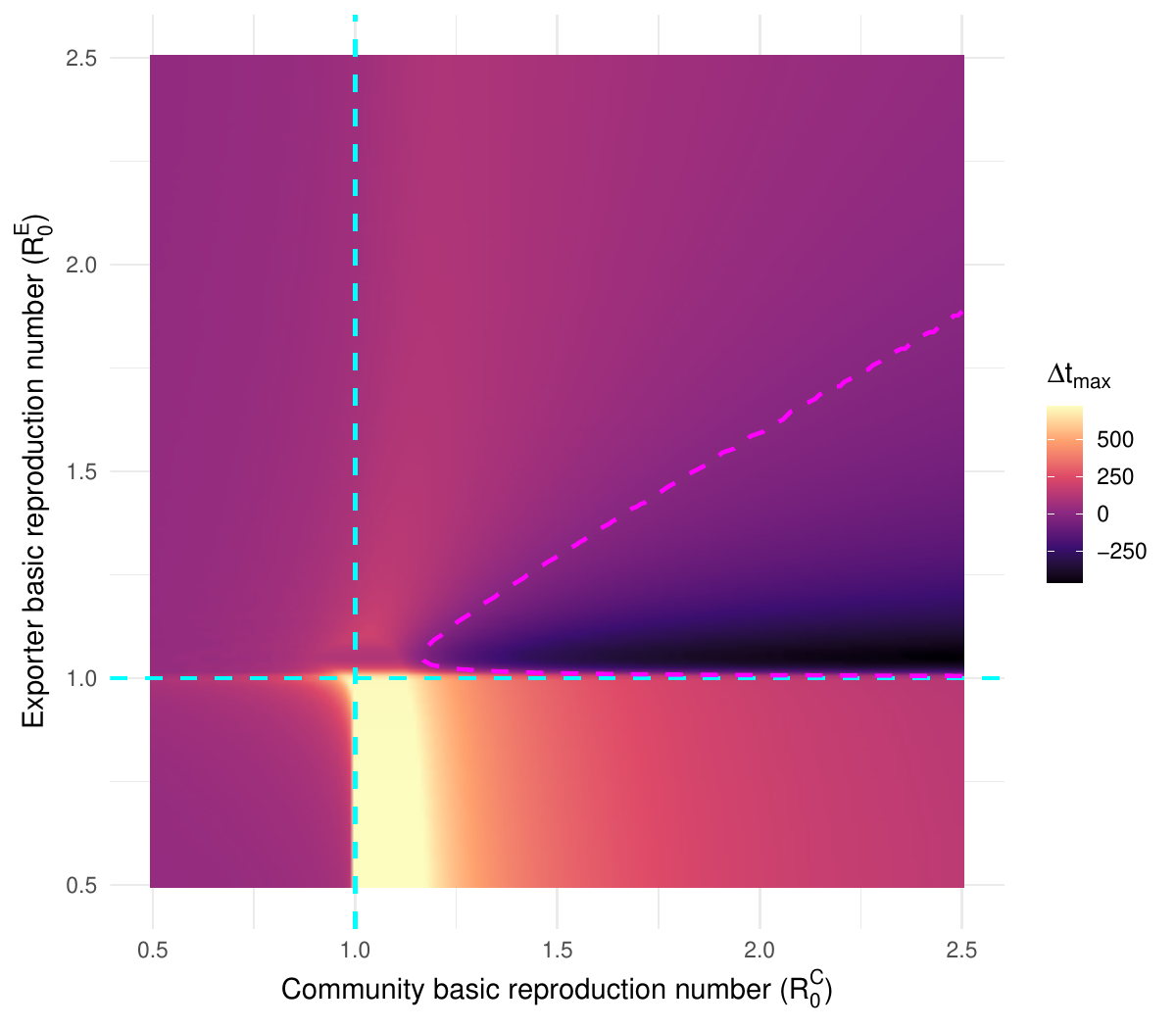}
    \caption{$\R_0^E$ versus $\R_0^C$}
    \label{fig:ode-heatmap-r0-timing}
  \end{subfigure}
  \begin{subfigure}[b]{0.48\textwidth}
    \centering
    \includegraphics[width=\linewidth]{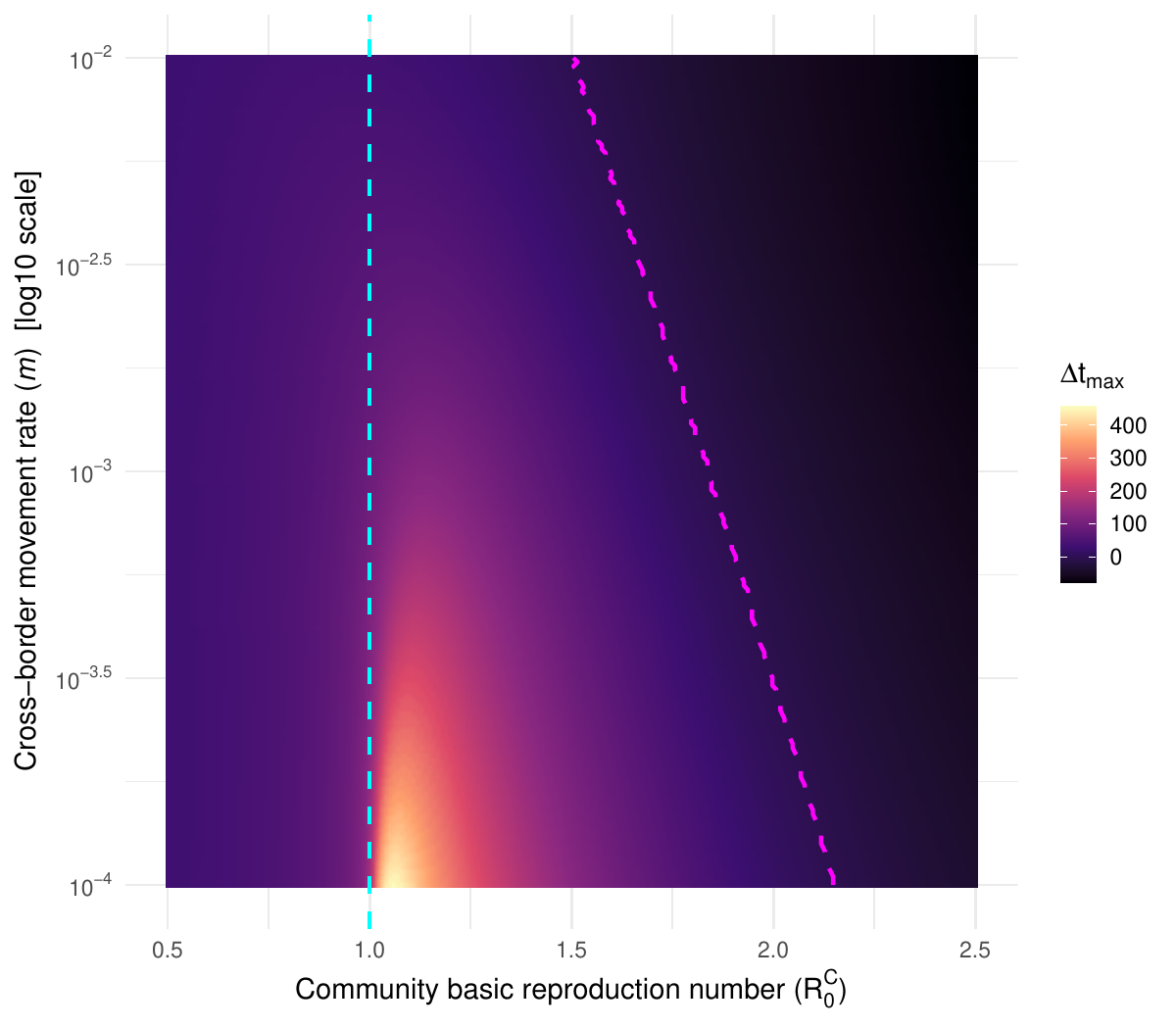}
    \caption{$m_X$ versus  $\R_0^C$}
    \label{fig:ode-heatmap-mov-timing}
  \end{subfigure}
  \caption{Peak timing differences $\Delta t_{\max}=t_{\max}^C - t_{\max}^E$ between the Exporter and the Importer, in the deterministic model, as a function of the local control $\R_0^C$ and (a) $\R_0^E$ and (b) the movement rate $m_X$.
  The magenta curve is the level curve $\Delta t_{\max}=0$, while the cyan curves show where $\R_0^X=1$, $X\in\{E,C\}$.}
  \label{fig:ode-heatmaps-R0C-vs-R0E-m}
\end{figure}

In Figure~\ref{fig:ode-heatmaps-R0C-vs-R0E-m}, we use the deterministic model to study the synchronisation delay (or peak timing offset) between the Exporter and the Importer.
We record the peak times, shown for the Importer divided between introduction layer and community in Figure~\ref{fig:ode-heatmaps-peak-value-R0C-vs-R0E-movement} and compute the time difference $\Delta t_{\max}=t_{\max}^C - t_{\max}^E$.
In both figures, the $x$-axis shows the effect of local control measures in the Importer, as proxied by $\R_0^C$.
Interestingly, we observe in Figure~\ref{fig:ode-heatmap-r0-timing} the existence of a critical region (delimited by the red curve), where for values of $\R_0^E$ slightly larger than 1 and increasing values of $\R_0^C$, $\Delta t_{\max}$ becomes increasingly negative, i.e., the epidemic peak in the Importer \emph{precedes} the one in the Exporter.

Note that in Figure~\ref{fig:ode-heatmap-r0-timing}, the part below $\R_0^E=1$ is not really relevant: in the Exporter, having $\R_0^E<1$ means that the peak occurs at time $t=0$.
It follows that $\Delta t_{\max}$ depends almost exclusively on $\R_0^C$.

\subsection{Continuous time Markov chain (full model)}
\label{sec:results-exporter-importer-CTMC}

\begin{figure}
    \centering
    \begin{subfigure}{0.49\textwidth}
        \includegraphics[width=\linewidth]{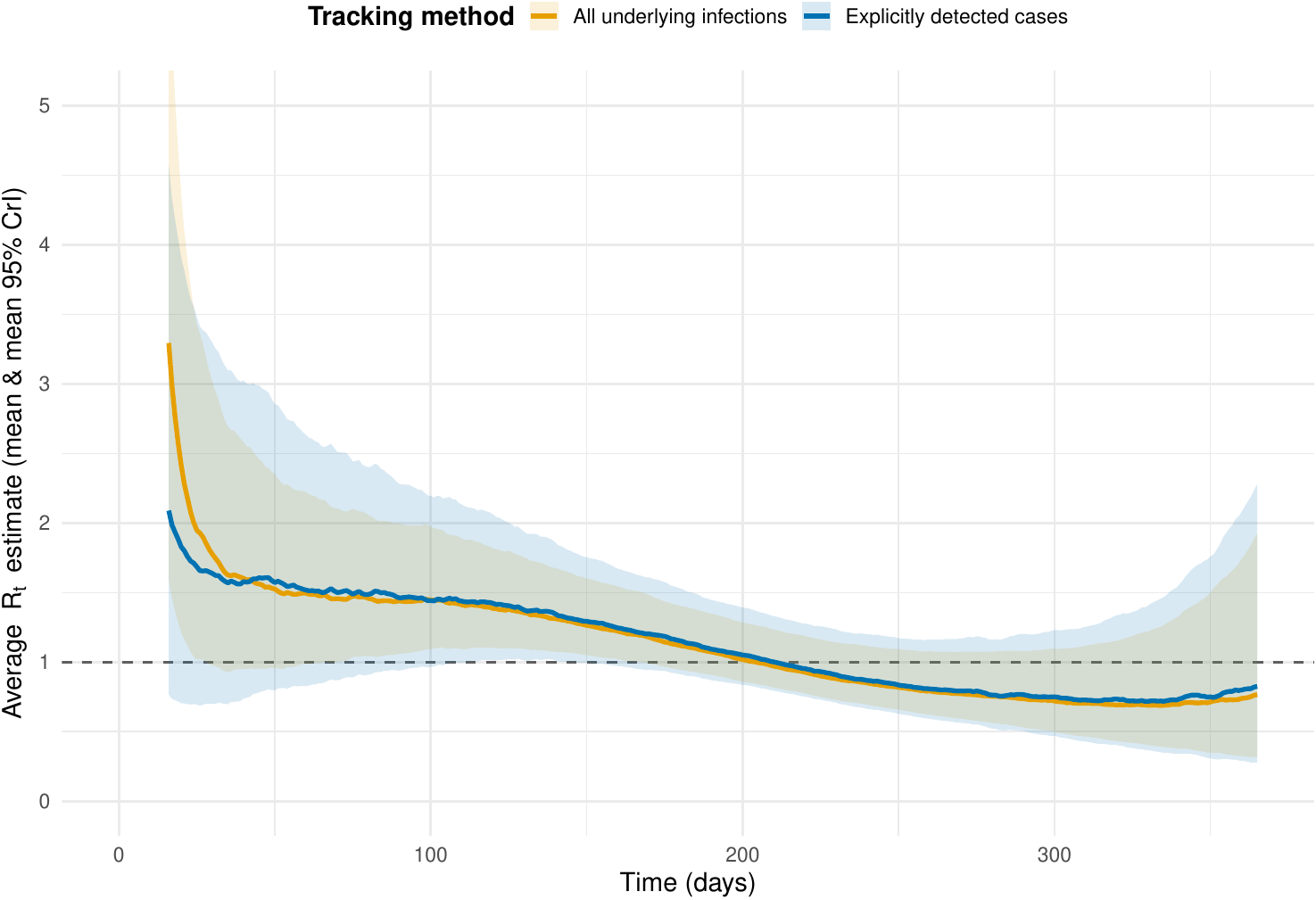}
        \caption{Exporter}
        \label{fig:Rt-exporter}
    \end{subfigure}
    \begin{subfigure}{0.49\textwidth}
        \includegraphics[width=\linewidth]{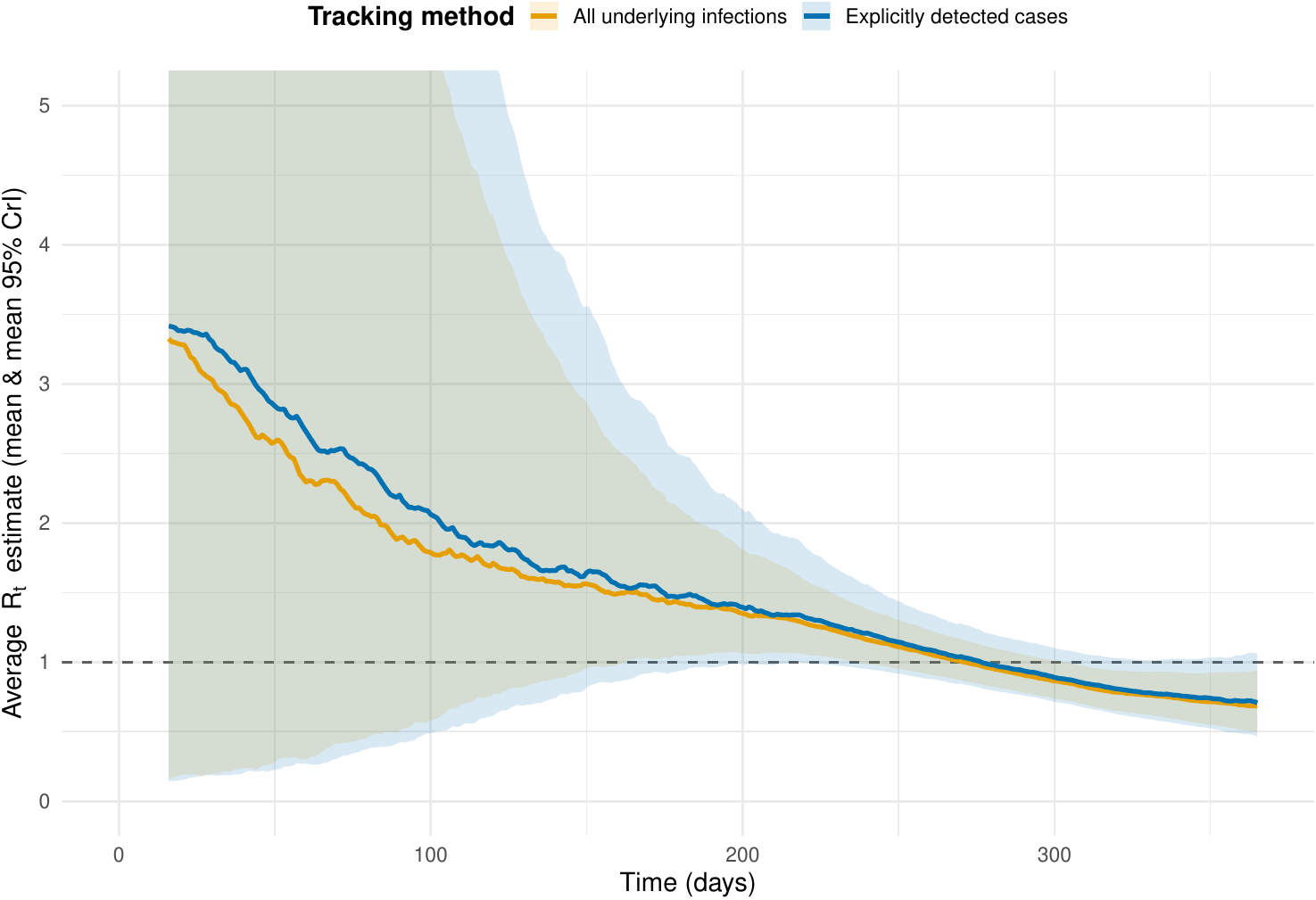}
        \caption{Importer}
        \label{fig:Rt-importer}
    \end{subfigure}
    \caption{Mean $\R_t$ estimation from 1,000 simulations having had at least one introduction event for (a) the Exporter and (b) the Importer.
    Estimation using the \texttt{EpiEstim} package.
    Yellow curve: all sources of infections; blue curve: only detected infections.}
    \label{fig:Rt}
\end{figure}
In Figure~\ref{fig:Rt}, we show the result of using the \texttt{R} package \texttt{EpiEstim}, which computes the value of the time-dependent $\R_t$ based on epidemiological data.
One thousand simulations of the Exporter (Figure~\ref{fig:Rt-exporter}) and the Importer (Figure~\ref{fig:Rt-importer}) are run and the value of $\R_t$ is computed using the package, with the 95\% credible interval (CrI) of values shown.
We observe that in the Exporter, $\R_t^E$ starts near $\R_0^E=1.5$ with a wide credible interval early on, then decreases monotonically as susceptible individuals are depleted and crosses below 1 around day 210 (around the mean epidemic peak in the Exporter, not shown), with a tighter credible interval during the main phase of the epidemic that widens again late in the outbreak as incidence becomes small.
In the Importer, $\R_t^I$ starts above 3 with a very wide credible interval at the start of local spread when case numbers are low and driven by stochastic introductions, then decreases and crosses below 1 around day 270 with a much tighter credible interval during the main phase of spread.
Our main purpose for using \texttt{EpiEstim} is that this sort of tool might be used by public health officials to take intervention decisions.
We return to this later when we consider interventions in Section~\ref{sec:travel-control}.

\begin{figure}
    \centering
    \includegraphics[width=0.6\textwidth]{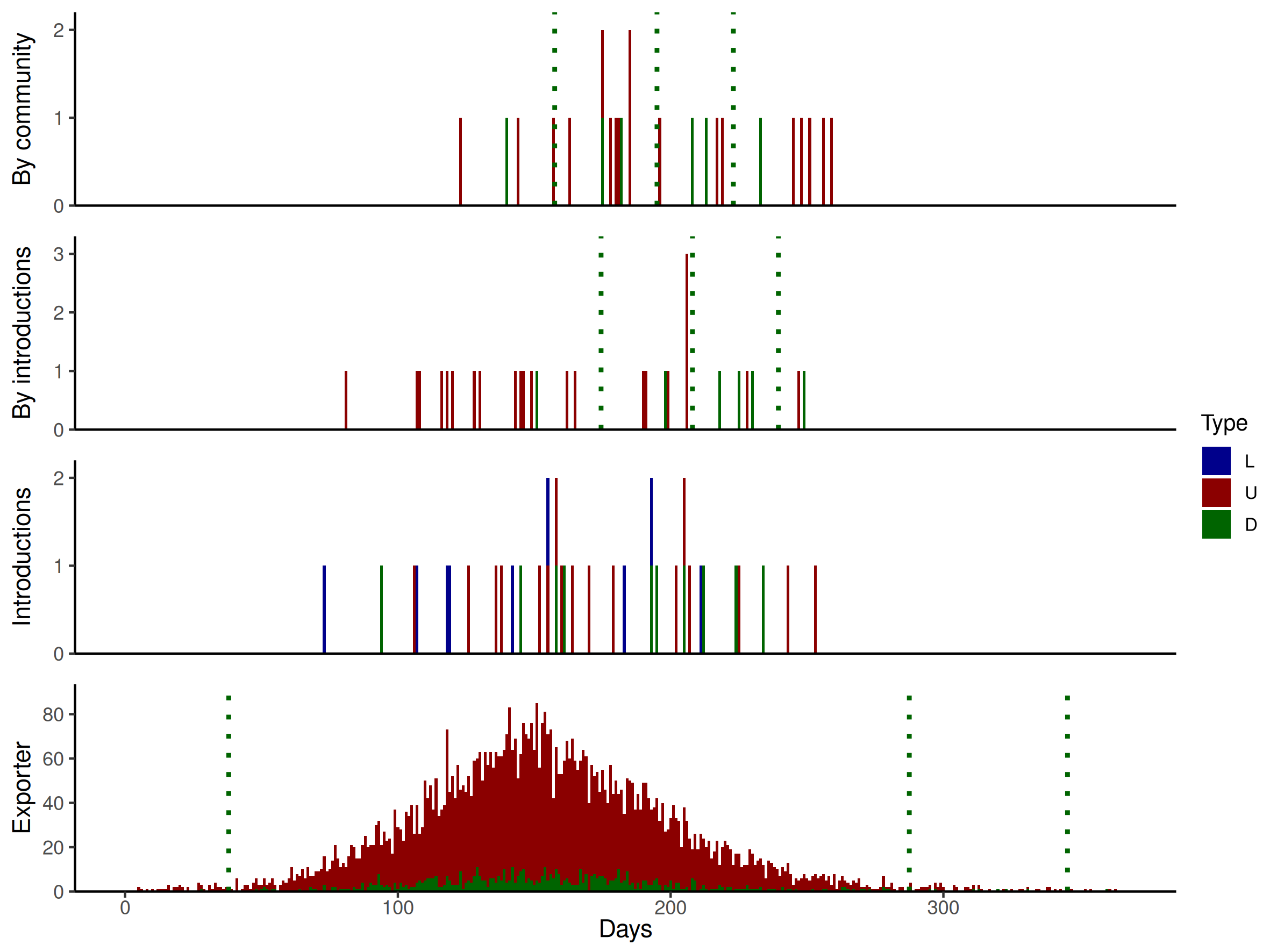}
    \caption{Sample realisation of the full CTMC model with $\R_0^E=1.5$, $\R_0^C=0.8$ and a total initial population of 10,000 in each location, showing the incidence in various components of the model.
    From the bottom panel going up: incidence in the Exporter, incidence of introductions (individuals infected in the Exporter arriving in the Importer), incidence in the Importer generated by introductions and incidence in the importer generated by individuals in the Importer.
    Vertical green dashed lines indicate the separation between \emph{observed} outbreaks.}
    \label{fig:transmission-scenario-R0E_15_R0C_08_mvt_00005_seed_349}
\end{figure}

Figure~\ref{fig:transmission-scenario-R0E_15_R0C_08_mvt_00005_seed_349} shows the summary of a single sample realisation of the full continuous-time Markov chain focusing on transitions of importance in the model. 
Other examples can be found in Appendix~\ref{app:additional-computational-stuff} (Figures~\ref{fig:transmission-scenarios}).
In all these figures, we show, from bottom to top panel, the daily number of events under consideration in the Exporter, arrivals into the introduction layer, then infections in the community originating from the introduction layer (third panel) and from the community (top panel).
Except for the introduction layer, we focus on new infectious cases (detected or undetected).
In the introduction layer, we indicate what type of introduction took place and thus also show arrival of latent individuals.
In all but the introduction layer, vertical green dotted bars indicate the midpoint between two successive \emph{observable} outbreaks (i.e., when more than $\tau$ days elapse between two successive detected cases; see Section~\ref{sec:model-types-epidemic} and Appendix~\ref{app:characterisation-outbreaks}).

In the realisation in Figure~\ref{fig:transmission-scenario-R0E_15_R0C_08_mvt_00005_seed_349}, a few cases transit from the Exporter to the Importer, but the reproduction number in the Importer ($\R_0^C=0.8$) is not favourable to a sustained outbreak there and transmission chains fizzle out, with gaps exceeding $\tau$ between detections resulting in 4 distinct observable outbreaks in the community originating from within the community.
Note that since simulations are stochastic, this is by no means the only scenario that happens.
In Figure~\ref{fig:transmission-scenario-R0E_15_R0C_08_mvt_00005_seed_84}, under the same $\R_0^C=0.8$, the outbreak in the Importer is much more pronounced.

When the Importer has looser control ($\R_0^C=1.5$), the outbreak there becomes much stronger, as local spread is further reinforced by any additional introduction.
We see this clearly in Figures~\ref{fig:transmission-scenario-R0E_15_R0C_15_mvt_00001_seed_18} and \ref{fig:transmission-scenario-R0E_15_R0C_15_mvt_0001_seed_96}, in which, it should be noted, disease translocation occurs despite a very low number of introductions.
Figure~\ref{fig:transmission-scenario-R0E_15_R0C_15_mvt_0001_seed_96}, in particular, tells a cautionary tale.
There, the Exporter experiences four minor outbreaks.
A single detected individual then travels to the Importer, where they infect one individual who in turn triggers a very sizeable outbreak.

\subsection{Multitype branching process approximation (full model)}
\label{sec:results-exporter-importer-MBPA}

The calculation of the theoretical probability of extinction for the joint Exporter-Importer branching process approximation (Theorem~\ref{th:exists-fp-exp-imp}) is delineated in Appendix~\ref{app:CTMC-MBPA-exporter-importer}.

\begin{figure}[htbp]
    \centering
    \includegraphics[width=0.6\linewidth]{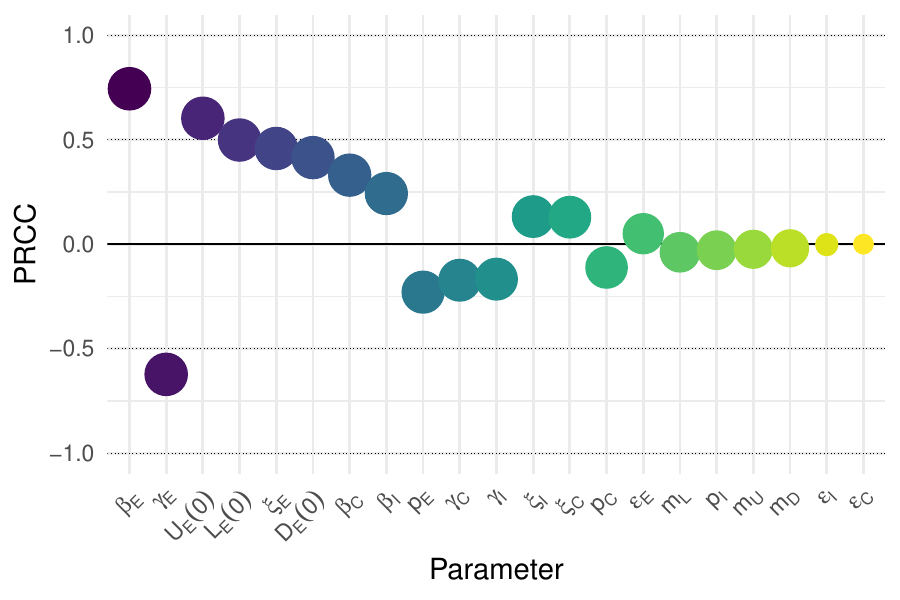}
    \caption{PRCC of the probability of an outbreak in the full model, as given by Theorem~\ref{th:exists-fp-exp-imp}.}
    \label{fig:mbpa_prcc_full_model_unconditional}
\end{figure}

In Figure~\ref{fig:mbpa_prcc_full_model_unconditional}, we show the partial rank correlation coefficient sensitivity of the probability of an outbreak in the full Exporter-Importer system \eqref{sec:model}.
Note that we added the initial conditions $L_E(0)$, $D_E(0)$ and $U_E(0)$, which we limit to small values in order for the MBP approximation to remain valid.
We do not limit our use of Theorem~\ref{th:exists-fp-exp-imp} to the case of $\R_0>1$, meaning that the sensitivity also includes the probability of no-outbreak.
We observe that besides $\beta_E$, $\gamma_E$ and $\xi_E$, which were the most influential parameters in Figure~\ref{fig:sensitivity-R0E}, we see here that the initial number of infected $U_E(0)$, $L_E(0)$ and $D_E(0)$ (in that order) also play an important role.

\section{Effect of travel control measures}
\label{sec:travel-control}
In the previous sections, we discuss control measures, but in a simple way, looking mostly at the effect of varying $\R_0^C$, which we use as a proxy for the intensity of the control effort in the Importer.
Here, we consider more precise and realistic control scenarios: pre-departure screening, quarantine and travel interruptions.

\subsection{Effect of pre-departure screening}
\label{sec:travel-control-screening}

We start by considering pre-departure screening, as this is incorporated explicitly into the model.
Pre-departure screening applies to all individuals outbound from the Exporter but its effect is on individuals in the $D_E$ and $U_E$ compartments.
That is, in the examples of Figures~\ref{fig:transmission-scenario-R0E_15_R0C_08_mvt_00005_seed_349} and Figures~\ref{fig:transmission-scenario-R0E_15_R0C_08_mvt_00005_seed_84}--\ref{fig:transmission-scenario-R0E_15_R0C_15_mvt_0001_seed_96}, the only introductions always allowed are the dark blue (latent) ones, with some dark red (undetected) ones also occurring when testing fails.
We assume that if screening is present, i.e., if $\sigma_E>0$, then $D_E$ individuals are altogether forbidden from travelling and thus set $m_D=0$.

\begin{figure}[htbp]
    \centering
    \begin{subfigure}{0.49\textwidth}
        \includegraphics[width=\linewidth]{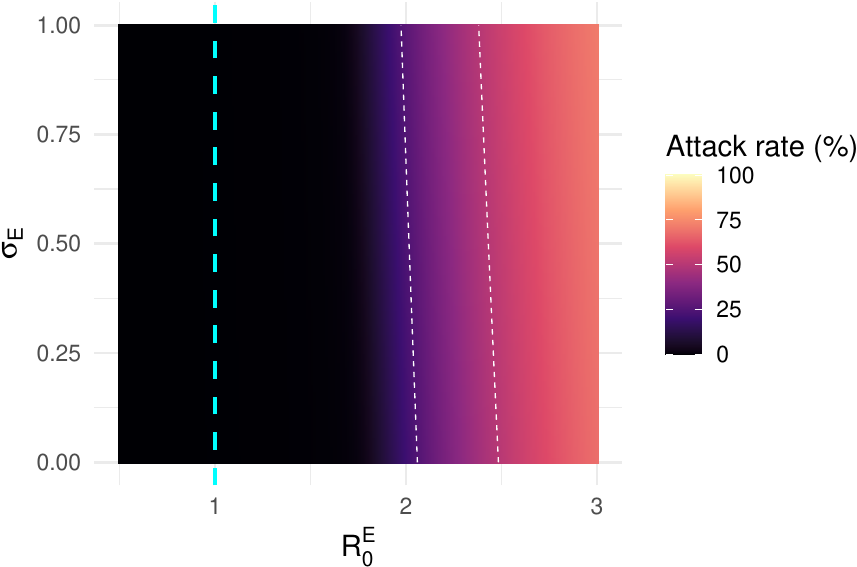}
        \caption{Exporter}
        \label{fig:attack-rate-R0E-sigmaE-exporter}
    \end{subfigure}
    \begin{subfigure}{0.49\textwidth}
        \includegraphics[width=\linewidth]{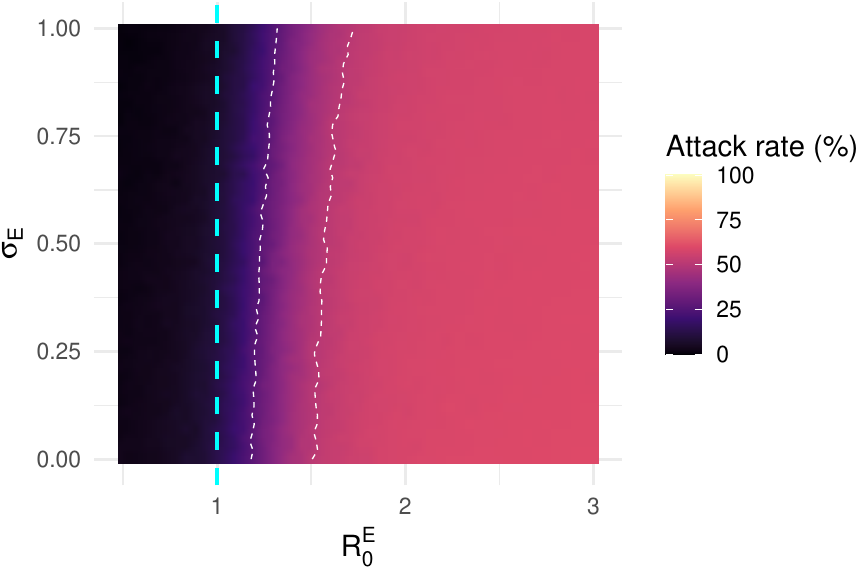}
        \caption{Importer}
        \label{fig:attack-rate-R0E-sigmaE-importer}
    \end{subfigure}
    \caption{Attack rate (in \%) in (a) the Exporter (deterministic final size) and (b) the Importer (mean over 1,000 CTMC simulations per grid point, $\R_0^C=1.5$) as a function of $\R_0^E$ and the efficiency $\sigma_E$ of the pre-departure test.
    Movement of detected individuals is zero.}
    \label{fig:attack-rate-R0E-sigmaE}
\end{figure}

First, note that from the perspective of the Exporter, one indirect public health cost of this measure is limited. 
Indeed, in Figure~\ref{fig:attack-rate-R0E-sigmaE-exporter} (computed from the deterministic final size relation), we see that regardless of the value of the pre-departure testing efficiency $\sigma_E$, the attack rate in the Exporter is entirely driven by $\R_0^E$, despite the fact that with perfect test efficacy, all outbound undetected individuals stay in the Exporter, thereby enhancing the epidemic there.
The same is true for the mean attack rate in the Importer across CTMC simulations (Figure~\ref{fig:attack-rate-R0E-sigmaE-importer}), which is also insensitive to $\sigma_E$ because latently infected ($L_E$) travellers are not detected by pre-departure screening and continue to seed the Importer.

This highlights an important tension between logistical efficiency, epidemiological efficacy and financial burden under the IHR \cite{WHO-IHR-2005}.
From a global transport network perspective, outbound (exit) screening at the source is vastly more efficient and less disruptive than inbound (entry) screening across destination countries, requiring intervention at far fewer airports and screening orders of magnitude fewer low-risk travellers per at-risk traveller \cite{KhanEckhardtBrownsteinNaqviEtAl2013}.
In addition, because air travel durations are typically short relative to the incubation period of most respiratory pathogens, few latently infected ($L_E$) travellers progress to detectable status in transit, meaning that entry screening upon arrival adds little detection benefit over exit screening \cite{KhanEckhardtBrownsteinNaqviEtAl2013}.
However, while the indirect epidemiological cost to the Exporter of retaining detected travellers is negligible (Figure~\ref{fig:attack-rate-R0E-sigmaE-exporter}), the direct operational and healthcare costs under the IHR fall entirely on the Exporter: under Articles~23, 32(c) and~40(1), the State Party conducting exit screening is prohibited from charging travellers for screening, supplementary medical examinations or isolation and quarantine, and must bear the cost of appropriate medical treatment, accommodation, food and water for travellers who are denied departure.
Exit screening thus places the full financial and clinical burden on the Exporter to protect foreign jurisdictions, even though, as shown in Figure~\ref{fig:attack-rate-R0E-sigmaE-importer}, the leakage of latently infected ($L_E$) travellers renders stand-alone screening insufficient to prevent an outbreak in a receptive Importer having $\R_0^C>1$.

\subsection{Effect of quarantine}
\label{sec:travel-control-quarantine}

Quarantine is the automatic isolation of incoming individuals, irrespective of their actual disease status.
In \cite{ArinoBajeuxPortetWatmough2020}, some computations were performed regarding quarantine in a model somewhat similar to the one used here in locations.
By considering the progression of individuals through compartments in the absence of contacts, as would be the case for an individual fully compliant with quarantine, \cite{ArinoBajeuxPortetWatmough2020} derived the efficacy of quarantine as a function of quarantine duration.

Here, we use a simpler approach: we lower the transmission coefficient $\beta_I$ of individuals in the introduction layer by a factor corresponding to quarantine stringency (so that a quarantine stringency of $s \in [0,1]$ replaces $\beta_I$ with $(1-s)\beta_I$).
If $s=1$ ($\beta_I=0$), then quarantine is perfect, as individuals in the introduction layer make no contacts with community members.
In this case, if this measure is implemented before any pathogen is introduced, then obviously there is no introduction.

\begin{figure}[htbp]
    \centering
    \begin{subfigure}{0.49\textwidth}
        \includegraphics[width=\linewidth]{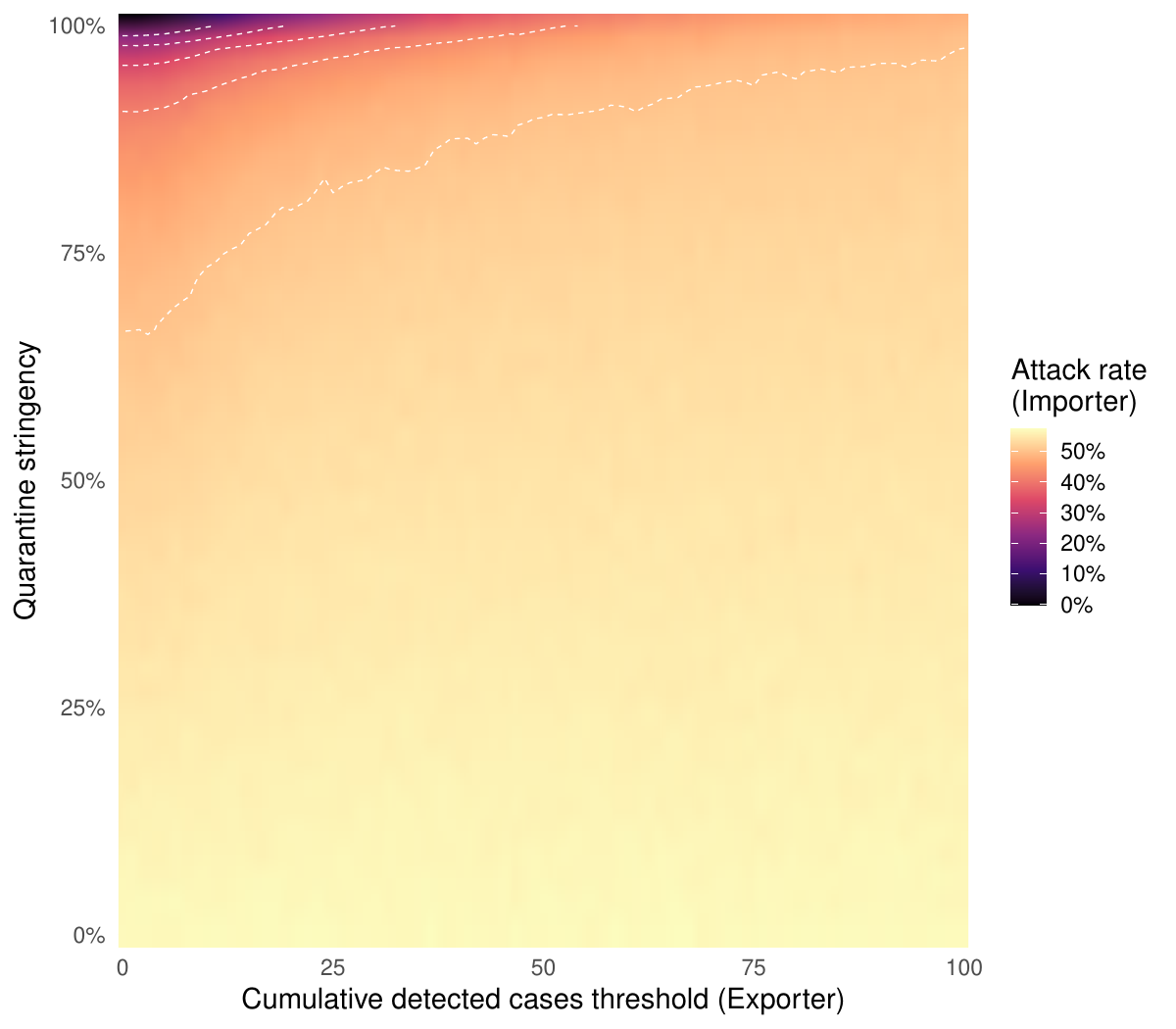}
        \caption{Trigger threshold vs.\ stringency}
        \label{fig:effect-quarantine-threshold}
    \end{subfigure}
    \begin{subfigure}{0.49\textwidth}
        \includegraphics[width=\linewidth]{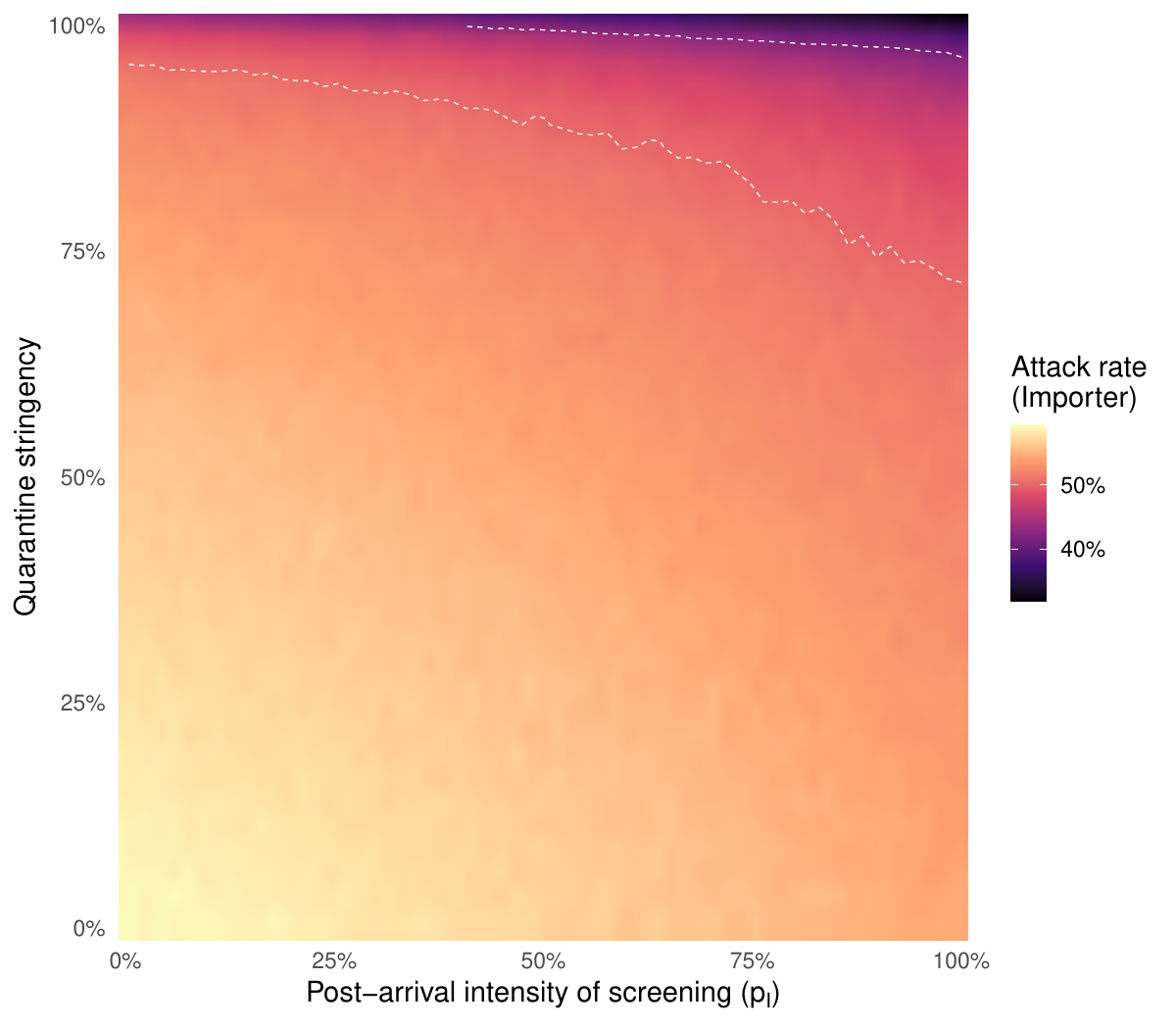}
        \caption{Post-arrival screening $p_I$ vs.\ stringency}
        \label{fig:effect-quarantine-pI}
    \end{subfigure}
    \caption{Mean attack rate (in \%) in the Importer ($\R_0^E=\R_0^C=1.5$, 10,000 CTMC simulations per grid point) as a function of quarantine stringency (fractional reduction in $\beta_I$) and (a) the cumulative number of detected cases in the Exporter used to trigger quarantine or (b) the post-arrival screening intensity $p_I$ when quarantine is triggered at 50 cumulative detected cases in the Exporter.}
    \label{fig:effect-quarantine}
\end{figure}

In Figure~\ref{fig:effect-quarantine}, we use CTMC simulations to examine how quarantine stringency interacts with the timing of implementation and with post-arrival screening.
In Figure~\ref{fig:effect-quarantine-threshold}, quarantine is triggered once cumulative detected cases in the Exporter reach a given threshold.
We observe that quarantine substantially reduces the mean attack rate in the Importer only when stringency is very high (above roughly $80\%$) and the measure is triggered early, at a low threshold of cumulative detected cases in the Exporter.
Once a larger number of cases have accumulated in the Exporter, the probability that at least one infected traveller has already seeded the Importer prior to quarantine implementation, or leaks through an imperfect quarantine, is high enough that an outbreak establishes in the community.
In Figure~\ref{fig:effect-quarantine-pI}, we fix the implementation trigger at 50 cumulative detected cases in the Exporter and vary the post-arrival screening intensity $p_I$ alongside quarantine stringency.
Because latently infected arrivals ($L_I$) progress to the detected compartment $D_I$ with probability $p_I$, where their infectiousness is already reduced by $\xi_I = 0.1$, increasing $p_I$ reinforces quarantine stringency: the largest reduction in the Importer attack rate occurs in the upper-right region where stringent quarantine is paired with high post-arrival case detection.

\subsection{Effect of travel interruptions}
\label{sec:travel-control-travel}
We now suppose that at some point after the start of spread in the Exporter, authorities in the Importer implement a complete travel ban, after which $m_L=m_U=m_D=0$.
Here, we use CTMC simulations to evaluate travel bans triggered by epidemiological indicators. 
(Figure~\ref{fig:interruption-travel} in Appendix~\ref{app:additional-computational-stuff} considers a travel ban implemented at a prescribed time after the start of spread in the Exporter.)

\begin{figure}[htbp]
    \centering
    \begin{subfigure}{0.49\textwidth}
        \includegraphics[width=\linewidth]{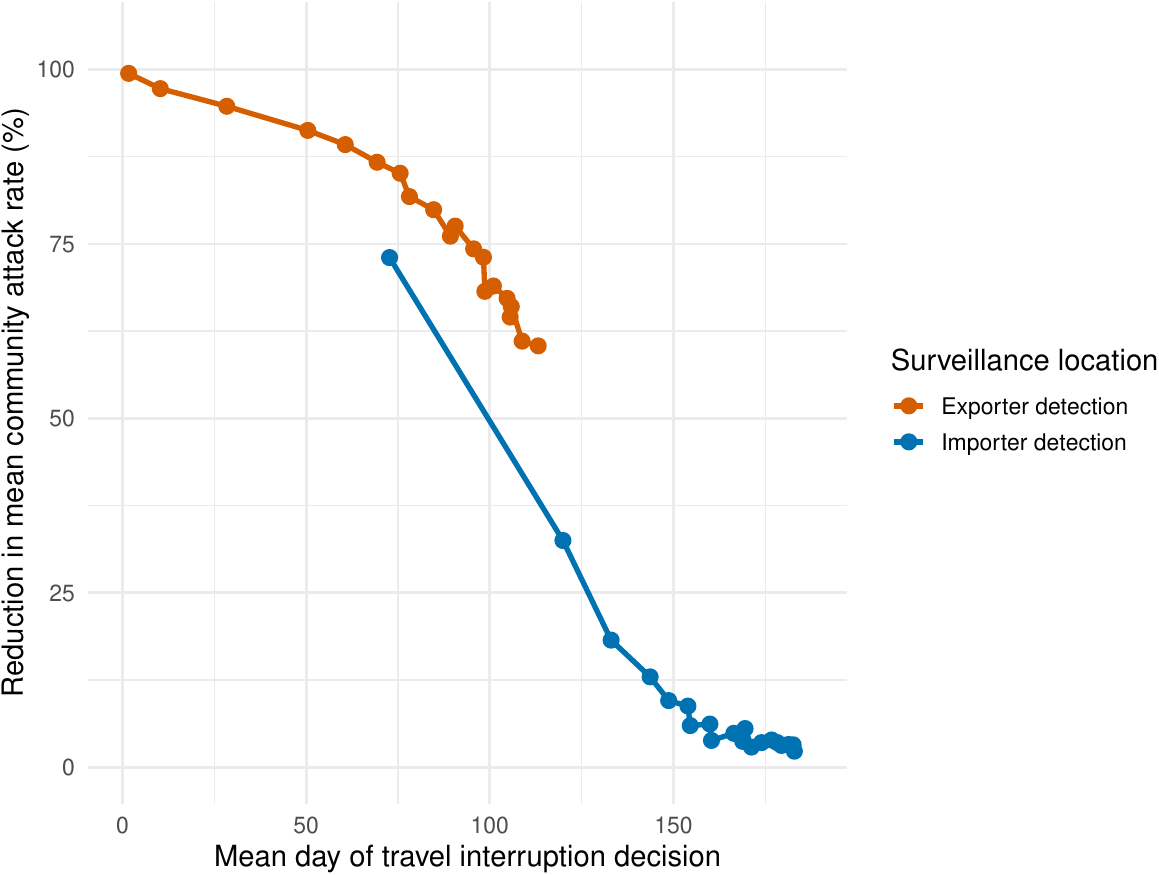}
        \caption{Surveillance in Exporter vs Importer}
        \label{fig:trigger-comparisons-locations}
    \end{subfigure}
    \begin{subfigure}{0.49\textwidth}
        \includegraphics[width=\linewidth]{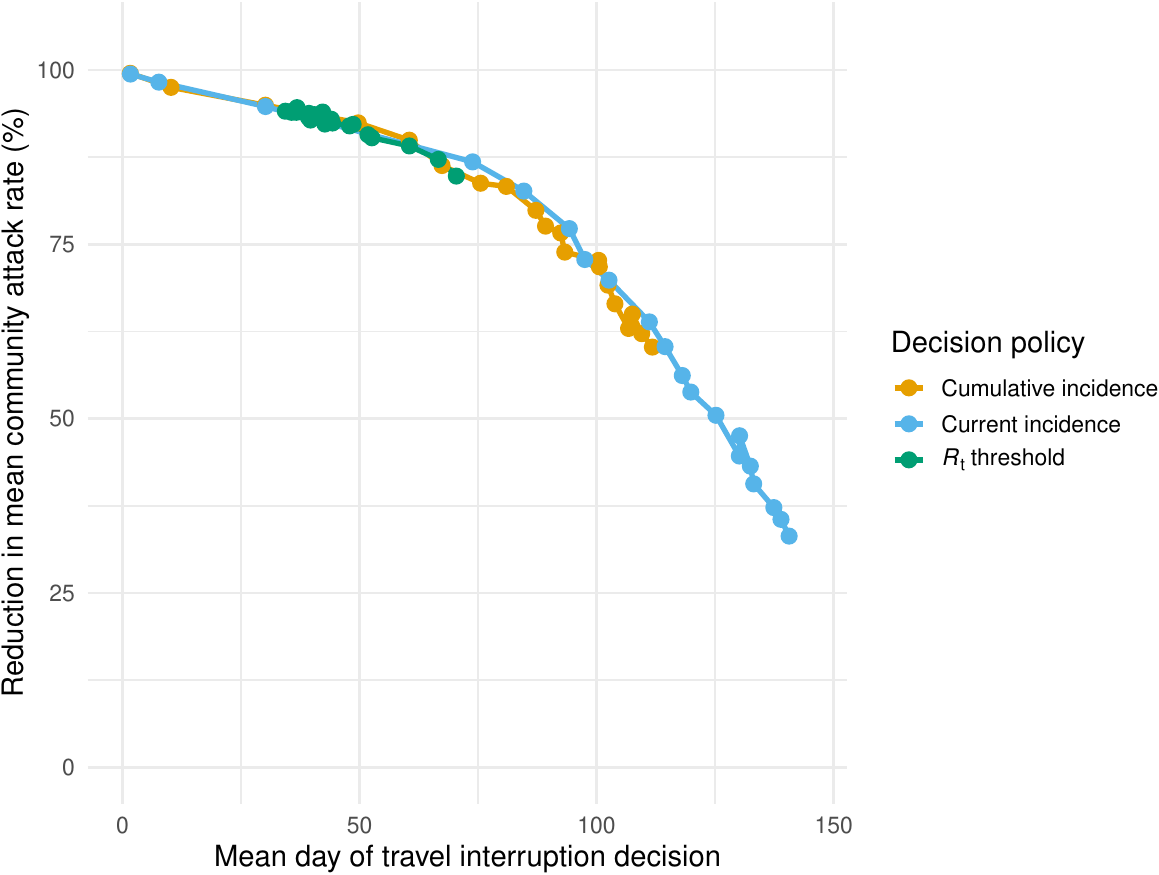}
        \caption{Comparison of triggers in Exporter}
        \label{fig:trigger-comparisons-types}
    \end{subfigure}
    \caption{Relative percentage reduction in the mean community attack rate in the Importer achieved by various travel interruption triggers ($\R_0^E=\R_0^C=1.5$, 5,000 CTMC simulations per threshold; reductions are measured relative to the baseline mean community attack rate of $49.4\%$ without travel restrictions).
    Decisions are taken when the monitored quantity reaches a prescribed threshold (see text), with the resulting mean day of travel interruption shown on the $x$-axis.
    (a) Travel interruption triggered by cumulative detected cases in the Exporter (orange curve) versus in the Importer (blue curve).
    (b) Travel interruption triggered by detected cases ($D_E$) in the Exporter using cumulative incidence (yellow curve), 7-day incidence (blue curve) or $\R_t$ threshold (green curve).}
    \label{fig:trigger-comparisons}
\end{figure}

We first examine whether travel interruptions should be triggered by surveillance in the Exporter or in the Importer.
In Figure~\ref{fig:trigger-comparisons-locations}, a complete travel ban is implemented once the cumulative number of detected cases reaches a given threshold (20 values sampled in $[1,100]$), tracked either in the Exporter ($D_E$, orange curve) or in the Importer ($D_I+D_C$, blue curve), and evaluated in terms of the relative percentage reduction from the unmitigated baseline mean community attack rate of $49.4\%$.
In both curves, the lowest threshold (1 detected case) corresponds to the top-leftmost point and increasing the threshold shifts the decision to later times and smaller reductions in the Importer attack rate.
Acting on cumulative detections in the Exporter clearly outperforms acting on detections in the Importer across all thresholds: even the most sensitive local trigger (1 detected case in the Importer) activates on average around day 73 (achieving a $73.0\%$ reduction, to a mean attack rate of $13.3\%$) and yields a smaller reduction than waiting for over 60 cumulative detected cases in the Exporter.
Furthermore, the reduction in attack rate when monitoring the Importer drops steeply between the first few threshold values (falling below $20\%$ once the threshold reaches 11 cases), whereas the loss of efficacy is much more gradual when monitoring the Exporter.
This illustrates that, although politically more delicate, decisions to interrupt travel should be based on epidemic intelligence in connected Exporters rather than waiting for local detections in the Importer.

Having established that triggers should be based on surveillance in the Exporter, we compare in Figure~\ref{fig:trigger-comparisons-types} three decision mechanisms evaluated using detected cases ($D_E$) in the Exporter, with 20 threshold values tested for each policy:
cumulative detected incidence sampled in $[1,100]$ (yellow curve, identical to the orange curve in Figure~\ref{fig:trigger-comparisons-locations}), current incidence over a trailing 7-day window sampled in $[1,50]$ (blue curve) and the instantaneous reproduction number $\R_t$ estimated over a trailing 7-day window using \texttt{EpiEstim} \cite{CoriFergusonFraserCauchemez2013} (once at least 12 cumulative cases have been detected) sampled in $[1,2.5]$ (green curve).
When plotted against the mean day on which the travel ban is triggered, all three Exporter-based indicators lie along the same efficacy curve: the relative reduction in the Importer attack rate is governed primarily by how early the trigger halts travel, regardless of which specific indicator triggers the decision on that day.
However, the three indicators span different portions of this curve: low thresholds on cumulative or 7-day incidence can trigger a ban within the first few days of spread in the Exporter (achieving reductions above $98\%$), whereas $\R_t$ requires a minimum accumulation of cases before it can be reliably estimated, so even its lowest threshold ($\R_t = 1$) activates on average around day 34 (achieving a reduction of about $94.1\%$).
On the other hand, $\R_t$-based triggers suffer from less ``dispersion'' and within the chosen $[1,2.5]$ range, always lead to a shutdown that achieves a high relative reduction in the attack rate (at least $84.8\%$, keeping the mean attack rate below $8\%$).

\begin{figure}[htbp]
    \centering
    \begin{subfigure}{0.49\textwidth}
        \includegraphics[width=\linewidth]{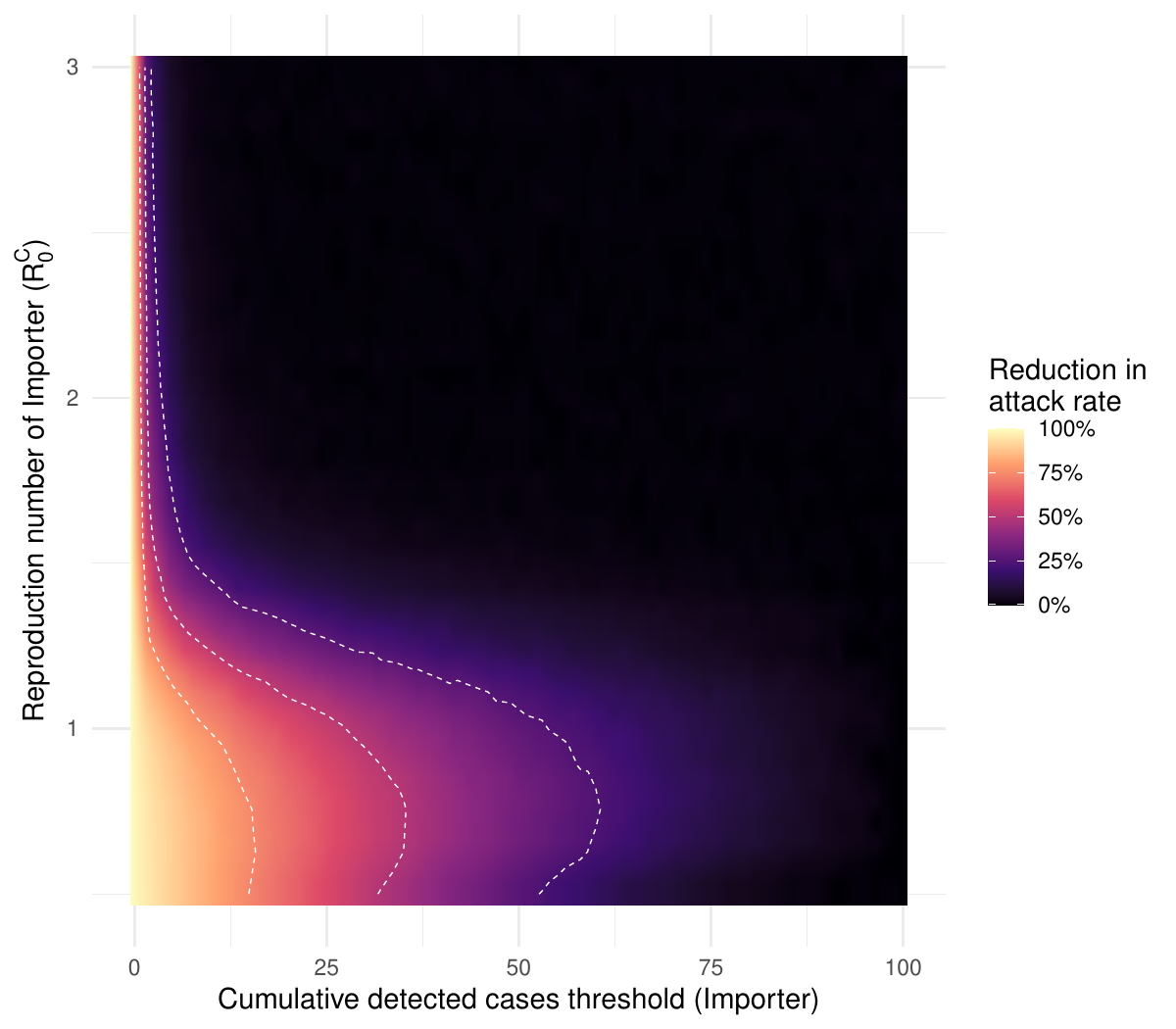}
        \caption{Detection in Importer}
        \label{fig:interruption-travel-threshold-importer}
    \end{subfigure}
    \begin{subfigure}{0.49\textwidth}
        \includegraphics[width=\linewidth]{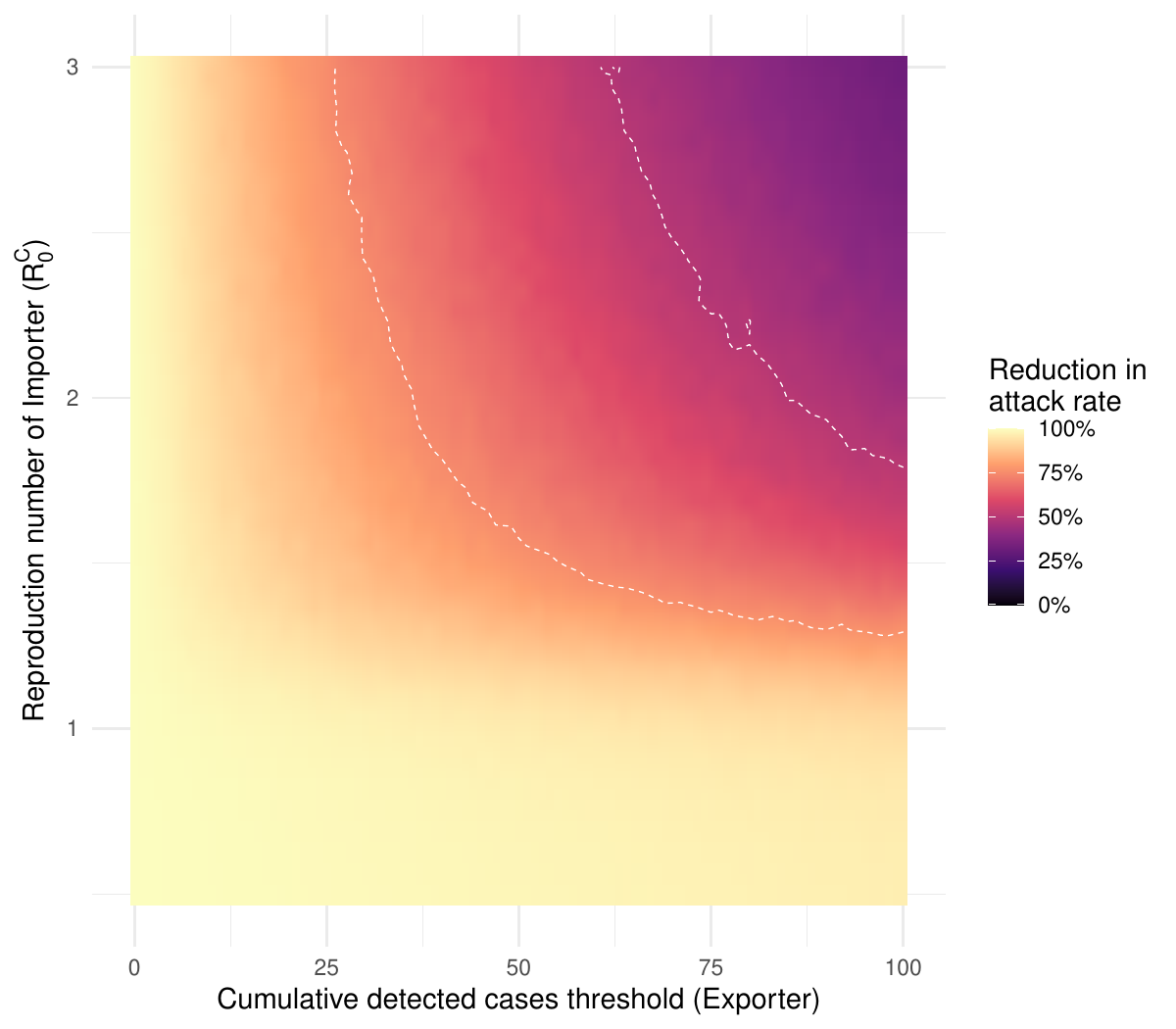}
        \caption{Detection in Exporter}
        \label{fig:interruption-travel-threshold-exporter}
    \end{subfigure}
    \caption{Relative percentage reduction in the mean community attack rate in the Importer ($\R_0^E=1.5$, 10,000 CTMC simulations per grid point) as a function of the local reproduction number $\R_0^C$ and the cumulative detected cases threshold used to trigger a complete travel interruption, when cumulative detections are monitored in (a) the Importer ($D_I+D_C$) and (b) the Exporter ($D_E$).
    Reductions at each $\R_0^C$ are computed relative to the baseline mean attack rate without travel restrictions, which increases from $0.7\%$ at $\R_0^C=0.5$ to $49.4\%$ at $\R_0^C=1.5$ and $86.9\%$ at $\R_0^C=3.0$.}
    \label{fig:interruption-travel-threshold}
\end{figure}

We further explore the contrast between Importer- and Exporter-based cumulative detection triggers across varying levels of local transmission control ($\R_0^C$) using CTMC simulations in Figure~\ref{fig:interruption-travel-threshold}, which extends the comparison of Figure~\ref{fig:trigger-comparisons-locations} ($\R_0^C=1.5$) to the full range $\R_0^C\in[0.5,3]$.
Both panels display the relative percentage reduction in the Importer mean community attack rate on the same scale, measured at each value of $\R_0^C$ against the corresponding unmitigated baseline, when the travel ban is triggered by cumulative detected cases in the Importer (Figure~\ref{fig:interruption-travel-threshold-importer}) versus in the Exporter (Figure~\ref{fig:interruption-travel-threshold-exporter}).
In Figure~\ref{fig:interruption-travel-threshold-importer}, we see that when travel interruptions rely on local detections in the Importer, the relative reduction drops rapidly toward zero as soon as the threshold exceeds a handful of cases, especially for larger values of $\R_0^C$.
By contrast, Figure~\ref{fig:interruption-travel-threshold-exporter} shows that triggering the travel interruption on cumulative detected cases in the Exporter sustains a substantial relative reduction in the Importer attack rate across a much wider range of thresholds across all $\R_0^C>1$, visually confirming the protection gained by acting on Exporter surveillance rather than waiting for local detections.

\subsection{Comparison of travel control measures}
\label{sec:travel-control-comparison}

Comparing the three classes of travel control measures examined in Sections~\ref{sec:travel-control-screening}--\ref{sec:travel-control-travel} highlights why no single border intervention suffices on its own when the Importer community is receptive to transmission ($\R_0^C > 1$).
Pre-departure screening in the Exporter (Figure~\ref{fig:attack-rate-R0E-sigmaE}) targets actively infectious travellers ($U_E$ and $D_E$), but because latently infected individuals ($L_E$) are not yet detectable upon departure, they continue to enter the Importer and seed community outbreaks regardless of the test sensitivity $\sigma_E$.
Post-arrival quarantine and screening in the introduction layer (Figure~\ref{fig:effect-quarantine}) address this limitation by restricting contacts of all incoming travellers during their sojourn in the introduction layer, including those who arrive in $L_I$ and subsequently progress to $D_I$ or $U_I$.
However, unless quarantine stringency is very high and implemented early (or reinforced by intense post-arrival screening $p_I$), even a small number of transmission events from the introduction layer into $S_C$ is sufficient to ignite a self-sustaining epidemic in the community when $\R_0^C > 1$.

Complete or partial travel interruptions (Figures~\ref{fig:trigger-comparisons} and~\ref{fig:interruption-travel-threshold}) cut off the inflow of all infected states ($L_E$, $D_E$ and $U_E$) simultaneously, but their standalone efficacy is very sensitive to the timing and location of the decision trigger.
Waiting for cases to be detected within the Importer before interrupting travel (Figures~\ref{fig:trigger-comparisons-locations} and~\ref{fig:interruption-travel-threshold-importer}) almost always acts too late, because local detection lags behind the arrival of silent latent and undetected introductions.
Triggering interventions based on surveillance in the Exporter (Figure~\ref{fig:interruption-travel-threshold-exporter}) provides a wider window of action, yet its benefit still erodes rapidly as the detection threshold increases.

The analytical final size relations \eqref{eq:cumulative-importations}--\eqref{eq:dLambda-dpI} derived in Appendix~\ref{app:R0-and-final-sizes-exporter-importer} (see Appendices~\ref{app:supercritical-vs-subcritical} and~\ref{app:system-vs-importer-travel}) further clarify why these three classes of measures behave differently when viewed from the perspective of the Importer alone ($\Delta S_C = S_C(0)-S_C(\infty)$) versus the coupled Exporter-Importer system as a whole ($\Delta S_E + \Delta S_C$).
First, as shown in Appendix~\ref{app:supercritical-vs-subcritical}, when $\R_0^C > 1$, equation \eqref{eq:supercritical-DeltaSC} implies that in the deterministic limit, any non-zero cumulative importation load $\Lambda_{E\to I} > 0$ produces nearly the same community final size $\Delta S_C$; thus, when the Importer community is receptive ($\R_0^C > 1$), travel measures reduce the expected attack rate in the CTMC realisations by driving importation counts low enough for introductions to go extinct stochastically before establishment (or by halting travel completely prior to the first successful introduction).
Second, at the system level (Appendix~\ref{app:system-vs-importer-travel}), equations \eqref{eq:dRcE-dm-dsigmaE}--\eqref{eq:dRcE-dsigmaE} show that movement reductions ($m_X \downarrow$) and pre-departure screening ($\sigma_E \uparrow$) retain infected individuals inside the Exporter (since departing travellers in $U_E$ intercepted by screening are placed into $D_E$ and remain in the Exporter), which slightly increases $\R_c^E$ and the Exporter final size $\Delta S_E$ whenever $\xi_E > 0$.
In addition, by \eqref{eq:cumulative-importations}, cumulative latent importations $\mathcal{I}_L$ do not depend explicitly on $\sigma_E$ and are therefore unaffected by pre-departure screening.
By contrast, equation \eqref{eq:dLambda-dpI} shows that post-arrival quarantine and screening ($p_I$ and $\xi_I$) in the introduction layer impose no upward feedback on $\Delta S_E$ while directly attenuating the contribution of $\mathcal{I}_L$ to $\Lambda_{E\to I}$, reducing epidemic burden both in the Importer and across the coupled system.
Consequently, in keeping with the ``no silver bullet'' observation, effective protection requires a layered combination of measures: pairing proactive, Exporter-informed border controls (strict quarantine and post-arrival testing $p_I$ to contain $L_I$ arrivals) with local public health measures that keep $\R_0^C$ near or below one, where \eqref{eq:subcritical-DeltaSC} guarantees that residual border leakages remain small and proportional to $\Lambda_{E\to I}$ rather than sparking major community outbreaks.
\section{Discussion}

The mathematical and computational results developed in this work clarify why border control measures sometimes succeed and sometimes fail, and highlight the decisive role played by the timing of interventions relative to the pathogen introduction process.
When the Importer community is receptive to transmission ($\R_0^C > 1$), the coupled Exporter--Importer dynamics (Sections~\ref{sec:results-importer} and~\ref{sec:results-exporter-importer}) show that successful introductions rapidly carry the Importer across the stochastic establishment phase \cite{ArinoMilliken2022b} into a self-sustaining local epidemic, after which further importations have negligible influence on the course of the outbreak.

The window during which border measures can meaningfully alter epidemic outcomes is narrow and precedes substantial local detection.
Latently infected ($L_E$) and undetected ($U_E$) travellers cross borders invisibly, and because of this, pre-departure screening ($\sigma_E$) alone cannot stop the seeding of outbreaks (Section~\ref{sec:travel-control-screening}), and cumulative case counts or $\R_t$ estimates within the Importer inherently lag behind actual introductions.
Triggering travel interruptions in response to local detections in the Importer (Section~\ref{sec:travel-control-travel}) consequently acts too late in most realisations, explaining why reactive travel bans during the COVID-19 pandemic largely failed to prevent the introduction of new variants \cite{GrepinHoLiEtAl2020,grepin2023effectiveness,habibi2020do} and imposed societal disruption without delivering the health protection required under Article 43 of the International Health Regulations \cite{LeePiper2025,WHO-IHR-2005}.

The jurisdictions where travel control worked well (such as Newfoundland and Labrador, Nunavut, Prince Edward Island, Yukon, northern Manitoba and several island nations \cite{CarinoQuillEtAl2022,DaSilva2023,gay2021pacific,HurfordMartignoniLoredoOstiAnokyeEtAl2022,McPheeKnowlesHoffmanKanary,MohammadiCojocaruArinoHurford2023,NAO2020PEI,Manitoba2020NorthernOrder,sun2023covid,VilchesAbdollahiCiprianoEtAl2022}) shared conditions that our model identifies as essential: geographic isolation with few, controllable entry points, standing or early Exporter-informed border protocols and strict post-arrival quarantine or testing ($p_I$) in the introduction layer (Section~\ref{sec:travel-control-quarantine}) capable of intercepting latent arrivals ($L_I$) before they interact with the community, all reinforced by local public health measures that kept $\R_0^C$ near or below one so that residual leakages fizzled out (Section~\ref{sec:travel-control-comparison}).

Beyond explaining the contrasting outcomes of past border interventions, our results address a critical operational gap when managing infectious diseases within zero-prevalence regions \cite{HurfordMartignoniLoredoOstiAnokyeEtAl2022,MartignoniArinoHurford2024}.
When a jurisdiction successfully eliminates community transmission, public risk perception naturally wanes and the political justification for maintaining local non-pharmaceutical interventions (such as social distancing or capacity limits, which act to keep $\R_0^C$ low) becomes increasingly fragile.
In addition, if there are no introductions, then these community measures can be relaxed without any consequences, which can support the overall economic viability of an elimination strategy \cite{MartignoniArinoHurford2024,Plank2026Joint}.
Due to the importance of undetected community cases, analyses that determine this probability \cite{PlankHartPolonskyEtAl2025} are necessary to inform the viability of relaxed community measures.
Similarly, as vaccination coverage increases in the Importer, public health officials need a quantitative basis to plan the safe rollback of local restrictions and border measures.

Without active community transmission to measure, officials cannot rely on standard epidemiological models calibrated to local case counts.
While statistical and mechanistic estimates of importation volumes and spillover counts provide an essential baseline for importation risk \cite{HurfordMartignoniLoredoOstiAnokyeEtAl2022,MilwidGabrieleRivetOgdenEtAl2024,MohammadiCojocaruArinoHurford2023}, importation rates alone do not translate directly into operational readiness.
By coupling the Exporter dynamics, the introduction layer and the Importer community in both deterministic and stochastic regimes, our framework connects importation rates with the risk and expected characteristics of an outbreak in the destination population, projecting not only the probability of an outbreak (Theorems~\ref{th:exists-fix-pt-importer} and~\ref{th:exists-fp-exp-imp}) but also its expected duration, peak timing, peak prevalence and final attack rate under varying combinations of travel control and local $\R_0^C$.
These projections allow policymakers to assess the risk of exceeding local healthcare capacity if an importation establishes, and demonstrate why even in jurisdictions where zero-prevalence control is achievable, relying on local case counts alone to trigger interventions is insufficient; proactive intelligence on incidence in connected Exporters is essential.
Given that the 2024 revisions to the International Health Regulations left the substantive rules governing travel measures under Article 43 largely unchanged, quantitative risk-based frameworks that integrate pathogen dynamics with exporter and importer characteristics are all the more necessary to guide coordinated border management in future emergencies \cite{LeePiper2025}.

Several simplifying assumptions of our modelling framework warrant discussion.
First, the model is simplified in the sense that only one of each type of compartment (latent, detected, undetected) is present, compared to the two used in \cite{ArinoBoelleMillikenPortet2021} or even more considered in \cite{ArinoBajeuxPortetWatmough2020}. 
While this makes sojourn times in groups of compartments exponentially distributed instead of Erlang distributed, it reduces model complexity.
Second, we consider only a single variant here.
Third, we ignore deaths and just consider individuals as having been \emph{removed} from being infectious for whatever reason.
Indeed, the aim here is not to fit data.

Even with these simplifications, our explicit separation of infectious individuals into detected ($D$) and undetected ($U$) compartments sheds light on debates surrounding underreporting and unexplained excess mortality in low-prevalence jurisdictions such as Newfoundland and Labrador.
Because every community infection ($S_C \to L_C$), even when transmitted by an undetected individual in $U_I$ or $U_C$, subsequently enters the detected compartment $D_C$ with probability $p_C$ upon completing latency, the incidence of detected cases is structurally coupled to the incidence of undetected cases by the ratio $p_C/(1-p_C)$.
Consequently, provided basic testing and surveillance are in place ($p_C > 0$), a large decoupled wave of undetected community transmission cannot persist silently without generating a commensurate surge in detected cases.
When $\R_0^C < 1$, undetected introductions generate only stuttering chains that quickly go extinct in both $U_C$ and $D_C$; when $\R_0^C > 1$, any established community spread grows exponentially and rapidly manifests in $D_C$.
It is therefore epidemiologically implausible to attribute substantial unexplained excess mortality in near-zero-incidence regions to widespread, unrecorded SARS-CoV-2 community transmission.

\section*{CRediT authorship contribution statement}
Julien Arino, Clotilde Djuikem and Amy Hurford contributed equally to this work across conceptualisation, methodology, formal analysis, investigation, software, visualisation, writing of the original draft and review and editing.

\section*{Funding}
JA and AH are supported in part by Natural Sciences and Engineering Research Council of Canada (NSERC) Discovery Grants.
Simulations were run on a cluster funded by a 2020 Public Health Agency of Canada (PHAC) contract to JA.

\section*{Data and code availability}
No empirical data were used in this study.
The computational code used to run the simulations and generate the figures is available from the corresponding author upon request.

\section*{Declaration of competing interest}
The authors declare that they have no known competing financial interests or personal relationships that could have appeared to influence the work reported in this paper.

\bibliographystyle{plain}
\bibliography{references}

\begin{thebibliography}{10}

\bibitem{ALLEN201399}
L.~J.~S. Allen and P.~{van den Driessche}.
\newblock Relations between deterministic and stochastic thresholds for disease
  extinction in continuous- and discrete-time infectious disease models.
\newblock {\em Mathematical Biosciences}, 243(1):99--108, 2013.

\bibitem{Andreasen2011}
V.~Andreasen.
\newblock The final size of an epidemic and its relation to the basic
  reproduction number.
\newblock {\em Bulletin of Mathematical Biology}, 73(10):2305--2321, 2011.

\bibitem{ArinoBrauerDriesscheWatmoughWu2007}
J.~Arino, F.~Brauer, P.~van~den Driessche, J.~Watmough, and J.~Wu.
\newblock {A final size relation for epidemic models}.
\newblock {\em Mathematical Biosciences and Engineering}, 4(2):159--175, 2007.

\bibitem{athreya1972branching}
K.~B. Athreya and P.~E. Ney.
\newblock {\em Branching Processes}.
\newblock Springer, 1972.

\bibitem{berman1979nonnegative}
A.~Berman and R.~J. Plemmons.
\newblock {\em Nonnegative Matrices in the Mathematical Sciences}.
\newblock Academic Press, New York, 1979.

\bibitem{BidariEtAl2016}
S.~Bidari, X.~Chen, D.~Peters, D.~Pittman, and P.~L. Simon.
\newblock Solvability of implicit final size equations for {SIR} epidemic
  models.
\newblock {\em Mathematical Biosciences}, 282:181--190, 2016.

\bibitem{Brauer2019}
F.~Brauer.
\newblock The final size of a serious epidemic.
\newblock {\em Bulletin of Mathematical Biology}, 81(3):869--877, 2019.

\bibitem{Harris1963}
T.~E. Harris.
\newblock {\em The Theory of Branching Processes}, volume 119 of {\em
  Grundlehren der mathematischen Wissenschaften}.
\newblock Springer, Berlin, Heidelberg, 1963.

\bibitem{MagalSeydiWebb2016}
P.~Magal, O.~Seydi, and G.~Webb.
\newblock Final size of an epidemic for a two-group {SIR} model.
\newblock {\em SIAM Journal on Applied Mathematics}, 76(5):2042--2059, 2016.

\bibitem{MagalSeydiWebb2018}
P.~Magal, O.~Seydi, and G.~Webb.
\newblock Final size of a multi-group {SIR} epidemic model: {I}rreducible and
  non-irreducible modes of transmission.
\newblock {\em Mathematical Biosciences}, 301:59--67, 2018.

\bibitem{VdDWatmough2002}
P.~van~den Driessche and J.~Watmough.
\newblock Reproduction numbers and sub-threshold endemic equilibria for
  compartmental models of disease transmission.
\newblock {\em Mathematical Biosciences}, 180:29--48, 2002.

\end{thebibliography}


\begin{thebibliography}{10}

\bibitem{ArinoBajeuxPortetWatmough2020}
J.~Arino, N.~Bajeux, S.~Portet, and J.~Watmough.
\newblock Quarantine and the risk of {COVID}-19 importation.
\newblock {\em Epidemiology \& Infection}, 148:e298, 2020.

\bibitem{ArinoBoelleMillikenPortet2021}
J.~Arino, P.-Y. Bo\"elle, E.~M. Milliken, and S.~Portet.
\newblock Risk of {COVID-19} variant importation -- {H}ow useful are travel
  control measures?
\newblock {\em Infectious Disease Modelling}, 6:875--897, 2021.

\bibitem{ArinoBrauerPvdDWatmoughWu2006}
J.~Arino, F.~Brauer, P.~van~den Driessche, J.~Watmough, and J.~Wu.
\newblock {Simple models for containment of a pandemic}.
\newblock {\em Journal of the Royal Society Interface}, 3(8):453--457, 2006.

\bibitem{ArinoMilliken2022b}
J.~Arino and E.~M. Milliken.
\newblock Effect of movement on the early phase of an epidemic.
\newblock {\em Bulletin of Mathematical Biology}, 84(11):128, 2022.

\bibitem{ArinoVdD2003a}
J.~Arino and P.~van~den Driessche.
\newblock A multi-city epidemic model.
\newblock {\em Mathematical Population Studies}, 10(3):175--193, 2003.

\bibitem{CarinoQuillEtAl2022}
M.~Carino, Z.~Quill, M.~Gabbs, E.~Sellers, J.~Hamilton, T.~Pinto, M.~Jetha,
  J.~Ho, O.~Garcia~Alecio, A.~Dart, and B.~Wicklow.
\newblock Impact of {COVID-19} pandemic on adolescents and young adults living
  with type 2 diabetes.
\newblock {\em Canadian Journal of Diabetes}, 46(2):145--151, 2022.

\bibitem{CoriFergusonFraserCauchemez2013}
A.~Cori, N.~M. Ferguson, C.~Fraser, and S.~Cauchemez.
\newblock A new framework and software to estimate time-varying reproduction
  numbers during epidemics.
\newblock {\em American Journal of Epidemiology}, 178(9):1505--1512, 2013.

\bibitem{DaSilva2023}
M.~Da~Silva.
\newblock Public health, internal borders, and the ends of federalism.
\newblock {\em Canadian Journal of Law and Society}, 38(3):409--429, 2023.

\bibitem{ermus2023great}
C.~Ermus.
\newblock {\em The Great Plague Scare of 1720: Disaster and Diplomacy in the
  Eighteenth-Century {Atlantic} World}.
\newblock Cambridge University Press, 2023.

\bibitem{gay2021pacific}
N.~Gay, C.~van Gemert, O.~E. Merilles, and P.~Georgel.
\newblock {Pacific Island} nations face an urgent need for actions and future
  research on {COVID-19}.
\newblock {\em The Lancet Regional Health -- Western Pacific}, 18:100326, 2022.

\bibitem{GovernmentofCanadaCBSA}
{Government of Canada}.
\newblock Timeline -- {CBSA} border measures 2021 ({U}pdated {April} 23, 2021),
  2021.
\newblock
  \url{https://www.publicsafety.gc.ca/cnt/trnsprnc/brfng-mtrls/prlmntry-bndrs/20210907/05-en.aspx}.

\bibitem{grepin2023effectiveness}
K.~A. Gr{\'e}pin, J.~Aston, and J.~Burns.
\newblock Effectiveness of international border control measures during the
  {COVID-19} pandemic: a narrative synthesis of published systematic reviews.
\newblock {\em Philosophical Transactions of the Royal Society A},
  381(2257):20230134, 2023.

\bibitem{GrepinHoLiEtAl2020}
K.~A. Gr{\'e}pin, T.-L. Ho, Z.~Liu, S.~Marion, J.~Piper, C.~Z. Worsnop, and
  K.~Lee.
\newblock Evidence of the effectiveness of travel-related measures during the
  early phase of the {COVID-19} pandemic: a rapid systematic review.
\newblock {\em BMJ Global Health}, 6(3):e004537, 2021.

\bibitem{habibi2020do}
R.~Habibi, G.~L. Burci, T.~C. de~Campos, D.~Chirwa, M.~Cin{\`a}, S.~Dagron,
  M.~Eccleston-Turner, L.~Forman, L.~O. Gostin, B.~M. Meier, S.~Negri, G.~Ooms,
  S.~Sekalala, A.~Taylor, A.~E. Yamin, and S.~J. Hoffman.
\newblock Do not violate the {International Health Regulations} during the
  {COVID-19} outbreak.
\newblock {\em The Lancet}, 395(10225):664--666, 2020.

\bibitem{HurfordMartignoniLoredoOstiAnokyeEtAl2022}
A.~Hurford, M.~M. Martignoni, J.~C. Loredo-Osti, F.~Anokye, J.~Arino, B.~S.
  Husain, B.~Gaas, and J.~Watmough.
\newblock Pandemic modelling for regions implementing an elimination strategy.
\newblock {\em Journal of Theoretical Biology}, 561:111378, 2023.

\bibitem{iooss2026sensitivity}
B.~Iooss, S.~Da~Veiga, A.~Janon, and G.~Pujol.
\newblock {\em {sensitivity}: Global Sensitivity Analysis of Model Outputs and
  Importance Measures}, 2026.
\newblock R package version 1.31.0.

\bibitem{KhanEckhardtBrownsteinNaqviEtAl2013}
K.~Khan, R.~Eckhardt, J.~S. Brownstein, R.~Naqvi, W.~Hu, D.~Kossowsky,
  D.~Scales, J.~Arino, M.~Macdonald, J.~Wang, J.~Sears, and M.~S. Cetron.
\newblock Entry and exit screening of airline travellers during the {A(H1N1)}
  2009 pandemic: a retrospective evaluation.
\newblock {\em Bulletin of the World Health Organization}, 91(5):368--376,
  2013.

\bibitem{kosciejew2023documentary}
M.~Kosciejew.
\newblock A documentary history of the immunity (or vaccine) passport: health
  certificates of public health, personal identity and power from the plague to
  the coronavirus pandemic.
\newblock {\em Social History of Medicine}, 36(1):110--138, 2023.

\bibitem{LauerGrantzBiEtAl2020}
S.~A. Lauer, K.~H. Grantz, Q.~Bi, F.~K. Jones, Q.~Zheng, H.~R. Meredith, A.~S.
  Azman, N.~G. Reich, and J.~Lessler.
\newblock The incubation period of coronavirus disease 2019 ({COVID-19}) from
  publicly reported confirmed cases: Estimation and application.
\newblock {\em Annals of Internal Medicine}, 172(9):577--582, 2020.

\bibitem{LeePiper2025}
K.~Lee and J.~Piper.
\newblock Latest revisions to the {International Health Regulations} will fail
  to prevent future travel chaos.
\newblock {\em BMJ Global Health}, 10(1):e017077, 2025.

\bibitem{MartignoniArinoHurford2024}
M.~M. Martignoni, J.~Arino, and A.~Hurford.
\newblock Is {SARS-CoV-2} elimination or mitigation best? {Regional} and
  disease characteristics determine the recommended strategy.
\newblock {\em Royal Society Open Science}, 11(6):240186, 2024.

\bibitem{McPheeKnowlesHoffmanKanary}
S.~McPhee-Knowles, B.~Hoffman, and L.~Kanary.
\newblock The {Yukon}'s experience with {COVID-19}: {Travel} restrictions,
  variants and spread among the unvaccinated.
\newblock {\em Canada Communicable Disease Report}, 48(1):17--21, 2022.

\bibitem{MilwidGabrieleRivetOgdenEtAl2024}
R.~M. Milwid, V.~Gabriele-Rivet, N.~H. Ogden, P.~Turgeon, A.~Fazil, D.~London,
  S.~de~Montigny, and E.~E. Rees.
\newblock A methodology for estimating {SARS-CoV-2} importation risk by air
  travel into {Canada} between {July} and {November} 2021.
\newblock {\em BMC Public Health}, 24(1):1088, 2024.

\bibitem{MohammadiCojocaruArinoHurford2023}
Z.~Mohammadi, M.~G. Cojocaru, J.~Arino, and A.~Hurford.
\newblock Importation models for travel-related {SARS-CoV-2} cases reported in
  {Newfoundland} and {Labrador} during the {COVID-19} pandemic.
\newblock {\em Royal Society Open Science}, 12(11):241902, 2025.

\bibitem{NAO2020PEI}
{North American Observatory on Health Systems and Policies}.
\newblock North {American} {COVID-19} policy response monitor: {Prince Edward
  Island}.
\newblock Technical report, University of Toronto, 2020.

\bibitem{PlankHartPolonskyEtAl2025}
M.~J. Plank, W.~S. Hart, J.~Polonsky, M.~Keita, S.~Ahuka-Mundeke, and R.~N.
  Thompson.
\newblock Estimation of end-of-outbreak probabilities in the presence of
  delayed and incomplete case reporting.
\newblock {\em Proceedings of the Royal Society B: Biological Sciences},
  292(2039):20242825, 2025.

\bibitem{Plank2026Joint}
M.~J. Plank, M.~Sushames, T.~Fisher-Taylor, A.~Afshari, R.~N. Thompson,
  A.~Hurford, and S.~Hendy.
\newblock Joint economic and epidemiological modelling of alternative pandemic
  response strategies.
\newblock {\em Journal of the Royal Society Interface}, 23(241):20260098,
  August 2026.

\bibitem{Manitoba2020NorthernOrder}
{Province of Manitoba}.
\newblock Order under {The Public Health Act} prohibiting travel to {Northern
  Manitoba} and remote communities, 2020.
\newblock Chief Provincial Public Health Officer, April 17, 2020.

\bibitem{PHAC2020CaseContact}
{Public Health Agency of Canada}.
\newblock Public health management of cases and contacts associated with
  coronavirus disease ({COVID-19}).
\newblock Technical report, Government of Canada, Ottawa, ON, 2020.

\bibitem{sobol2001global}
I.~M. Sobol.
\newblock Global sensitivity indices for nonlinear mathematical models and
  their {Monte Carlo} estimates.
\newblock {\em Mathematics and Computers in Simulation}, 55(1--3):271--280,
  2001.

\bibitem{sun2023covid}
X.~Sun, S.~Wandelt, and A.~Zhang.
\newblock Why are {COVID-19} travel bubbles a tightrope walk? {An}
  investigation based on the {Trans-Tasmanian} case.
\newblock {\em Communications in Transportation Research}, 3:100089, 2023.

\bibitem{VilchesAbdollahiCiprianoEtAl2022}
T.~N. Vilches, E.~Abdollahi, L.~E. Cipriano, M.~Haworth-Brockman, Y.~Keynan,
  H.~Sheffield, J.~M. Langley, and S.~M. Moghadas.
\newblock Impact of non-pharmaceutical interventions and vaccination on
  {COVID-19} outbreaks in {Nunavut}, {Canada}: a {Canadian Immunization
  Research Network (CIRN)} study.
\newblock {\em BMC Public Health}, 22(1):1042, 2022.

\bibitem{WeberKopfVaha}
E.~Weber, A.~Kopf, and M.~Vaha.
\newblock {COVID-19} pandemics in the {Pacific} {Island} countries and
  territories.
\newblock In R.~Akhtar, editor, {\em Coronavirus ({COVID-19}) Outbreaks,
  Environment and Human Behaviour: International Case Studies}, pages 25--47.
  Springer, 2021.

\bibitem{wikipediaCOVDzero}
{Wikipedia}.
\newblock {Zero-COVID}, 2024.
\newblock Retrieved on May 30, 2024.

\bibitem{WHO-IHR-2005}
{World Health Organization}.
\newblock {\em International Health Regulations (2005)}.
\newblock World Health Organization, 2nd edition, 2008.

\bibitem{who2022tuvalu}
{World Health Organization}.
\newblock {World Health Organization} coronavirus ({COVID-19}) dashboard, 2022.

\end{thebibliography}

\clearpage
\newpage
\appendix
\counterwithin{equation}{section}
\counterwithin{figure}{section}
\counterwithin{table}{section}
\setcounter{equation}{0}
\setcounter{theorem}{0}
\renewcommand{\theequation}{\thesection.\arabic{equation}}
\renewcommand{\thetheorem}{\thesection.\arabic{theorem}}
\setcounter{section}{0}
\setcounter{page}{1}

\begin{center}
    \Large
    Introduced cases, spread of infection in a community and travel control measures \\[0.2cm]
    Supplementary material \\[0.5cm]
    \large
    Julien Arino\,\orcidlink{0000-0001-6409-5027}$^{1}$, Clotilde Djuikem\,\orcidlink{0000-0003-4815-743X}$^{1}$ and Amy Hurford\,\orcidlink{0000-0001-6461-5686}$^{2}$ \\[0.3cm]
    \normalsize
    $^{1}$Department of Mathematics, University of Manitoba, Winnipeg, Manitoba, Canada \\
    $^{2}$Departments of Biology and Mathematics \& Statistics, Memorial University of Newfoundland and Labrador, St John's, Newfoundland and Labrador, Canada
\end{center}
\vskip0.5cm

\begin{abstract}
We give here more details about some of the content of the main paper.
To distinguish between content of the paper itself and content in this Supplemental material, sections, equations and results here are numbered using or starting with letters.
References are also specific to this part.
\end{abstract}

\section{Reproduction numbers and final sizes}
\label{app:R0-and-final-sizes}
\subsection{In the Exporter}
\label{app:R0-and-final-sizes-exporter}
We use the method in \citeapp{ArinoBrauerDriesscheWatmoughWu2007} with $\bx=(L_E,D_E,U_E)$, $D=1$, $\beta(x,y,z)=\beta_E$, $\Pi=(1,0,0)^T$, $\bb=(0,\xi_E,1)$ and 
\begin{equation}\label{eq:V-exporter}
\bV_E =
\begin{pmatrix}
\varepsilon_E+m_L & 0 & 0 \\
-p_E\varepsilon_E & \gamma_E+m_D & -\sigma_Em_U \\
-(1-p_E)\varepsilon_E & 0 & \gamma_E+m_U
\end{pmatrix}.
\end{equation}
From this, it follows by \citeapp[Theorem 2.1]{ArinoBrauerDriesscheWatmoughWu2007} that the control reproduction number in the Exporter is
\[
\R_c^E = \beta_E \bb\bV_E^{-1}\Pi DS_E(0) =  \bb\bV_E^{-1}\Pi \beta_ES_E(0).
\]
Because of the structure of $\bb$ and $\Pi$, $\bb\bV_E^{-1}\Pi=\xi_E\bV_E^{-1}(2,1)+\bV_E^{-1}(3,1)$, where $\bV_E^{-1}(i,j)$ denotes the $(i,j)$ entry in $\bV_E^{-1}$.
We find
\[
\bV_E^{-1}(2,1) = \frac{p_E\varepsilon_E(\gamma_E+m_U)+(1-p_E)\sigma_Em_U\varepsilon_E}
{(\varepsilon_E+m_L)(\gamma_E+m_D)(\gamma_E+m_U)}
\]
and
\[
\bV_E^{-1}(3,1) = \frac{(1-p_E)\varepsilon_E}{(\varepsilon_E+m_L)(\gamma_E+m_U)}.
\]
From this, we obtain the expression \eqref{eq:RcE} for $\R_c^E$,
\[
\R_c^E =
\frac{\varepsilon_E}{\varepsilon_E+m_L}
\left(
\frac{\xi_E p_E}{\gamma_E+m_D}
+
\frac{1-p_E}{\gamma_E+m_U}
\left[1+\frac{\xi_E\sigma_Em_U}{\gamma_E+m_D}\right]
\right)
\beta_ES_E(0),
\]
from which we deduce the expression \eqref{eq:R0E} for $\R_0^E = \left.\R_c^E\right|_{\sigma_E=0}$ when $\sigma_E = 0$.
The method in \citeapp{ArinoBrauerDriesscheWatmoughWu2007} also gives the final size relation; since there is a single susceptible class, we can use \citeapp[Eq. (10)]{ArinoBrauerDriesscheWatmoughWu2007}, giving
\[
\ln\left(
\frac{S_E(0)}{S_E(\infty)}
\right)
=
\frac{\R_c^E}{S_E(0)}
\left(S_E(0)-S_E(\infty)\right)
+\beta_E\bb\bV_E^{-1}\bx(0),
\]
where $S_E(\infty)$ denotes the value of $S_E(t)$ as $t\to\infty$.
By the theory in \citeapp{ArinoBrauerDriesscheWatmoughWu2007}, we know that this limit exists and is unique.
Here,
\begin{equation}
\label{eq:final-size-exporter-bVx}
\begin{aligned}
\bb\bV_E^{-1}\bx(0) &=
\left[\bV_E^{-1}(2,1)L_E(0)+
\bV_E^{-1}(2,2)D_E(0)+
\bV_E^{-1}(2,3)U_E(0)\right]\xi_E \\
&\quad +\bV_E^{-1}(3,1)L_E(0)
+\bV_E^{-1}(3,3)U_E(0).
\end{aligned}
\end{equation}
Besides $\bV_E^{-1}(2,1)$ and $\bV_E^{-1}(3,1)$ already made explicit above, we have
\[
\bV_E^{-1}(2,2)=\frac{1}{\gamma_E+m_D},
\]
\[
\bV_E^{-1}(2,3)=\frac{\sigma_Em_U}
{(\gamma_E+m_D)(\gamma_E+m_U)}
\]
and
\[
\bV_E^{-1}(3,3)=
\frac{1}{\gamma_E+m_U}.
\]
Using \eqref{eq:final-size-exporter-bVx}, we find the final size relation \eqref{eq:final-size-Exporter}.

\subsection{In the Exporter when $m_D=0$}
\label{app:R0-and-final-sizes-exporter-mD0}
When detected individuals cannot travel ($m_D=0$), the transition matrix $\bV_E$ in \eqref{eq:V-exporter} is identical except its $(2,2)$ entry simplifies to $\gamma_E$.
Consequently, substituting $m_D=0$ into the prior computations gives the control reproduction number
\[
\left.\R_c^E\right|_{m_D=0} = 
\frac{\varepsilon_E}{\varepsilon_E+m_L}
\left(
\frac{\xi_E p_E}{\gamma_E}
+
\frac{1-p_E}{\gamma_E+m_U}
\left[1+\frac{\xi_E\sigma_Em_U}{\gamma_E}\right]
\right)
\beta_ES_E(0).
\]
The final size relation and $\bV_E^{-1}$ components remain the same as in Appendix~\ref{app:R0-and-final-sizes-exporter}, substituting $m_D = 0$.

\subsection{In the Importer without travel}
\label{app:R0-and-final-sizes-importer}
It is not possible to compute an $\R_0$ for the Importer by itself in presence of connection to the Exporter.
Indeed, the inflows of $L_E$, $D_E$ and $U_E$ individuals precludes the existence of a disease-free equilibrium, despite the fact that they are known to go to zero.
We can nonetheless reason in the absence of connection, i.e., when $m_L=m_D=m_U=0$.
To use the method in \citeapp{ArinoBrauerDriesscheWatmoughWu2007}, order infected variables as 
\[
\bx=(L_I,D_I,U_I,L_C,D_C,U_C)
\]
and set $D=1$ since $S_C$ is the only susceptible compartment, $\beta(x,y,z)=\beta_{C}$, $\Pi=(0,0,0,1,0,0)^T$ the fraction of the susceptible compartment going into the different infected compartments ($L_I$ only here), 
\[
\bb=\left(
0,\frac{\beta_{I}\xi_{I}}{\beta_{C}},\frac{\beta_{I}}{\beta_{C}},0,\xi_{C},1
\right)
\] 
the relative horizontal transmissions and 
\begin{equation}\label{eq:V-importer}
\bV_I =
\begin{pmatrix}
\varepsilon_I & 0 & 0 & 0 & 0 & 0 \\
-p_I\varepsilon_I & \gamma_I & 0 & 0 & 0 & 0 \\
-(1-p_I)\varepsilon_I & 0 & \gamma_I & 0 & 0 & 0 \\
0 & 0 & 0 & \varepsilon_C & 0 & 0 \\
0 & 0 & 0 & -p_C\varepsilon_C & \gamma_C & 0 \\
0 & 0 & 0 & -(1-p_C)\varepsilon_C & 0 & \gamma_C 
\end{pmatrix}    
\end{equation}
the other flows within infected compartments.
From \citeapp[Theorem 2.1]{ArinoBrauerDriesscheWatmoughWu2007},
\[
\R_0^C = \beta_{C}\bb\bV_I^{-1}\Pi DS_C(0)
=\bb\bV_I^{-1}\Pi\beta_CS_C(0).
\]
Because of the structure of $\bb$, $\bV_I^{-1}$ and $\Pi$, $\bb\bV_I^{-1}\Pi=\xi_C\bV_I^{-1}(5,4)+\bV_I^{-1}(6,4)$, with
\[
\bV_I^{-1}(5,4) = \frac{p_C}{\gamma_C}
\]
and
\[
\bV_I^{-1}(6,4) = \frac{1-p_C}{\gamma_C}.
\]
The expression \eqref{eq:R0C} follows.
As for the Exporter, we use the method in \citeapp{ArinoBrauerDriesscheWatmoughWu2007} to obtain a final size relation; since there is a single susceptible class, we can use \citeapp[Eq. (10)]{ArinoBrauerDriesscheWatmoughWu2007}, giving
\[
\ln\left(
\frac{S_C(0)}{S_C(\infty)}
\right)
=
\frac{\R_0^C}{S_C(0)}
\left(S_C(0)-S_C(\infty)\right)
+\beta_C\bb\bV_I^{-1}\bx(0),
\]
where $S_C(\infty)$ denotes the value of $S_C(t)$ as $t\to\infty$.
We simplify this general expression by noting that in all the cases we consider, $L_C(0)=D_C(0)=U_C(0)$ since we are interested in the consequences of introductions into the Importer.
Doing so, the expression \eqref{eq:final-size-importer-bVx} follows.

\subsection{$\mathcal{R}_0$ and final size in the full Exporter-Importer system}
\label{app:R0-and-final-sizes-exporter-importer}
It is not possible to use the method of \citeapp{ArinoBrauerDriesscheWatmoughWu2007} with the full system because infections are ``localised'' to the locations.
This also means that a final size relation cannot be obtained directly from \citeapp{ArinoBrauerDriesscheWatmoughWu2007}.
To compute the basic reproduction number for the full system, we first switch to using the more general next generation matrix method of \citeapp{VdDWatmough2002}.
Ordering infected variables as $(L_E,D_E,U_E,L_I,D_I,U_I,L_C,D_C,U_C)$, the vector of new infections $\mathcal{F}$ and the vector of transfers between compartments $\mathcal{V}$ take the form
\[
    \mathcal{F}=\begin{pmatrix}
        \beta_E(\xi_ED_E+U_E)S_E\\
        0\\
        0\\
        0\\
        0\\
        0\\
        \left[\beta_{I}(\xi_{I}D_I+U_I)+\beta_{C}(\xi_{C}D_C+U_C)\right]S_C\\
        0\\
        0
    \end{pmatrix}
\]
and
\[
    \mathcal{V}=\begin{pmatrix}
         (\varepsilon_E+m_L)L_E\\
  -p_E\varepsilon_EL_E-\sigma_Em_UU_E+(\gamma_E+m_D)D_E\\
    -(1-p_E)\varepsilon_EL_E+(\gamma_E+m_U)U_E\\
      -m_LL_E+\varepsilon_IL_I\\
  -m_DD_E-p_I\varepsilon_IL_I
    +\gamma_ID_I\\
  -(1-\sigma_E)m_UU_E-(1-p_I)\varepsilon_IL_I
    +\gamma_IU_I\\
    \varepsilon_CL_C\\
 - p_C\varepsilon_CL_C+\gamma_CD_C\\
   -(1-p_C)\varepsilon_CL_C
    +\gamma_CU_C
    \end{pmatrix},
\]
respectively.
It follows that the matrices $F$ and $V$ used in the computation of the next generation matrix $FV^{-1}$ take the form
\[
\bF=
\begin{pmatrix}
F_{11} & 0 \\
0 & F_{22} 
\end{pmatrix},
\]
with
\[
F_{11} = \begin{pmatrix}
    0 & \beta_E\xi_E & \beta_E \\
    0 & 0 & 0 \\
    0 & 0 & 0
\end{pmatrix}S_E(0)
\]
and
\[
F_{22} = \begin{pmatrix}
    0 & 0 & 0 & 0 & 0 & 0 \\
    0 & 0 & 0 & 0 & 0 & 0 \\
    0 & 0 & 0 & 0 & 0 & 0 \\
    0 & \beta_{I}\xi_{I} & \beta_{I} & 0 & \beta_{C}\xi_{C} & \beta_{C} \\
    0 & 0 & 0 & 0 & 0 & 0 \\
    0 & 0 & 0 & 0 & 0 & 0 
\end{pmatrix}S_C(0),
\]
while
\[
\bV=
\begin{pmatrix}
    \bV_E & 0 \\
    V_{21} & \bV_I
\end{pmatrix},
\]
with
\[
V_{21} =
\begin{pmatrix}
    -m_L & 0 & 0 \\
    0 & -m_D & 0 \\
    0 & 0 & -(1-\sigma_E)m_U \\
    0 & 0 & 0 \\
    0 & 0 & 0 \\
    0 & 0 & 0
\end{pmatrix}.
\]
As a block lower triangular matrix, $\bV$ is easy to invert, with
\begin{equation}\label{eq:Vinv-full}
\bV^{-1}=
\begin{pmatrix}
    \bV_E^{-1} & 0 \\
    -\bV_I^{-1}V_{21}\bV_E^{-1} & \bV_I^{-1}
\end{pmatrix}.
\end{equation}
As a consequence,
\[
\bF\bV^{-1}=
\begin{pmatrix}
F_{11} & 0 \\
0 & F_{22} 
\end{pmatrix}
\begin{pmatrix}
    \bV_E^{-1} & 0 \\
    -\bV_I^{-1}V_{21}\bV_E^{-1} & \bV_I^{-1}
\end{pmatrix}
=
\begin{pmatrix}
F_{11}\bV_E^{-1} & 0 \\
\bigstar & F_{22}\bV_I^{-1}
\end{pmatrix}
\]
and thus
\[
    \mathcal{R}_0 = \max\left\{
        \rho\left(F_{11}\bV_E^{-1}\right),
        \rho\left(F_{22}\bV_I^{-1}\right)
    \right\},
\]
where $\rho(\cdot)$ is the spectral radius.
As a consequence, we get the expression \eqref{eq:R0}.

Regarding the final size of the epidemic in the full model \eqref{sys:general-model}, the method of \citeapp{ArinoBrauerDriesscheWatmoughWu2007} used in Appendices~\ref{app:R0-and-final-sizes-exporter} and~\ref{app:R0-and-final-sizes-importer} requires that all susceptible compartments experience a common scalar force of infection up to relative susceptibility factors, i.e., that the new infection matrix $\bF$ have rank one.
In the coupled Exporter-Importer system, however, transmission is spatially localised: $S_E$ is only infected by $D_E$ and $U_E$, whereas $S_C$ is only infected by $D_I$, $U_I$, $D_C$ and $U_C$.
Consequently, $\bF$ has rank two and the next generation matrix $\bF\bV^{-1}$ above is block lower triangular (reducible), so the integral of the force of infection cannot be eliminated using a single scalar relation.
Nevertheless, we can derive a triangular pair of final size equations for $(S_E(\infty), S_C(\infty))$ using the multi-group final size framework of \citeapp{Andreasen2011} and, more specifically, the approach developed by \citeapp{MagalSeydiWebb2018} for multi-group epidemic models with non-irreducible (reducible) modes of transmission (see also \citeapp{BidariEtAl2016,Brauer2019,MagalSeydiWebb2016}).

To do so, partition the infected compartments into the Exporter vector $\bx_E(t) = (L_E(t), D_E(t), U_E(t))^T$ and the Importer vector $\bx_{IC}(t) = (L_I(t), D_I(t), U_I(t), L_C(t), D_C(t), U_C(t))^T$, and let $\Pi_E = (1,0,0)^T \in \IR^3$ and $\Pi_{IC} = (0,0,0,1,0,0)^T \in \IR^6$ denote the vectors indicating the compartments into which new infections enter.
Observing that $-\dot{S}_E(t) = \beta_E(\xi_E D_E(t) + U_E(t))S_E(t)$ and $-\dot{S}_C(t) = [\beta_I(\xi_I D_I(t) + U_I(t)) + \beta_C(\xi_C D_C(t) + U_C(t))]S_C(t)$, the nine infected equations in \eqref{sys:general-model} take the block linear form
\[
\frac{d}{dt}\begin{pmatrix} \bx_E(t) \\ \bx_{IC}(t) \end{pmatrix}
=
\begin{pmatrix} -\dot{S}_E(t)\,\Pi_E \\ -\dot{S}_C(t)\,\Pi_{IC} \end{pmatrix}
-
\begin{pmatrix} \bV_E & 0 \\ V_{21} & \bV_I \end{pmatrix}
\begin{pmatrix} \bx_E(t) \\ \bx_{IC}(t) \end{pmatrix}.
\]
Since $\bx_E(\infty) = \b0$ and $\bx_{IC}(\infty) = \b0$, integrating from $t=0$ to $t=\infty$ and multiplying by the block lower triangular inverse $\bV^{-1}$ in \eqref{eq:Vinv-full} yields
\begin{equation}\label{eq:integral-xE}
\int_0^\infty \bx_E(t)\,dt = \Delta S_E\,\bV_E^{-1}\Pi_E + \bV_E^{-1}\bx_E(0)
\end{equation}
and
\begin{equation}\label{eq:integral-xIC}
\int_0^\infty \bx_{IC}(t)\,dt = \Delta S_C\,\bV_I^{-1}\Pi_{IC} + \bV_I^{-1}\left(\bx_{IC}(0) + \mathcal{I}_{E\to I}\right),
\end{equation}
where $\Delta S_E := S_E(0)-S_E(\infty)$, $\Delta S_C := S_C(0)-S_C(\infty)$ and
\[
\mathcal{I}_{E\to I} := (-V_{21}) \int_0^\infty \bx_E(t)\,dt
=
\begin{pmatrix}
m_L \int_0^\infty L_E(t)\,dt \\
m_D \int_0^\infty D_E(t)\,dt \\
(1-\sigma_E)m_U \int_0^\infty U_E(t)\,dt \\
0 \\ 0 \\ 0
\end{pmatrix}
=:
\begin{pmatrix}
\mathcal{I}_L \\ \mathcal{I}_D \\ \mathcal{I}_U \\ 0 \\ 0 \\ 0
\end{pmatrix}
\]
is the vector of total cumulative importations into $L_I$, $D_I$ and $U_I$ over $[0,\infty)$.
Notice that $\mathcal{I}_{E\to I}$ enters \eqref{eq:integral-xIC} in the exact same manner as the initial condition $\bx_{IC}(0)$.

Now let $\bb_E = (0, \xi_E, 1)$ and $\bb_{IC} = (0, \beta_I\xi_I/\beta_C, \beta_I/\beta_C, 0, \xi_C, 1)$ denote the relative horizontal transmission vectors from Appendices~\ref{app:R0-and-final-sizes-exporter} and~\ref{app:R0-and-final-sizes-importer}.
Integrating $-\dot{S}_E(t)/S_E(t)$ and $-\dot{S}_C(t)/S_C(t)$ over $[0,\infty)$ gives
\[
\ln\left(\frac{S_E(0)}{S_E(\infty)}\right) = \beta_E \bb_E \int_0^\infty \bx_E(t)\,dt,
\qquad
\ln\left(\frac{S_C(0)}{S_C(\infty)}\right) = \beta_C \bb_{IC} \int_0^\infty \bx_{IC}(t)\,dt.
\]
Substituting \eqref{eq:integral-xE} into the first relation recovers the isolated Exporter final size equation \eqref{eq:final-size-Exporter}, which depends only on $S_E(\infty)$ and can therefore be solved first to determine $S_E(\infty)$ and $\Delta S_E$.
Using \eqref{eq:integral-xE} and the entries of $\bV_E^{-1}$ from Appendix~\ref{app:R0-and-final-sizes-exporter}, the cumulative importations $(\mathcal{I}_L, \mathcal{I}_D, \mathcal{I}_U)$ are then given explicitly in terms of $\Delta S_E$ by
\begin{equation}\label{eq:cumulative-importations}
\begin{aligned}
\mathcal{I}_L &= \frac{m_L}{\varepsilon_E + m_L}\left(\Delta S_E + L_E(0)\right), \\
\mathcal{I}_D &= \frac{m_D}{\gamma_E + m_D}\left[
  \frac{\varepsilon_E}{\varepsilon_E + m_L}\left(p_E + \frac{\sigma_E m_U(1-p_E)}{\gamma_E + m_U}\right)\left(\Delta S_E + L_E(0)\right)
  + D_E(0) + \frac{\sigma_E m_U}{\gamma_E + m_U} U_E(0)
\right], \\
\mathcal{I}_U &= \frac{(1-\sigma_E)m_U}{\gamma_E + m_U}\left[
  \frac{(1-p_E)\varepsilon_E}{\varepsilon_E + m_L}\left(\Delta S_E + L_E(0)\right) + U_E(0)
\right].
\end{aligned}
\end{equation}
Finally, substituting \eqref{eq:integral-xIC} into the equation for $\ln(S_C(0)/S_C(\infty))$ and assuming the Importer is initially disease-free ($\bx_{IC}(0) = \b0$), we obtain the final size relation for $S_C(\infty)$ in the coupled model,
\begin{equation}\label{eq:final-size-Importer-coupled}
\ln\left(\frac{S_C(0)}{S_C(\infty)}\right)
= \frac{\R_0^C}{S_C(0)}\left(S_C(0) - S_C(\infty)\right)
+ \Lambda_{E\to I}(\Delta S_E),
\end{equation}
where the effective cumulative importation load exerted on the Importer community is
\begin{equation}\label{eq:Lambda-E-to-I}
\Lambda_{E\to I}(\Delta S_E)
:= \frac{\beta_I}{\gamma_I}\left[(p_I\mathcal{I}_L + \mathcal{I}_D)\xi_I + (1-p_I)\mathcal{I}_L + \mathcal{I}_U\right].
\end{equation}
Notice that \eqref{eq:final-size-Importer-coupled}--\eqref{eq:Lambda-E-to-I} have the same form as \eqref{eq:final-size-Importer}--\eqref{eq:final-size-importer-bVx} for the isolated Importer, with the initial introductions $(L_I(0), D_I(0), U_I(0))$ replaced by the total cumulative importations $(\mathcal{I}_L, \mathcal{I}_D, \mathcal{I}_U)$.

Equations \eqref{eq:cumulative-importations}--\eqref{eq:Lambda-E-to-I} allow to characterise both the dependence of the Importer final size $\Delta S_C$ on $\R_0^C$ and the system-level impact of the three travel control mechanisms on $(\Delta S_E, \Delta S_C)$.

\subsubsection{Supercritical versus subcritical regimes in the Importer}
\label{app:supercritical-vs-subcritical}
Because the movement rates $m_X$ are small compared to the recovery and progression rates, the cumulative importation term satisfies $\Lambda_{E\to I}(\Delta S_E) \ll 1$.
When $\R_0^C > 1$, let $\Delta S_C^0 > 0$ denote the unique positive root of the unforced final size equation $\ln(S_C(0)/(S_C(0)-\Delta S_C^0)) = (\R_0^C/S_C(0))\Delta S_C^0$.
Expanding \eqref{eq:final-size-Importer-coupled} to first order in $\Lambda_{E\to I}(\Delta S_E) > 0$ around $\Delta S_C^0$ yields
\begin{equation}\label{eq:supercritical-DeltaSC}
\Delta S_C \approx \Delta S_C^0 + \frac{S_C(0) - \Delta S_C^0}{1 - \R_0^C\left(1 - \Delta S_C^0/S_C(0)\right)}\,\Lambda_{E\to I}(\Delta S_E),
\end{equation}
so that for any $\Lambda_{E\to I}(\Delta S_E) > 0$, the deterministic final size $\Delta S_C$ remains close to the unforced community epidemic size $\Delta S_C^0$ and is largely insensitive to partial reductions in $\Lambda_{E\to I}(\Delta S_E)$.
By contrast, when $\R_0^C < 1$ and $\Lambda_{E\to I}(\Delta S_E) \ll 1 - \R_0^C$, $\Delta S_C/S_C(0)$ is small; linearising $\ln(S_C(0)/S_C(\infty)) = -\ln(1 - \Delta S_C/S_C(0)) \approx \Delta S_C/S_C(0)$ in \eqref{eq:final-size-Importer-coupled} gives
\begin{equation}\label{eq:subcritical-DeltaSC}
\Delta S_C \approx \frac{S_C(0)\,\Lambda_{E\to I}(\Delta S_E)}{1 - \R_0^C},
\end{equation}
which shows that in a controlled Importer ($\R_0^C < 1$), the cumulative number of community infections scales linearly with the effective importation load $\Lambda_{E\to I}(\Delta S_E)$, amplified by the factor $(1 - \R_0^C)^{-1}$.

\subsubsection{System-level versus Importer-level effects of travel measures}
\label{app:system-vs-importer-travel}
Differentiating the Exporter control reproduction number $\R_c^E$ in \eqref{eq:RcE} with respect to the movement rates $m_X$ ($X \in \{L, D, U\}$) and the pre-departure screening parameter $\sigma_E$ shows that
\begin{equation}\label{eq:dRcE-dm-dsigmaE}
\frac{\partial \R_c^E}{\partial m_L} < 0, \qquad
\frac{\partial \R_c^E}{\partial m_D} < 0, \qquad
\left.\frac{\partial \R_c^E}{\partial m_U}\right|_{\sigma_E=0} < 0
\end{equation}
and
\begin{equation}\label{eq:dRcE-dsigmaE}
\frac{\partial \R_c^E}{\partial \sigma_E} = \frac{\varepsilon_E}{\varepsilon_E + m_L}\frac{(1-p_E)\xi_E m_U}{(\gamma_E + m_U)(\gamma_E + m_D)}\beta_E S_E(0) \ge 0,
\end{equation}
with strict inequality in \eqref{eq:dRcE-dsigmaE} whenever $\xi_E > 0$.
Because emigration removes infected individuals from the Exporter before recovery, restricting travel ($m_X \downarrow$) or intercepting undetected departing travellers into $D_E$ via pre-departure screening ($\sigma_E \uparrow$) retains infectious individuals in the Exporter and increases $\R_c^E$ and $\Delta S_E$.
Furthermore, in \eqref{eq:cumulative-importations}, $\mathcal{I}_L = [m_L/(\varepsilon_E + m_L)](\Delta S_E + L_E(0))$ has no explicit dependence on $\sigma_E$ (and increases slightly with $\sigma_E$ through $\Delta S_E$), so pre-departure screening cannot reduce latent importations.
Conversely, the post-arrival quarantine and detection parameters $p_I$ and $\xi_I$ do not appear in $\R_c^E$ or $\Delta S_E$, while in \eqref{eq:Lambda-E-to-I},
\begin{equation}\label{eq:dLambda-dpI}
\frac{\partial \Lambda_{E\to I}}{\partial p_I} = -\frac{\beta_I}{\gamma_I}(1-\xi_I)\mathcal{I}_L < 0 \quad (\xi_I < 1), \qquad
\frac{\partial \Lambda_{E\to I}}{\partial \xi_I} = \frac{\beta_I}{\gamma_I}\left(p_I\mathcal{I}_L + \mathcal{I}_D\right) > 0,
\end{equation}
confirming that post-arrival control in the introduction layer reduces both $\Delta S_C$ and the system-wide final size $\Delta S_E + \Delta S_C$, and is the only border measure aside from a complete travel ban ($m_L = 0$) that attenuates the contribution of latent arrivals $\mathcal{I}_L$.



\section{Continuous-time Markov chains and their multitype branching process approximations}
\label{app:CTMC-MBPA}

We use a multitype branching process approximation (MBPA) to approximate the CTMC \( \mathbf{X}_t \) near the disease-free equilibrium of the deterministic system \eqref{sys:general-model}. 
This approximation is valid during the early phase of the epidemic, when the number of infected individuals is small and the susceptible population remains approximately constant.
Under this assumption, the infection dynamics become nearly linear and the CTMC can be approximated by a branching process.

Let \( Z(t) = (Z_k(t))_k \) denote the associated continuous-time multitype branching process, where each \( Z_k(t) \), with \( k \in \mathcal{I}:=\{L_E, D_E, U_E,L_I, D_I, U_I,L_C, D_C, U_C\} \), represents the number of infected individuals of type \( k \) at time \( t \geq 0 \).
The set of all types defines the structure of the process.

Each individual of type \( k \) evolves independently and generates offspring according to a type-dependent probabilistic mechanism derived from the infinitesimal generator of the CTMC.

We define the probability generating function (p.g.f.) for type \( k \in \mathcal{I} \) at time \( t \) as
\begin{equation}\label{eq:F-pgf-general}
F_k(t, \bz) = \mathbb{E} \left[ \bz^{Z(t)} \,\middle|\, Z(0) = e_k \right],
\end{equation}
where \( \bz = (z_k)_k \in [0,1]^9 \), \( e_k \) is the unit vector corresponding to type \( k \) and \( \bz^{Z(t)} := \prod _{k\in \mathcal{I}} z_k^{Z_k(t)}  \).

Let \( F(t, \bz) = (F_k(t, \bz))_k \) denote the full vector of generating functions.
Standard results from the theory of branching processes \citeapp[Chapter V, Section 7]{athreya1972branching} imply that \( F(t, \bz) \) satisfies the backward Kolmogorov system
\begin{equation}
\frac{\partial}{\partial t} F_k(t, \bz) = \omega_k \left[ f_k(F(t, \bz)) - F_k(t, \bz) \right], \quad k \in \mathcal{I},
\end{equation}
with initial condition \( F_k(0, \bz) = z_k \), and where \( \omega_k \) denotes the exit rate from type \( k \).
The function \( f_k(\bz) \) is the \emph{offspring generating function} for type \( k \), defined as
\begin{equation}\label{eq:gen-form-offpgf}
f_k(\bz) = \sum_{\mathbf{r} \in \mathbb{N}^9} \mathbb{P}_k(\mathbf{r}) \, \bz^{\mathbf{r}},
\end{equation}
where \( \mathbb{P}_k(\mathbf{r}) \) is the probability that a single individual of type \( k \) produces \( \mathbf{r} = (r_k)_k \) offspring of each type during its infectious period.

This approximation allows analytical computation of extinction probabilities and facilitates the analysis of early epidemic dynamics, particularly in structured populations.

We use MBPA to approximate the CTMC~\eqref{eq:CTMC} near the disease-free equilibrium of~\eqref{sys:general-model}, assuming $S_E = S_E(0)$ and $S_C = S_C(0)$ are constant.
Denoting $\bz=(z_k)_{k \in \mathcal{I}}$, the offspring probability-generating function (p.g.f.) is given by
\begin{subequations}\label{eq:pgf-exporter-importer}
\begin{equation}\label{eq:pgf-exporter-importer-F}
    f(\bz)=\bigl(f_{L_E}(\bz),f_{D_E}(\bz),f_{U_E}(\bz),f_{L_I}(\bz),
    f_{D_I}(\bz),f_{U_I}(\bz),f_{L_C}(\bz),f_{D_C}(\bz),f_{U_C}(\bz)\bigr),
\end{equation}
where
\begin{align}
f_{L_E}(\bz) &= \frac{p_E \varepsilon_E z_{D_E}+\left(1-p_E\right) \varepsilon_E z_{U_E}+m_L z_{L_I}}{\varepsilon_E+m_L}, \\
f_{D_E}(\bz) &= \frac{\beta_E \xi_E S_E(0) {z_{L_E}}{z_{D_E}}+\gamma_E+m_D z_{D_I}}{\beta_E \xi_E S_E(0) +\gamma_E+m_D},\\
f_{U_E}(\bz) &= \frac{\beta_E S_E(0) z_{L_E}z_{U_E}+\gamma_E+\sigma_E m_U z_{D_E}+ (1-\sigma_E)m_U z_{U_I}}{\beta_E S_E(0)+\gamma_E+m_U}, \\
f_{L_I}(\bz) &= p_I z_{D_I}+(1-p_I)  z_{U_I}, \\
f_{D_I}(\bz) &= \frac{\beta_{I} \xi_{I} S_{C}(0) z_{L_C}z_{D_I}+\gamma_{I}}{\beta_{I} \xi_{I}S_{C}(0)+\gamma_I}, \\
f_{U_I}(\bz) &= \frac{\beta_{I} S_C(0) z_{L_C}z_{U_I}+\gamma_I}{\beta_{I}S_C(0)+\gamma_I}, \\
f_{L_C}(\bz) &= p_C z_{D_C}+\left(1-p_C\right) z_{U_C}, \\
f_{D_C}(\bz) &= \frac{\beta_{C} \xi_{C} S_C(0) z_{L_C}z_{D_C}+\gamma_C}{\beta_{C} \xi_{C}S_C(0)+\gamma_C}, \\ 
f_{U_C}(\bz) &= \frac{\beta_{C}S_C(0)z_{L_C}z_{U_C}+\gamma_C}{\beta_{C}S_C(0)+\gamma_C}.
\end{align}
\end{subequations}

\subsection{In the Exporter}
\label{app:CTMC-MBPA-exporter}
The branching process in Section~\ref{sec:model-MBPA} uses here \( Z(t) = (Z_{L_E}(t), Z_{D_E}(t), Z_{U_E}(t)) \).
We define the probability generating function (p.g.f.) for type \( k \in \{L_E, D_E, U_E\} \) at time \( t \) as
\begin{equation}\label{eq:F-pgf-exporter}
F_k(t, \bz) = \mathbb{E} \left[ \bz^{Z(t)} \,\middle|\, Z(0) = e_k \right],
\end{equation}
where \( \bz = (z_{L_E}, z_{D_E}, z_{U_E}) \in [0,1]^3 \), \( e_k \) is the unit vector corresponding to type \( k \) and \( \bz^{Z(t)} := z_{L_E}^{Z_{L_E}(t)} z_{D_E}^{Z_{D_E}(t)} z_{U_E}^{Z_{U_E}(t)} \).
Assuming $S_E = S_{E}(0)$ is constant, the offspring generating functions are given by
\begin{subequations}\label{eq:pgf-exporter}
\begin{equation}\label{eq:pgf-exporter-F}
    f_E(\bz)=(f_{L_E}(\bz),f_{D_E}(\bz),f_{U_E}(\bz))
\end{equation}
where
\begin{align}
    f_{L_E}(\bz) &= \frac{p_E \varepsilon_E z_{D_E}+\left(1-p_E\right) \varepsilon_E z_{U_E}+m_L}{\varepsilon_E+m_L}, \\
    f_{D_E}(\bz) &= \frac{\beta_E \xi_E S_E(0) {z_{L_E}}{z_{D_E}}+\gamma_E+m_D }{\beta_E \xi_E S_E(0) +\gamma_E+m_D},\\
    f_{U_E}(\bz) &= \frac{\beta_E S_E(0) z_{L_E}z_{U_E}+\gamma_E+\sigma_E m_U z_{D_E}+(1-\sigma_E)m_U }{\beta_E S_E(0)+\gamma_E+m_U}.
\end{align}
\end{subequations}

\begin{theorem}\label{th:exists-fix-pt-exporter}
The multitype branching process associated with the Exporter is positive regular and nonsingular. 
Let the initial state be given by \( L_E(0) = \ell_{E0} \), \( D_E(0) = d_{E0} \) and \( U_E(0) = u_{E0} \). 
Then, the probabilities of extinction and outbreak are given by
\begin{subequations}
\label{eq:prob-ext-exporter}
\begin{align}
    \mathbb{P}_{\text{ext}}^E &= \left(z_{L_E}\right)^{\ell_{E0}} \left(z_{D_E}\right)^{d_{E0}} \left(z_{U_E}\right)^{u_{E0}}, \\
    \mathbb{P}_{\text{outbreak}}^E &= 1 - \mathbb{P}_{\text{ext}}^E,
\end{align}
\end{subequations}
where
\[
\bz := (z_{L_E}, z_{D_E}, z_{U_E})
\]
is a fixed point in \( [0,1]^3 \) of the offspring distribution~\eqref{eq:pgf-exporter}.
In addition, the following alternative holds (with $\R_c^E$ reducing to $\R_0^E$ when $\sigma_E=0$):
\begin{itemize}
    \item if \( \R_c^E \leq 1 \), then the unique fixed point is \( \bz = \mathbf{1} \), and hence \( \mathbb{P}_{\text{ext}}^E = 1 \);
    \item if \( \R_c^E > 1 \), then in addition to \( \bz = \mathbf{1} \), there exists a unique fixed point \( \bz \in (0,1)^3 \) such that \( f_E(\bz) = \bz \).
\end{itemize}
\end{theorem}

\begin{proof}[Proof of Theorem~\ref{th:exists-fix-pt-exporter}]
We analyse the Jacobian matrix of the offspring generating function~\eqref{eq:pgf-exporter}, given by
\[
Df_E(\bz)=
\begin{pmatrix}
0 & \dfrac{p_E \varepsilon_E}{\varepsilon_E + m_L} & \dfrac{(1 - p_E)\varepsilon_E}{\varepsilon_E + m_L} \\
\dfrac{S_E(0) \beta_E \xi_E z_{D_E}}{S_E(0) \beta_E \xi_E + \gamma_E + m_D} & \dfrac{S_E(0) \beta_E \xi_E z_{L_E}}{S_E(0) \beta_E \xi_E + \gamma_E + m_D} & 0 \\
\dfrac{S_E(0) \beta_E z_{U_E}}{S_E(0) \beta_E + \gamma_E + m_U} & \dfrac{\sigma_E m_U}{S_E(0) \beta_E + \gamma_E + m_U} & \dfrac{S_E(0) \beta_E z_{L_E}}{S_E(0) \beta_E + \gamma_E + m_U}
\end{pmatrix}.
\]

\begin{enumerate}
\item[(i)] The matrix \( Df_E(\bz) \) is nonnegative and monotone for \( \bz \in [0,1)^3 \).
By \citeapp[Theorem 2.3 p.113]{berman1979nonnegative}, this implies that the process is nonsingular.

\item[(ii)] The matrix of first moments is defined as \( \mathbb{M}_E = Df_E(\mathbf{1}) \).  
One can verify that \( \mathbb{M}_E^2 \) has all entries strictly positive, hence \( \mathbb{M}_E \) is primitive by \citeapp[Corollary 3.1.4]{berman1979nonnegative}.
\end{enumerate}

From (i) and (ii), we conclude that the branching process is positive, regular and nonsingular.
Therefore, the classical threshold behaviour applies: by the Threshold Theorem \citeapp{ALLEN201399} and \citeapp[Theorem 7.1 p. 41]{Harris1963}.
This concludes the proof.
\end{proof}

\subsection{The Importer in isolation}
\label{app:CTMC-MBPA-importer}
As in the deterministic case of Section~\ref{sec:results-importer} and Appendix~\ref{app:R0-and-final-sizes-importer}, we suppose here that movement rates are zero, i.e., $m_L=m_D=m_U=0$.
In this case, a multitype branching process approximation of the continuous time Markov chain can be used.

The branching process in Section~\ref{sec:model-MBPA} uses here \( Z(t) = (Z_{L_I}(t), Z_{D_I}(t), Z_{U_I}(t), Z_{L_C}(t), Z_{D_C}(t), Z_{U_C}(t)) \).
We define the probability generating function (p.g.f.) for type \( k \in \mathcal{J}:= \{L_I, D_I, U_I, L_C, D_C, U_C\} \) at time \( t \) as
\begin{equation}\label{eq:F-pgf-importer}
F_k(t, \bz) = \mathbb{E} \left[ \bz^{Z(t)} \,\middle|\, Z(0) = e_k \right],
\end{equation}
where \( \bz = (z_k)_k \in [0,1]^6 \), \( e_k \) is the unit vector corresponding to type \( k \in \mathcal{J}\).
Assuming $S_C = S_{C}(0)$ is constant, the offspring generating functions are given by
\begin{subequations}\label{eq:pgf-importer}
\begin{equation}\label{eq:pgf-importer-F}
    f_I(\bz)=(f_{L_I}(\bz),f_{D_I}(\bz),f_{U_I}(\bz),f_{L_C}(\bz),f_{D_C}(\bz),f_{U_C}(\bz))
\end{equation}
where
\begin{align}
    f_{L_I}(\bz) &= p_I z_{D_I}+(1-p_I)  z_{U_I}, \\
f_{D_I}(\bz) &= \frac{\beta_{I} \xi_{I} S_{C}(0) z_{L_C}z_{D_I}+\gamma_{I}}{\beta_{I} \xi_{I}S_{C}(0)+\gamma_I}, \\
f_{U_I}(\bz) &= \frac{\beta_{I} S_C(0) z_{L_C}z_{U_I}+\gamma_I}{\beta_{I}S_C(0)+\gamma_I}, \\
f_{L_C}(\bz) &= p_C z_{D_C}+\left(1-p_C\right) z_{U_C}, \\
f_{D_C}(\bz) &= \frac{\beta_{C} \xi_{C} S_C(0) z_{L_C}z_{D_C}+\gamma_C}{\beta_{C} \xi_{C}S_C(0)+\gamma_C}, \\ 
f_{U_C}(\bz) &= \frac{\beta_{C}S_C(0)z_{L_C}z_{U_C}+\gamma_C}{\beta_{C}S_C(0)+\gamma_C}.
\end{align} 
\end{subequations}

\begin{theorem}\label{th:exists-fix-pt-importer}
The 6-type branching process associated with the Importer in isolation is nonsingular and reducible, coupling the transient introduction layer $\mathcal{J}_I := \{L_I, D_I, U_I\}$ to the positive regular community subsystem $\mathcal{J}_C := \{L_C, D_C, U_C\}$.
Assume $\beta_I > 0$, and let the initial state be given by \( x(0)=(x_k(0))_{k \in \mathcal{J}}\). 
Then, the probabilities of extinction and outbreak are given by
\begin{subequations}
\label{eq:prob-ext-importer}
\begin{align}
    \mathbb{P}_{\text{ext}}^I &= \prod_{k \in \mathcal{J}}\left(z_{k}\right)^{x_k(0)} , \\
    \mathbb{P}_{\text{outbreak}}^I &= 1 - \mathbb{P}_{\text{ext}}^I,
\end{align}
\end{subequations}
where
\[
\bz := (z_k)_{k \in \mathcal{J}}
\]
is the minimal fixed point in \( [0,1]^6 \) of the offspring distribution~\eqref{eq:pgf-importer}.
In addition, the following alternative holds:
\begin{itemize}
    \item if \( \mathcal{R}_0^C \leq 1 \), then the unique fixed point in $[0,1]^6$ is \( \bz = \mathbf{1} \), and hence \( \mathbb{P}_{\text{ext}}^I = 1 \);
    \item if \( \mathcal{R}_0^C > 1 \), then in addition to \( \bz = \mathbf{1} \), there exists a unique fixed point \( \bz \in (0,1)^6 \) such that \( f_I(\bz) = \bz \).
    Specifically, $\bz_C := (z_{L_C}, z_{D_C}, z_{U_C}) \in (0,1)^3$ is the unique interior fixed point of the community subsystem, and $(z_{L_I}, z_{D_I}, z_{U_I}) \in (0,1)^3$ is uniquely determined by
    \begin{equation}\label{eq:importer-layer-explicit}
    z_{D_I} = \frac{\gamma_I}{\beta_I \xi_I S_C(0)(1 - z_{L_C}) + \gamma_I}, \quad
    z_{U_I} = \frac{\gamma_I}{\beta_I S_C(0)(1 - z_{L_C}) + \gamma_I}, \quad
    z_{L_I} = p_I z_{D_I} + (1 - p_I) z_{U_I}.
    \end{equation}
\end{itemize}
\end{theorem}

\begin{proof}[Proof of Theorem~\ref{th:exists-fix-pt-importer}]
The Jacobian matrix of the Importer offspring generating function~\eqref{eq:pgf-importer} has the block upper-triangular form
\[Df_I(\bz)=\begin{pmatrix}
 \mathbb{A}_{11}(\bz) &  \mathbb{A}_{12}(\bz)\\
    \b0 &  \mathbb{A}_{22}(\bz)\\
\end{pmatrix},\]
where
\[\mathbb{A}_{11}(\bz)=\begin{pmatrix}0 & p_{I} & 1 - p_{I}\\0 & \frac{S_C(0) \beta_{I} \xi_{I} z_{L_C}}{S_C(0) \beta_{I} \xi_{I} + \gamma_{I}} & 0\\0 & 0 & \frac{S_C(0) \beta_{I} z_{L_C}}{S_C(0) \beta_{I} + \gamma_{I}}\end{pmatrix},
\]
\[\mathbb{A}_{12}(\bz)=\begin{pmatrix}0 & 0 & 0\\\frac{S_C(0) \beta_{I} \xi_{I} z_{D_I}}{S_C(0) \beta_{I} \xi_{I} + \gamma_{I}} & 0 & 0\\\frac{S_C(0) \beta_{I} z_{U_I}}{S_C(0) \beta_{I} + \gamma_{I}} & 0 & 0\end{pmatrix},
\; \mathbb{A}_{22}(\bz)= \begin{pmatrix}0 & p_{C} & 1 - p_{C}\\\frac{S_C(0) \beta_{C} \xi_{C} z_{D_C}}{S_C(0) \beta_{C} \xi_{C} + \gamma_{C}} & \frac{S_C(0) \beta_{C} \xi_{C} z_{L_C}}{S_C(0) \beta_{C} \xi_{C} + \gamma_{C}} & 0\\\frac{S_C(0) \beta_{C} z_{U_C}}{S_C(0) \beta_{C} + \gamma_{C}} & 0 & \frac{S_C(0) \beta_{C} z_{L_C}}{S_C(0) \beta_{C} + \gamma_{C}}\end{pmatrix}.
\]
Because the lower-left block is zero (community infections never transition back into the introduction layer $\mathcal{J}_I$), the 6-type process is reducible, and the fixed-point system $f_I(\bz) = \bz$ decouples into two sequential 3-dimensional problems:
\begin{enumerate}
\item[(i)] \textbf{Community subsystem $\mathcal{J}_C$:} The last three equations $f_C(\bz_C) := (f_{L_C}(\bz_C), f_{D_C}(\bz_C), f_{U_C}(\bz_C)) = \bz_C$ depend exclusively on $\bz_C = (z_{L_C}, z_{D_C}, z_{U_C}) \in [0,1]^3$.
For all $\bz_C \in [0,1)^3$, $\mathbb{A}_{22}(\bz_C)$ is nonnegative and monotone, and the community mean matrix $\mathbb{M}_C = \mathbb{A}_{22}(\mathbf{1})$ satisfies $\mathbb{M}_C^2 > \b0$, so $\mathbb{M}_C$ is irreducible and primitive \citeapp[Corollary 3.1.4]{berman1979nonnegative}.
Thus the 3-type community branching process is nonsingular and positive regular with basic reproduction number $\mathcal{R}_0^C$.
By the classical multitype Threshold Theorem \citeapp{ALLEN201399,Harris1963}, if $\mathcal{R}_0^C \le 1$, then $\bz_C = \mathbf{1}$ is the unique fixed point of $f_C$ in $[0,1]^3$; if $\mathcal{R}_0^C > 1$, then in addition to $\mathbf{1}$ there exists a unique interior fixed point $\bz_C \in (0,1)^3$.

\item[(ii)] \textbf{Introduction layer $\mathcal{J}_I$:} For any fixed $z_{L_C} \in [0,1]$, the first three equations $z_{L_I} = f_{L_I}(\bz)$, $z_{D_I} = f_{D_I}(\bz)$ and $z_{U_I} = f_{U_I}(\bz)$ are linear in $(z_{L_I}, z_{D_I}, z_{U_I})$.
Solving the second and third equations for $z_{D_I}$ and $z_{U_I}$ and substituting into the first equation yields~\eqref{eq:importer-layer-explicit} directly.
When $\mathcal{R}_0^C \le 1$, substituting $z_{L_C} = 1$ into~\eqref{eq:importer-layer-explicit} gives $z_{D_I} = z_{U_I} = z_{L_I} = 1$, so $\bz = \mathbf{1}$ is the unique fixed point in $[0,1]^6$.
When $\mathcal{R}_0^C > 1$, substituting the unique community fixed point $z_{L_C} \in (0,1)$ into~\eqref{eq:importer-layer-explicit} with $\beta_I > 0$ yields $z_{D_I}, z_{U_I}, z_{L_I} \in (0,1)$ uniquely.
\end{enumerate}
This concludes the proof.
\end{proof}

\subsection{Full Exporter-Importer model}
\label{app:CTMC-MBPA-exporter-importer}
Using the probability generating function \eqref{eq:pgf-exporter-importer}, we obtain Theorem~\ref{th:exists-fp-exp-imp}, which is proved below.

\begin{theorem}\label{th:exists-fp-exp-imp}
The 9-type branching process associated with the full Exporter-Importer model with offspring generating function~\eqref{eq:pgf-exporter-importer} is nonsingular and reducible, coupling the Exporter subsystem $\mathcal{I}_E := \{L_E, D_E, U_E\}$ to the transient introduction layer $\mathcal{J}_I := \{L_I, D_I, U_I\}$ and the community subsystem $\mathcal{J}_C := \{L_C, D_C, U_C\}$.
Assume $\beta_I > 0$ and $m_L + m_D + (1-\sigma_E)m_U > 0$, and let the initial state be $\bx(0)=(x_k(0))_{k \in \mathcal{I}}$.
Then the probabilities of overall extinction and of a major outbreak (in either the Exporter or the Importer community) are given by
\begin{subequations}
\label{eq:prob-ext-multitype}
\begin{align}
\IP_{\text{ext}} &= \prod_{k \in \mathcal{I}} z_k^{x_k(0)}, \\
\IP_{\text{outbreak}} &= 1 - \IP_{\text{ext}},
\end{align}
\end{subequations}
where
\begin{equation*}\label{eq:def-z}
    \bz := (\bz_E, \bz_I, \bz_C) = \left(z_{L_E}, z_{D_E}, z_{U_E}, z_{L_I}, z_{D_I}, z_{U_I}, z_{L_C}, z_{D_C}, z_{U_C}\right)
\end{equation*}
is the minimal fixed point in $[0,1]^9$ of the offspring p.g.f.~\eqref{eq:pgf-exporter-importer}.
In addition, defining $\R_0 := \max\{\R_c^E, \R_0^C\}$, the following cases hold for the minimal fixed point $\bz \in [0,1]^9$:
\begin{itemize}
    \item \textbf{Case 1 ($\R_c^E \leq 1$ and $\R_0^C \leq 1$, i.e., $\R_0 \leq 1$):} the unique fixed point in $[0,1]^9$ is $\bz = \mathbf{1}$, where $\mathbf{1}$ is the unit vector of size $9$, and hence $\IP_{\text{ext}} = 1$;
    \item \textbf{Case 2 ($\R_c^E > 1$ and $\R_0^C \leq 1$):} the minimal fixed point in $[0,1]^9$ is $\bz = (\bz_E, \mathbf{1}, \mathbf{1})$, where $\bz_C = \bz_I = \mathbf{1} \in \IR^3$ and $\bz_E \in (0,1)^3$ is the unique interior fixed point of the isolated Exporter p.g.f.~\eqref{eq:pgf-exporter} given by Theorem~\ref{th:exists-fix-pt-exporter};
    \item \textbf{Case 3 ($\R_0^C > 1$, for any $\R_c^E > 0$):} the minimal fixed point in $[0,1]^9$ satisfies $\bz = (\bz_E, \bz_I, \bz_C) \in (0,1)^9$, where $(\bz_I, \bz_C) \in (0,1)^6$ is the unique interior fixed point of the Importer subsystem given by Theorem~\ref{th:exists-fix-pt-importer}, and $\bz_E \in (0,1)^3$ is the unique fixed point in $[0,1]^3$ of the Exporter equations $(f_{L_E}(\bz), f_{D_E}(\bz), f_{U_E}(\bz)) = \bz_E$ in~\eqref{eq:pgf-exporter-importer} for the given $\bz_I$.
\end{itemize}
\end{theorem}
\begin{proof}[Proof of Theorem~\ref{th:exists-fp-exp-imp}]
The Jacobian matrix of~\eqref{eq:pgf-exporter-importer} has the block upper-triangular form
\[Df(\bz)=\begin{pmatrix}
    \mathbb{B}_{11}(\bz) &  \mathbb{B}_{12}(\bz) & \b0\\
    \b0 &  \mathbb{B}_{22}(\bz) &  \mathbb{B}_{23}(\bz)\\
     \b0 &  \b0 &  \mathbb{B}_{33}(\bz)\\
\end{pmatrix},\]
where
\[\mathbb{B}_{11}(\bz)= \begin{pmatrix}0 & \frac{p_{E} \varepsilon_E}{ \varepsilon_E+m_L} & \frac{\varepsilon_E \left(1 - p_{E}\right)}{ \varepsilon_E+m_L}\\\frac{S_E(0) \beta_E \xi_E z_{D_E}}{S_E(0) \beta_E \xi_E + \gamma_E + m_D} & \frac{S_E(0) \beta_E \xi_E z_{L_E}}{S_E(0) \beta_E \xi_E + \gamma_E + m_D} & 0\\\frac{S_E(0) \beta_E z_{U_E}}{S_E(0) \beta_E + \gamma_E + m_U} & \frac{\sigma_E m_U}{S_E(0) \beta_E + \gamma_E + m_U} & \frac{S_E(0) \beta_E z_{L_E}}{S_E(0) \beta_E + \gamma_E + m_U}\end{pmatrix},
\]
\[\mathbb{B}_{12}(\bz)=\begin{pmatrix}\frac{m_L}{ \varepsilon_E+m_L} & 0 & 0\\0 & \frac{m_D}{S_E(0) \beta_E \xi_E + \gamma_E + m_D} & 0\\0 & 0 & \frac{(1-\sigma_E)m_U}{S_E(0) \beta_E + \gamma_E + m_U}\end{pmatrix},
\]
and $\mathbb{B}_{22}(\bz)=\mathbb{A}_{11}(\bz)$, $\mathbb{B}_{23}(\bz)=\mathbb{A}_{12}(\bz)$ and $\mathbb{B}_{33}(\bz)= \mathbb{A}_{22}(\bz)$ as defined in the proof of Theorem~\ref{th:exists-fix-pt-importer}.
Because $Df(\bz)$ is block upper-triangular, the 9-type branching process is reducible, and the fixed-point system $f(\bz) = \bz$ decouples from bottom to top:
\begin{enumerate}
\item[(i)] \textbf{Importer blocks $(\mathcal{J}_I, \mathcal{J}_C)$:} The last six equations of $f(\bz) = \bz$ depend only on $(\bz_I, \bz_C) \in [0,1]^6$ and coincide with $f_I(\bz_I, \bz_C) = (\bz_I, \bz_C)$ in~\eqref{eq:pgf-importer}.
By Theorem~\ref{th:exists-fix-pt-importer}, when $\R_0^C \le 1$, the unique solution in $[0,1]^6$ is $\bz_I = \bz_C = \mathbf{1}$; when $\R_0^C > 1$, the minimal solution in $[0,1]^6$ is the unique interior fixed point $(\bz_I, \bz_C) \in (0,1)^6$ satisfying~\eqref{eq:importer-layer-explicit}.

\item[(ii)] \textbf{Exporter block $\mathcal{I}_E$ when $\R_0^C \le 1$:} Substituting $\bz_I = \mathbf{1}$ into $(f_{L_E}(\bz), f_{D_E}(\bz), f_{U_E}(\bz)) = \bz_E$ in~\eqref{eq:pgf-exporter-importer} reduces the first three equations to $f_E(\bz_E) = \bz_E$ in~\eqref{eq:pgf-exporter}.
Cases~1 and~2 then follow directly from Theorem~\ref{th:exists-fix-pt-exporter}.

\item[(iii)] \textbf{Exporter block $\mathcal{I}_E$ when $\R_0^C > 1$:} Here the minimal Importer solution satisfies $\bz_I = (z_{L_I}, z_{D_I}, z_{U_I}) \in (0,1)^3$.
Solving the equations $z_{D_E} = f_{D_E}(\bz)$ and $z_{U_E} = f_{U_E}(\bz)$ in~\eqref{eq:pgf-exporter-importer} for $z_{D_E}$ and $z_{U_E}$ as functions of $z_{L_E} \in [0,1]$ yields
\[
z_{D_E}(z_{L_E}) = \frac{\gamma_E + m_D z_{D_I}}{\beta_E \xi_E S_E(0)(1 - z_{L_E}) + \gamma_E + m_D}, \quad
z_{U_E}(z_{L_E}) = \frac{\gamma_E + \sigma_E m_U z_{D_E}(z_{L_E}) + (1-\sigma_E) m_U z_{U_I}}{\beta_E S_E(0)(1 - z_{L_E}) + \gamma_E + m_U}.
\]
Both $z_{D_E}(z_{L_E})$ and $z_{U_E}(z_{L_E})$ are strictly increasing and strictly convex on $[0,1]$, mapping $[0,1]$ into $(0,1]$.
Substituting them into $z_{L_E} = f_{L_E}(\bz)$ reduces the system to the scalar fixed-point equation $z_{L_E} = \Psi(z_{L_E})$, where
\[
\Psi(z_{L_E}) := \frac{p_E \varepsilon_E z_{D_E}(z_{L_E}) + (1-p_E) \varepsilon_E z_{U_E}(z_{L_E}) + m_L z_{L_I}}{\varepsilon_E + m_L}
\]
is strictly increasing and strictly convex on $[0,1]$.
Since $\Psi(0) > 0$ and, under $m_L + m_D + (1-\sigma_E)m_U > 0$ with $\bz_I \in (0,1)^3$, $\Psi(1) < 1$, the convex function $g(z_{L_E}) := \Psi(z_{L_E}) - z_{L_E}$ satisfies $g(0) > 0 > g(1)$ and therefore has a unique root $z_{L_E} \in (0,1)$, which uniquely determines $\bz_E \in (0,1)^3$.
\end{enumerate}
This concludes the proof.
\end{proof}

\section{Characterisation of epidemic events and outbreaks}
\label{app:characterisation-outbreaks}

While not part of the model proper, the following notions are used to characterise epidemic realisations and outbreaks in the continuous-time Markov chain.

\subsection{Underlying and observable epidemics}
First, we distinguish between \emph{underlying} (or \emph{unobservable}) and \emph{observable} epidemics.
The underlying epidemic is the entire stochastic process across all infected compartments ($L_X$, $D_X$ and $U_X$ for $X\in\{E,I,C\}$).
The observable epidemic is the one observed by public health authorities and is represented by numbers in the detected infected compartments $D_E$, $D_I$ and $D_C$.
Unless otherwise specified, the underlying epidemic is the primary object of study, whereas intervention decisions are triggered based on the observable epidemic.

\subsection{Introductions}
\emph{Introductions} or \emph{importations} occur when an individual having acquired the infection in the Exporter travels to the Importer.
The individual may be detected or undetected at the time of travel as well as at the time of arrival in the Importer.
We adopt the same terminology as in \cite{ArinoBoelleMillikenPortet2021,ArinoBajeuxPortetWatmough2020} and distinguish between \emph{unsuccessful} and \emph{successful} introductions, where success is judged from the perspective of the pathogen.
An introduction is \emph{unsuccessful} if the introduced case does not lead to any subsequent case in the Importer.
If an introduced case infects at least one subsequent case in the Importer, then the introduction is \emph{successful}.

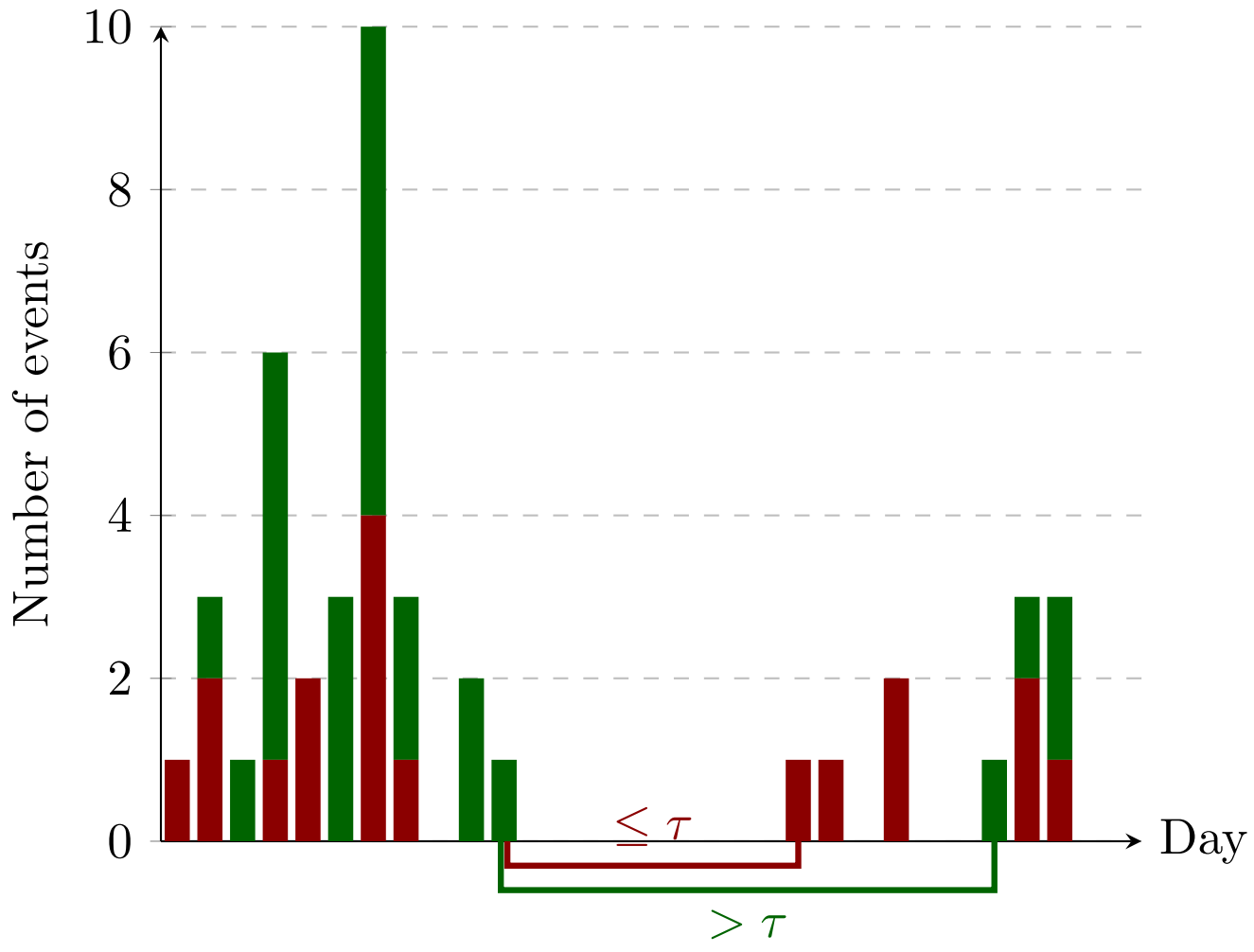
\begin{figure}[htbp]
    \centering
    \begin{tikzpicture}
    \begin{axis}[
        xlabel={Day},                 
        xlabel style={at={(ticklabel* cs:1)}, anchor=west},
        ylabel={Number of events},    
        ymin=0,                       
        xmin=0.5,
        xmax=30.5,
        ybar stacked,
        bar width=5pt,
        enlarge y limits={0.15},
        xtick=\empty,
        xticklabels=\empty,
        ymajorgrids=true, 
        grid style={dashed, color=gray!50}, 
        axis x line=bottom,
        axis y line=left,
        clip=false,
    ]

    \addplot+[ybar, fill=MyDarkRed, draw=none] table {
        x y
        1 1
        2 2
        3 0
        4 1
        5 2
        6 0
        7 4
        8 1
        10 0
        11 0
        20 1
        21 1
        23 2
        26 0 
        27 2 
        28 1 
    };

    \addplot+[ybar, fill=MyDarkGreen, draw=none] table {
        x y
        1 0
        2 1
        3 1
        4 5
        5 0
        6 3
        7 6
        8 2
        10 2
        11 1
        20 0
        21 0
        23 0
        26 1
        27 1
        28 2
    };

    \draw[MyDarkRed, very thick] 
        (axis cs:11.1, 0) -- (axis cs:11.1, -0.3) -- node[midway, above] {$\leq\tau$}  (axis cs:20, -0.3) -- (axis cs:20, 0);
    \draw[MyDarkGreen, very thick] 
        (axis cs:10.9, 0) -- (axis cs:10.9, -0.6) --  node[midway, below] {$>\tau$}  (axis cs:26, -0.6) -- (axis cs:26, 0);
    \end{axis}
    \end{tikzpicture}
    \caption{Sample detected (green) and undetected (red) transitions out of the latent compartment illustrating times between observable cases (green bracket) and between any types of cases (red bracket).
    Here, because the gap between detected cases exceeds the threshold $\tau$ while undetected cases bridge the gap within $\tau$, there are two observable outbreaks during a single unobservable outbreak.}
    \label{fig:inter-outbreak-times}
\end{figure}

\subsection{Outbreaks}
An \emph{outbreak} corresponds to a period of sustained transmission of the pathogen.
To make explicit what we mean by ``sustained'', we set a threshold $\tau$ for the maximum time allowed between two successive monitored events for them to be considered part of the same outbreak; if the time elapsed between two consecutive events exceeds $\tau$, the first outbreak is deemed to have ended and a new outbreak begins at the next event.
Depending on which events are monitored, we distinguish three types of outbreaks:
\begin{enumerate}
    \item \emph{Observable outbreaks} monitor only detection events.
    In the Exporter, an observable outbreak comprises consecutive transitions of the form $L_E\to D_E$ such that no two consecutive transitions are separated by more than $\tau$ time units.
    In the Importer, transitions $L_I\to D_I$ and $L_C\to D_C$ (together with detected importations $D_E\to D_I$) are used.
    In Figures~\ref{fig:transmission-scenario-R0E_15_R0C_08_mvt_00005_seed_349} and~\ref{fig:transmission-scenarios}, vertical green dashed bars mark the midpoint between the end of one observable outbreak and the start of the next.
    \item \emph{Unobservable outbreaks} (or general activation outbreaks) monitor all case activation events, whether detected or undetected.
    In the Exporter, this corresponds to transitions out of $L_E$ into either $D_E$ or $U_E$.
    In the Importer, transitions $L_I\to\{D_I,U_I\}$ and $L_C\to\{D_C,U_C\}$ within the introduction layer and the community are complemented by infectious importation events $D_E\to D_I$ and $U_E\to U_I$.
    Figure~\ref{fig:inter-outbreak-times} illustrates why a single unobservable outbreak can be perceived as multiple distinct observable outbreaks when detection is imperfect.
    \item \emph{Infection outbreaks} focus on actual infection events, i.e., transitions from $S$ to $L$ ($S_E\to L_E$ in the Exporter and $S_C\to L_C$ in the Importer).
    This is the truest measure of transmission incidence; $S\to L$ transitions cannot simply be pooled with $L\to D$ transitions without counting detected cases twice.
\end{enumerate}

\subsection{Further characterisation of outbreaks}
In order to characterise outbreaks, we also consider their peak, size, duration and frequency.
The \emph{peak of an outbreak} occurs when the transitions being monitored to determine the existence of an outbreak reach their maximum.
The \emph{size of an outbreak} is the total number of monitored transitions that occur during the outbreak; the \emph{duration of an outbreak} is the time between the first and last monitored transitions in the outbreak.
The \emph{time between two outbreaks} is the time elapsed between the last monitored transition of an outbreak and the first monitored transition of the following outbreak.
The \emph{frequency of outbreaks} is the inverse of the mean of the time between outbreaks over the period of study.

\section{A few more computational results}
\label{app:additional-computational-stuff}

\begin{figure}
    \centering
    \begin{subfigure}{0.49\textwidth}
        \includegraphics[width=\textwidth]{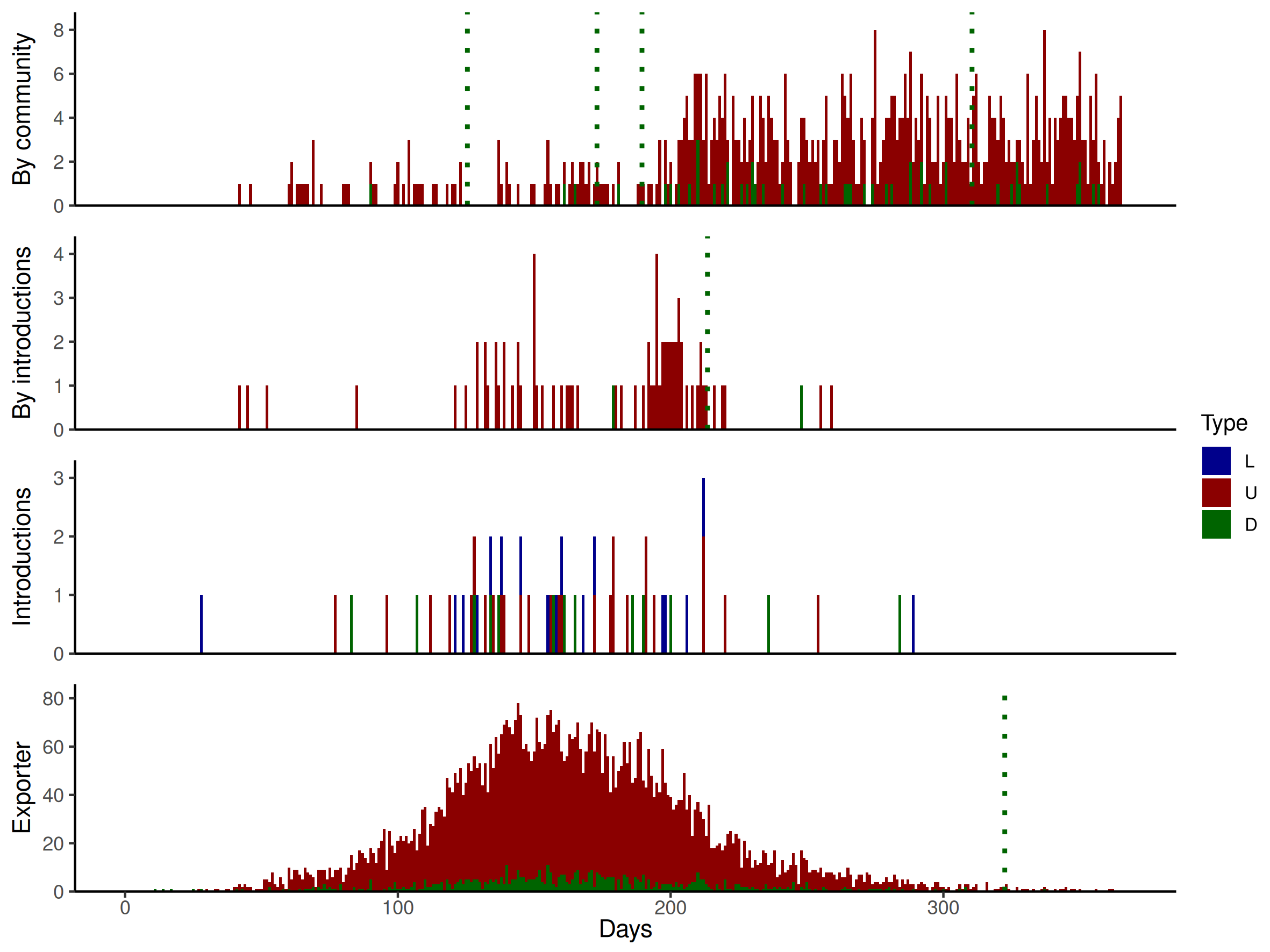}
        \caption{$\R_0^C=0.8$ and $m_X=0.0005$}
        \label{fig:transmission-scenario-R0E_15_R0C_08_mvt_00005_seed_84}
    \end{subfigure}
    \begin{subfigure}{0.49\textwidth}
        \includegraphics[width=\textwidth]{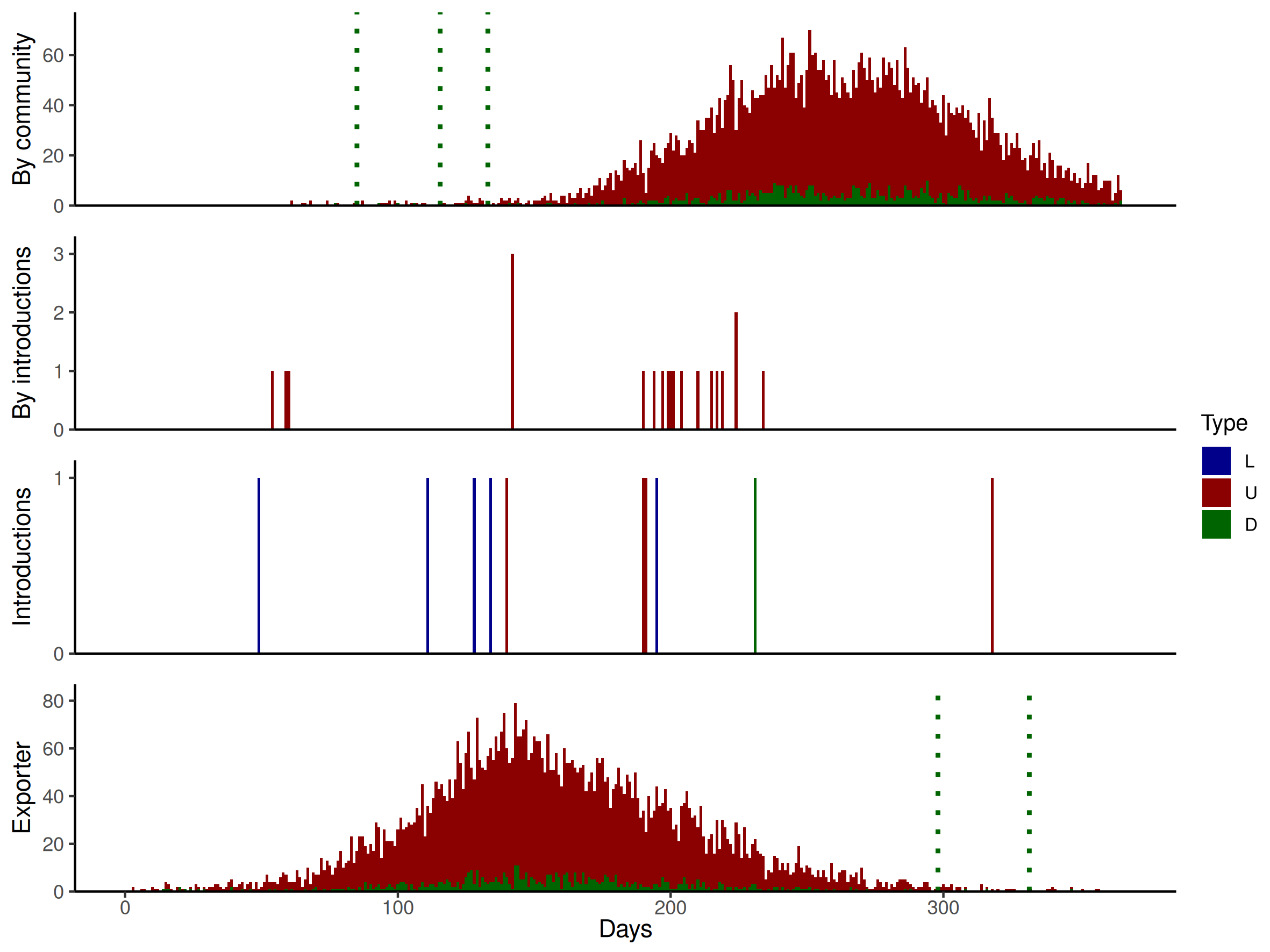}
        \caption{$\R_0^C=1.5$ and $m_X=0.0001$}
        \label{fig:transmission-scenario-R0E_15_R0C_15_mvt_00001_seed_18}
    \end{subfigure} \\
    \begin{subfigure}{0.49\textwidth}
        \includegraphics[width=\textwidth]{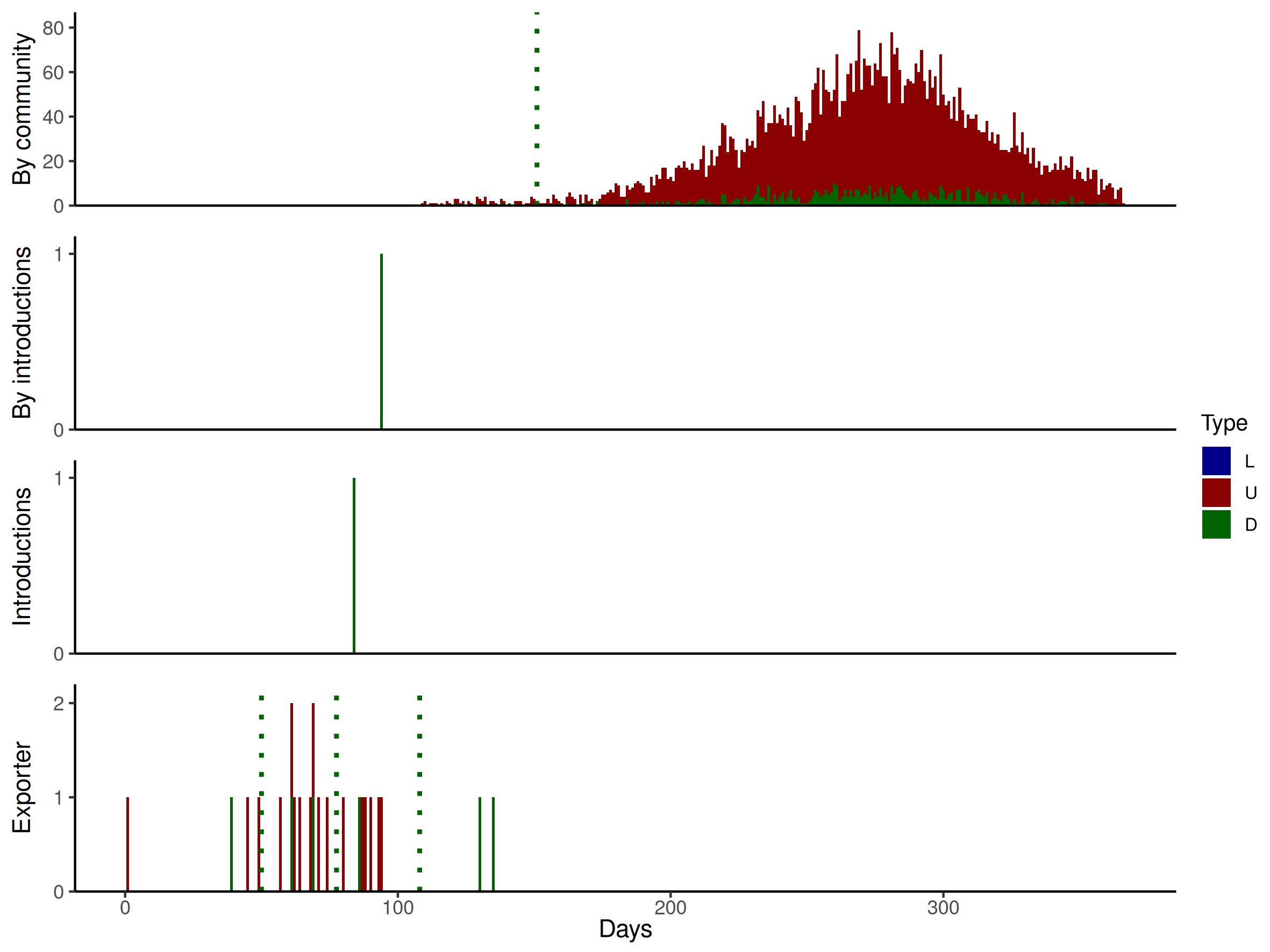}
        \caption{$\R_0^C=1.5$ and $m_X=0.001$.}
        \label{fig:transmission-scenario-R0E_15_R0C_15_mvt_0001_seed_96}
    \end{subfigure}
    \caption{Sample realisations of the full CTMC model with an initial total population of 10,000 in each location, $\R_0^E=1.5$ and (a) $\R_0^C=0.8$ and $m_X=0.0005$, (b) $\R_0^C=1.5$ and $m_X=0.0001$ and (c) $\R_0^C=1.5$ and $m_X=0.001$.}
    \label{fig:transmission-scenarios}
\end{figure}

\begin{figure}[htbp]
  \centering
  \begin{subfigure}[b]{0.48\textwidth}
    \centering
    \includegraphics[width=\linewidth]{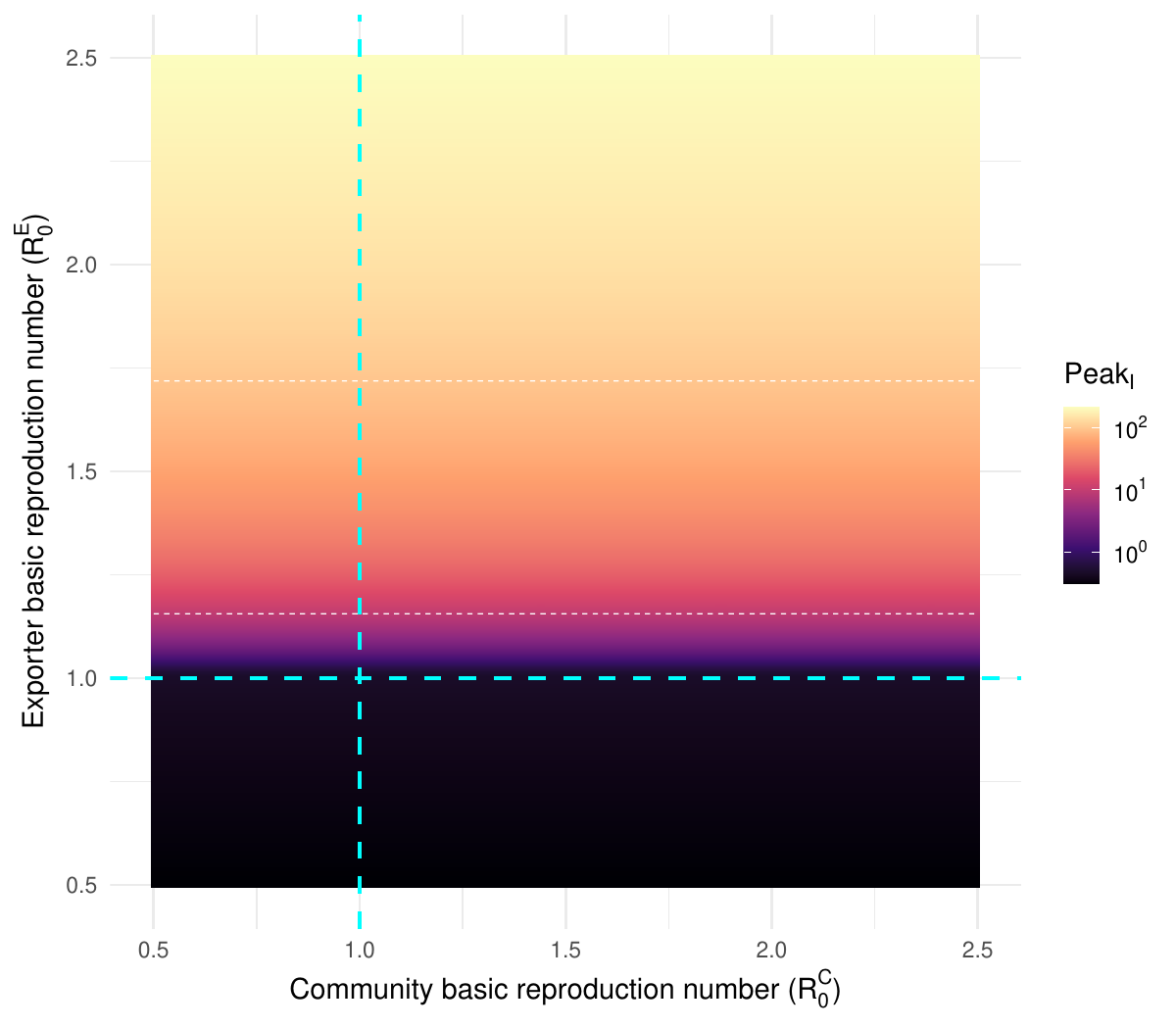}
    \caption{Introduction layer}
    \label{fig:ode-heatmap-r0-community}
  \end{subfigure}
  \hfill
  \begin{subfigure}[b]{0.48\textwidth}
    \centering
    \includegraphics[width=\linewidth]{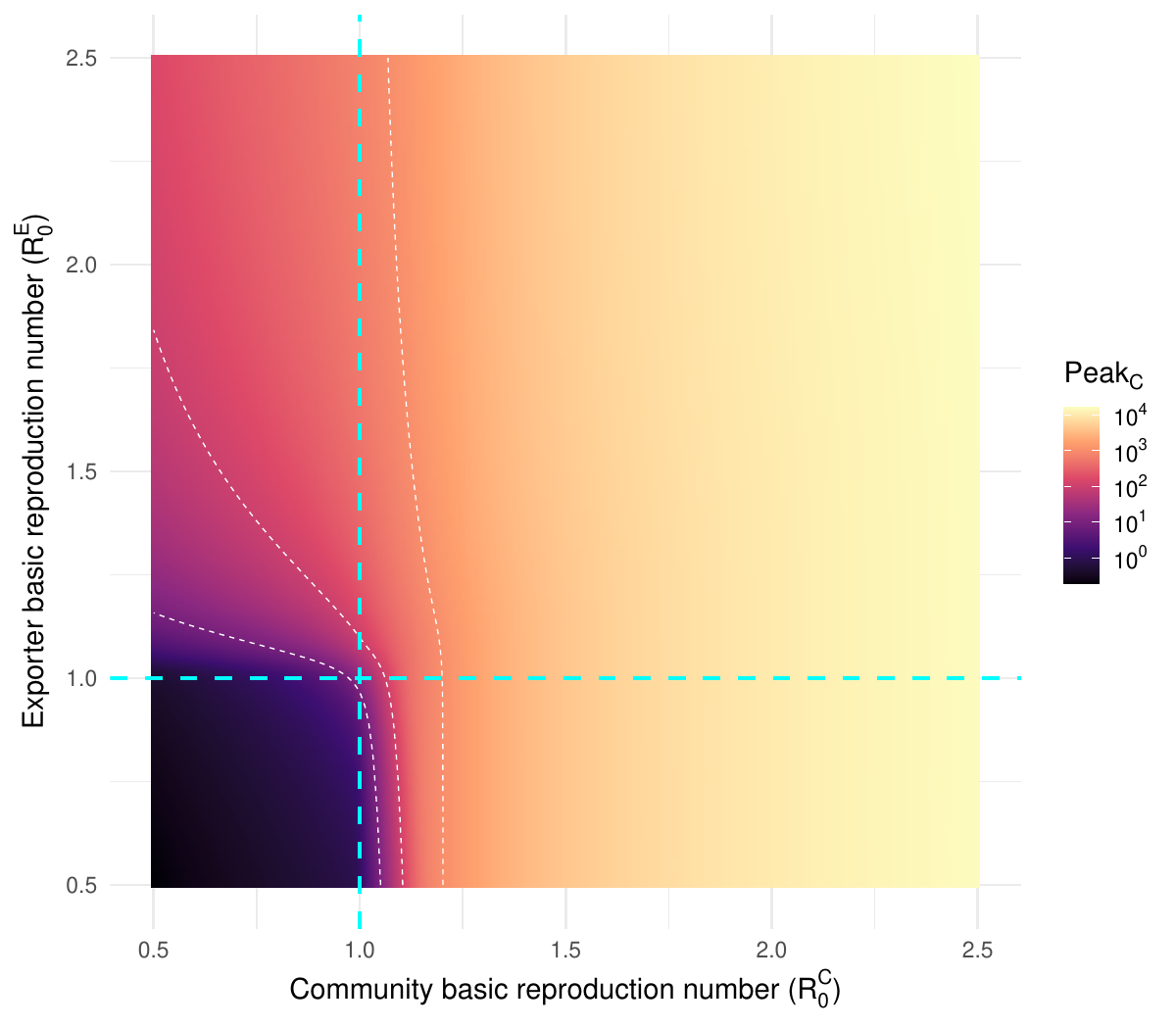}
    \caption{Community}
    \label{fig:ode-heatmap-r0-intro}
  \end{subfigure} \\
  \begin{subfigure}[b]{0.48\textwidth}
    \centering
    \includegraphics[width=\linewidth]{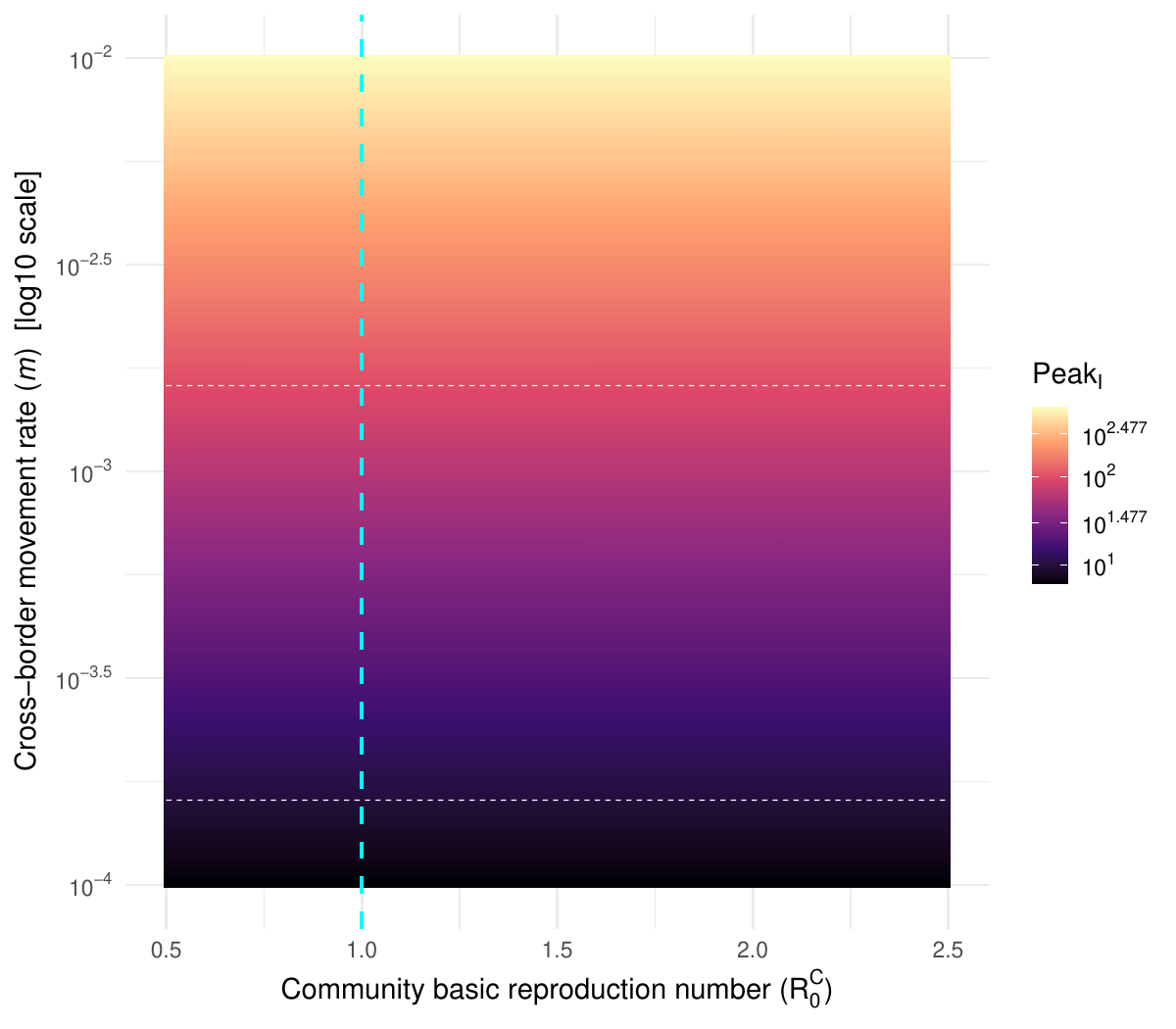}
    \caption{Introduction layer}
    \label{fig:ode-heatmap-mov-community}
  \end{subfigure}
  \hfill
  \begin{subfigure}[b]{0.48\textwidth}
    \centering
    \includegraphics[width=\linewidth]{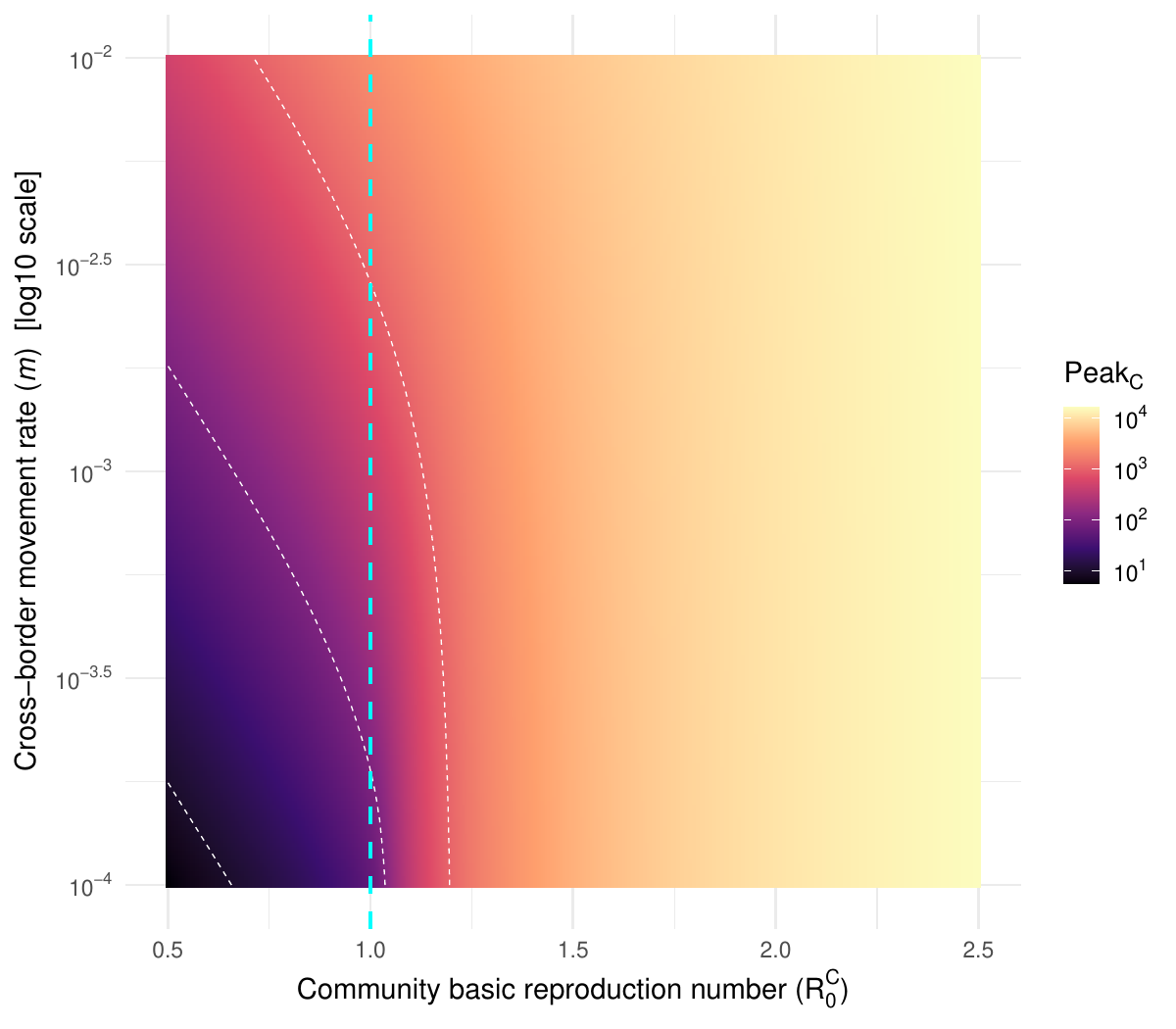}
    \caption{Community}
    \label{fig:ode-heatmap-mov-intro}
  \end{subfigure}
  \caption{Deterministic peak prevalence in the Importer as a function of $\R_0^C$ and (a,b) $\R_0^E$ and (c,d) $m_X$ in (a,c) the introduction layer and (b,d) the community.
  The deterministic model \eqref{sys:general-model} is used.}
  \label{fig:ode-heatmaps-peak-value-R0C-vs-R0E-movement}
\end{figure}

\begin{figure}[htbp]
    \centering
    \begin{subfigure}{0.49\textwidth}
        \includegraphics[width=\linewidth]{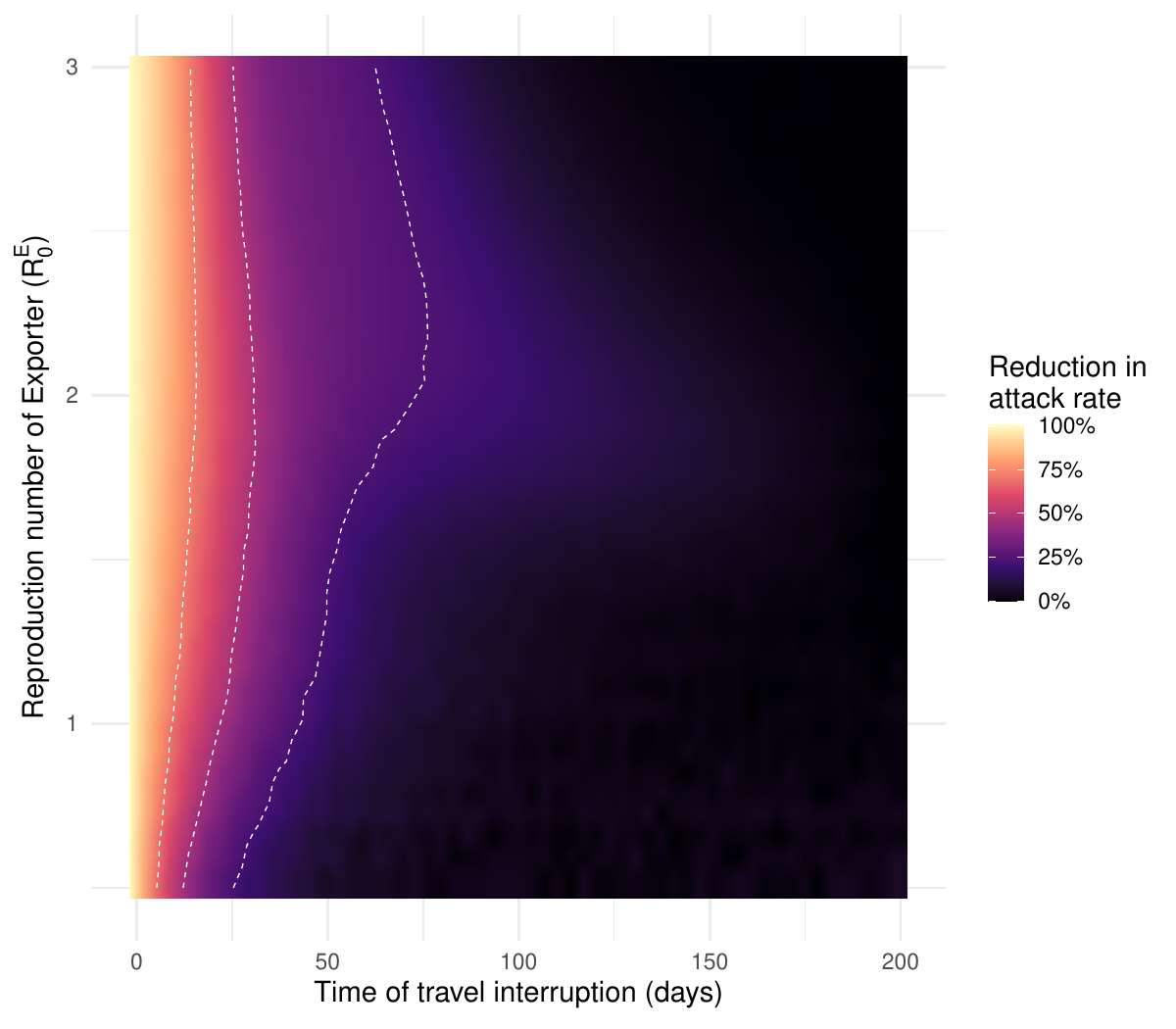}
        \caption{$\R_0^E$}
        \label{fig:interruption-travel-R0E}
    \end{subfigure}
    \begin{subfigure}{0.49\textwidth}
        \includegraphics[width=\linewidth]{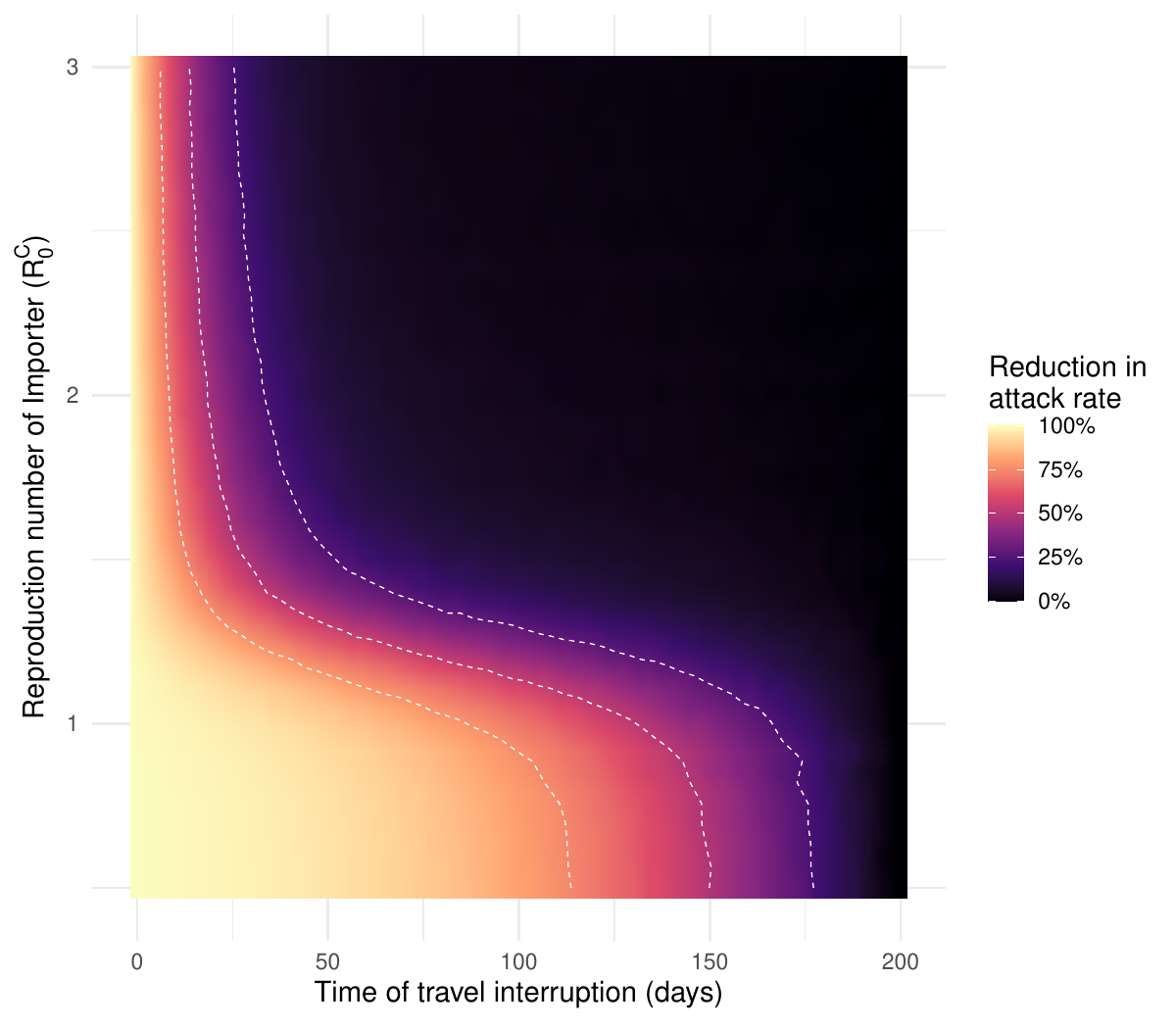}
        \caption{$\R_0^C$}
        \label{fig:interruption-travel-R0C}
    \end{subfigure}
    \caption{Relative percentage reduction in the mean attack rate in the Importer (10,000 CTMC simulations per grid point) as a function of the timing of total travel interruption and (a) $\R_0^E$ (with $\R_0^C=1.5$; baseline mean attack rate without travel restrictions increases from $21.3\%$ at $\R_0^E=0.5$ to $56.5\%$ at $\R_0^E\approx 1.5$ and $83.4\%$ at $\R_0^E=3.0$) and (b) $\R_0^C$ (with $\R_0^E=1.5$; baseline mean attack rate without travel restrictions increases from $3.3\%$ at $\R_0^C=0.5$ to $57.6\%$ at $\R_0^C\approx 1.5$ and $92.4\%$ at $\R_0^C=3.0$).
    Both locations have 10K individuals.}
    \label{fig:interruption-travel}
\end{figure}

This section collects additional numerical illustrations complementing Sections~\ref{sec:results-exporter-importer} and~\ref{sec:travel-control}: further sample realisations of the full CTMC model (Figure~\ref{fig:transmission-scenarios}), deterministic peak prevalence heatmaps in the Importer introduction layer and community (Figure~\ref{fig:ode-heatmaps-peak-value-R0C-vs-R0E-movement}) and the effect of interrupting travel at a fixed time after the start of spread in the Exporter (Figure~\ref{fig:interruption-travel}).

In Figure~\ref{fig:interruption-travel}, we use CTMC simulations to investigate the effect of a complete interruption of travel at a prescribed time after the start of the outbreak in the Exporter on the relative percentage reduction in the mean attack rate in the Importer.
In Figure~\ref{fig:interruption-travel-R0E}, we investigate the situation that prevails if the Importer is not doing a fantastic job controlling the disease ($\R_0^C=1.5$) and is subjected to varying conditions in the Exporter ($\R_0^E$).
As $\R_0^E$ increases, the epidemic in the Exporter grows faster and introductions into the Importer occur earlier, which narrows the time window during which a travel ban can achieve a substantial reduction in the Importer attack rate.
In Figure~\ref{fig:interruption-travel-R0C}, the Importer does have control over its reproduction number ($\R_0^C$) but not over the epidemic in the Exporter ($\R_0^E=1.5$).
When $\R_0^C \le 1$, local transmission chains cannot sustain themselves, so halting travel even relatively late still prevents a large fraction of the small importation-driven burden.
Conversely, as soon as $\R_0^C > 1$, delaying the travel interruption beyond the first few weeks allows introductions to seed local community spread, causing the relative reduction in the Importer attack rate to drop rapidly to zero.

\bibliographystyleapp{plain}
\bibliographyapp{references}

\end{document}